%% file: main.tex
\documentclass[12pt]{article}

\input{toggles}                              %

\input{preamble_files/preamble_main}         %
\input{preamble_files/preamble_notation}     %

\begin{document}

\maketitle %

\singlespacing

\vspace{-.5cm}

\begin{abstract}
\vspace{-.4cm}
\footnotesize

\noindent The missingness-at-random (MAR) assumption underlying many missing-data methods describes abstract statistical independences, whereas practitioners reason more naturally about the causal mechanisms that generate missingness. To bridge this gap, we provide a mechanistic characterization of MAR in terms of paths in a directed acyclic graph (DAG) that is modified to allow certain edges to cease to operate when an incomplete variable is missing, a property we call ``context-specific causal irrelevance (CSCI)''. Under a faithfulness condition, MAR holds if and only if the modified DAG contains no \textit{MAR-violating paths}. Without the allowance for CSCI, the latter condition reduces to a previously established d-separation condition, VMAR; thus, MAR without VMAR requires specific kinds of CSCI, as can occur with monotone dropout. These results shift the burden on investigators from passively invoking MAR to justifying, on substantive grounds, the specific mechanisms it requires.

\noindent We then consider identification of a target law $p(v)$ of variables $V$ by a MAR-assuming analysis, such as multiple imputation, that includes complete auxiliary variables $Z$ (i.e., variables not in $V$). Barring a perfect cancellation, such an analysis recovers $p(v)$ if {\it and only if} MAR holds with respect to $V$ and $Z$ jointly. We show that if there exists any set of auxiliaries such that $(V,Z)$ obeys MAR, then the set of complete candidate auxiliaries that are ancestors of the target variables or their missingness indicators is a sufficient set. Contrary to standard recommendations, not all potential auxiliary variables should be used in target-law identification: even complete auxiliaries can introduce MAR violations that would not otherwise have occurred.

\noindent We illustrate these results through simulations and a real-data analysis of sexual identity change in the Stockholm Public Health Cohort, where our results indicate that a candidate auxiliary variable, income, should be excluded; its inappropriate inclusion attenuated effect estimates toward the null.
\end{abstract}

\vspace{0.5cm}

\pagebreak

\bodyspacing

\section{Introduction}

The missingness-at-random (MAR) assumption underlies many common methods for handling missing data, such as multiple imputation \autocite{rubin2004multiple}, inverse-probability weighting (IPW),\autocite{robins1995analysis,wooldridge2007inverse} and full-information maximum likelihood \autocite{carpenter2021missing}. MAR states that for any given missingness pattern, the probability of that pattern's occurrence is conditionally independent of the variables that are missing for that pattern given the variables that are observed for that pattern.\footnote{There are multiple definitions of MAR; this definition is sometimes called \textit{everywhere-MAR} and is the version that establishes repeated-sampling properties of estimators, such as consistency \autocite{seaman2013meant}.} If this condition fails, the missingness mechanism is termed \textit{missing not at random} (MNAR). MAR and MNAR are defined relative to a particular set of variables $S$; when needed for clarity, we write $\MARS$ and $\MNARS$.

The MAR assumption refers to abstract statistical independences involving missingness patterns, making it difficult to interpret in practice.\autocite{seaman2013meant,mealli2015clarifying,chen2022pattern,ccvmi,doretti2018missing} An arguably more natural conceptualization would refer to the data-generating mechanism (DGM), that is, to the causal relationships among variables, including missingness indicators, that determine their joint distribution.\autocite{mohan2021graphical,ccvmi,lee2023assumptions,nabi2022causal} Indeed, among applied papers that discuss missingness assumptions at all, many do so in these mechanistic terms (e.g., reasons a clinician would order a certain test, or reasons participants would be lost to follow-up, such as moving home).\autocite{mainzer2024gap} Such DGMs can be conveniently represented using causal graphical models, such as directed acyclic graphs (DAGs). Such representations are useful even if the estimand of interest is not causal.

A DAG encodes independence restrictions that must hold for all joint distributions compatible with it via the d-separation criterion.\autocite{pearl2009causality} A sufficient graphical condition for $\MARS$ is $\VMARS$, defined as d-separation of all incomplete variables in $S$ from the missingness indicators of all incomplete variables in $S$, conditional on all complete variables in $S$. (A variable is \textit{complete} if it is observed with probability 1 in the DGM, and otherwise is \textit{incomplete}.) VMAR plays an important role in recent identification results for missing data.\autocite{nabi2022causal,mohan2021graphical}

However, not all forms of MAR satisfy VMAR. For example, monotone MAR is not necessarily VMAR, since incomplete variables can be associated with missingness indicators conditional on complete variables (Figure \ref{fig:monotone_MAR}). The independence restrictions required to have MAR without VMAR cannot be represented by d-separation in the DAG alone.\autocite{nabi2022causal} %
A natural question, then, is: What does MAR require of the causal processes that generate the variables, including missingness indicators? Relevant methodological work dates back to at least 1997, when Robins \& Gill described a broad class of DGMs that satisfy MAR.\autocite{robins1997non} This class, called \textit{randomized monotone MAR}, includes standard monotone MAR as a special case. However, to our knowledge, conditions on the causal DGM that are equivalent to MAR --- not merely sufficient for MAR --- remain unknown. Such conditions would be useful both epistemically, clarifying the meaning of MAR itself, and also practically, guiding reasoning about how credible MAR-assuming analyses are for a given study and its plausible missingness mechanisms.

The first part of this paper bridges the theoretical gap between VMAR and MAR. 
Under a faithfulness condition, we show that $\MARS$ holds if and \textit{only} if, for each incomplete variable $S^*$: (1) no path between $S^*$ and its own missingness indicator is unblocked conditional on the other variables in $S$; and (2) every path between $S^*$ and another variable's missingness indicator that is unblocked conditional on the other variables in $S$ ceases to operate when $S^*$ is missing. Condition (2) is a type of context-specific causal irrelevance (CSCI) that will be formalized below. Determinism among missingness indicators, as in monotone MAR, can give rise to CSCI such that condition (2) holds. Together, conditions (1) and (2) amount to the absence of what we will call MAR-violating paths; absent CSCI, they reduce to VMAR($S$).

Even when $\MARS$ fails, it is sometimes possible to identify the joint distribution $p(v)$ of a subset $V \subset S$. In this setting, we call $V$ the \textit{target variables} and $p(v)$ the \textit{target law}.\autocite{bhattacharya2020identification} 
The second part of this paper studies MAR-assuming target-law identification with \textit{auxiliary variables}, meaning variables disjoint from $V$ that are added to an imputation or IPW model in an effort to render MAR more plausible. The standard practice is to assume MAR holds with respect to $V$ and the chosen auxiliaries $Z$ jointly, use a corresponding MAR-assuming estimator, and then obtain $p(v)$ by marginalizing over $Z$. We address the question: for which data-generating mechanisms does such an analysis recover $p(v)$? We focus on complete auxiliaries, returning to incomplete auxiliaries in the Discussion. The auxiliaries are chosen from a set of candidate auxiliary variables, meaning non-latent variables that are neither target variables nor missingness indicators. (A variable is latent if it is observed with probability zero.) Existing recommendations vary: some methodologists advocate including all candidate auxiliary variables \autocite{little2012prevention,perkins2018principled,collins2001comparison}, while others recommend including only complete variables associated with missingness \autocite{carpenter2021missing}. However, either approach can introduce bias that is avoidable with a suitable choice of auxiliary variables. The few existing papers that have studied this problem provide compelling examples for specific DGMs.\autocite{curnow2023multiple,thoemmes2015graphical,thoemmes2014cautious}

Recovering $p(v)$ is in principle less demanding than recovering the joint law of $V$ and $Z$. However, we show that with complete auxiliaries, and barring certain perfect cancellations, the MAR-assuming analysis recovers $p(v)$ if and only if MAR($V \cup Z$) holds. Thus, the mechanistic characterization from the first part of the paper applies directly: under restrictions on CSCI, the analysis recovers $p(v)$ if and only if every path between an incomplete target variable and a missingness indicator of an incomplete target variable is blocked conditional on the complete target variables and the auxiliaries.

This characterization enables explicit construction of a \textit{sufficient set of auxiliary variables}, meaning a set $Z$ such that $(V, Z)$ obeys MAR. Whenever at least one set of candidate auxiliaries, complete or incomplete, is sufficient, then barring certain forms of CSCI, one such set is the complete candidates that causally affect at least one target variable or its missingness. Identifying this set requires only limited knowledge of the data-generating mechanism. A practical consequence is that, for target-law identification, it is never necessary, and in fact can be harmful, to include complete auxiliaries that are only non-causally associated with the target variables or their missingness indicators. (This paper focuses on consistency of estimators; auxiliary variables are, of course, also sometimes included to improve efficiency.)

This paper is organized as follows. Section \ref{sec:prelims} establishes notation, reviews DAGs, and formalizes CSCI. Section \ref{sec:equivalent_conditions_unrestricted} establishes the main results regarding MNAR-equivalent mechanistic conditions. Section \ref{sec:auxiliaries} addresses identification using complete auxiliary variables. Those results are illustrated through example DAGs and simulations (Supplement, Section \ref{supp-sec:sim_study}), which demonstrate the bias that can result from including all auxiliary variables in the imputation model per standard recommendations,\autocite{little2012prevention,perkins2018principled,collins2001comparison} and through a real-data analysis of changes in sexual identity over time in members of the Stockholm Public Health Cohort, which shows how our results can help identify auxiliary variables that mitigate, rather than introduce, bias.

\section{Preliminaries}
\label{sec:prelims}

We first fix notation and describe the DAGs used to represent missing-data mechanisms. We then review standard definitions of MAR and MNAR and formalize CSCI.

\subsection{Notation}

Table \ref{tb:notation} summarizes all key notation. Given a set $V$ of \textit{target variables}, the \textit{target law} is $p(v)$. We use $p(\cdot)$ generically for densities and probability mass functions. Let $W$ be a set of \textit{candidate auxiliary variables}, meaning all non-latent variables (other than missingness indicators, discussed below) that are not in $V$.  We will discuss missingness at random with respect to $V \cup W$ or arbitrary subsets thereof. $W$ may be empty; much of the classical missing-data literature focuses on this case \autocite{rubin1976inference}. We refer to $L := V \cup W$ as the \textit{substantive variables}; that is, all non-latent variables other than missingness indicators. Each member $L_j \in L$ denotes the underlying value of that variable, which is well-defined regardless of whether the analyst observes its value; the analyst instead observes a proxy that equals $L_j$ when $L_j$ is observed and is missing otherwise.\footnote{Some authors use distinct notation: $L_j^{(1)}$ for the underlying variable and $L_j$ for its proxy.\autocite{nabi2020full} We elide this distinction for brevity.} The terms \textit{complete} and \textit{incomplete} refer to these proxies: $L_j$ is \textit{complete} if its proxy is observed with probability 1, and otherwise is \textit{incomplete}. 

Given any subset $S \subseteq L$, let $S^+$ and $S^-$ be, respectively, the sets of complete and incomplete variables in $S$. (The set of complete variables may be empty.) The set of missingness indicators for incomplete variables in $S$ is $R(S) := \{ R_{S_j} : S_j \in S^- \}$, where $R_{S_j} := 1$ if $S_j$ is observed and $R_{S_j} := 0$ if it is missing. Note that complete variables do not have missingness indicators. Values of $R(S)$ are called \textit{missingness patterns}; specifically, $R(S) = 1$ will denote the complete-case pattern for $S$. Let $\Smis(r)$ and $\Sobs(r)$ refer to the sets of variables in $S$ that are, respectively, missing and observed for pattern $R(S) = r$, and let $\smis(r)$ and $\sobs(r)$ be specific values of $\Smis(r)$ and $\Sobs(r)$. 

We also make weak regularity assumptions ruling out certain perfect cancellations in the distribution of $(L, R(L))$ in which MNAR-induced errors in conditionals involving the MAR functional perfectly cancel with one another (formalized in the Supplement, Section \ref{supp-sec:regularity_assumptions}). Such cancellations are distinct from faithfulness violations. We also make a basic positivity assumption: 

\begin{assumption}[Complete-case positivity]
\label{ass:cc_positivity}
The complete-case pattern occurs with positive probability conditional on any value of the substantive variables, i.e., $p( R(L) = 1 \mid l ) > 0$ for all $l \in \supp(L)$.
\end{assumption}

This assumption is weaker than requiring every pattern to occur with positive probability given every value of $L$, and it permits deterministic relationships among the missingness indicators, as in monotone missingness.

Additionally, certain results will invoke the following weakened version of the intersection property; it is weaker in that, for a given $S \subseteq L$, it only needs to hold when the conditioning variables are members of $S$, not $R(L)$.

\begin{assumption}[Weak intersection with respect to $S$]
\label{ass:weak_intersection}
Let $S \subseteq L$. For all disjoint sets $A \subseteq (S \cup R(L))$, $B \subseteq S$, and $C \subseteq S$, and every value $a$ of $A$, the following holds. For any $B_i \in B$, let $B_{-i} := B \setminus \{B_i\}$. If, for each $B_i \in B$, $p(a \mid b, c) = p(a \mid b_{-i}, c)$ for all values $(b, c)$ of $(B, C)$, then $p(a \mid b, c) = p(a \mid c)$ for all $(b, c)$. 
\end{assumption}

A sufficient but not necessary condition for weak intersection is that the joint support of the variables in $S$ is the product of their marginal supports: $\supp(S) \; = \; \prod_{S_k \in S} \supp(S_k)$, where, for continuous variables, the support of a distribution is the set on which its density is positive. That sufficient condition rules out deterministic relationships among the members of $S$, but does not impose any constraints on $R(L)$ (Supplement, Lemma \ref{supp-lm:product_support_implies_weak_intersection}).

\begin{table}[!htbp]
\setstretch{1}
\resizebox{0.97\columnwidth}{!}{%
\begin{tabular}{@{}ll@{}}
\toprule
\textbf{Notation} & \textbf{Definition} \\ \midrule
\multicolumn{2}{@{}l}{\textit{Variables}} \\
$V$ & Target variables \\

$W$ & Candidate auxiliary variables \\

$Z \subseteq W$ & Auxiliary variables (i.e., candidates included in analysis), comprising complete $Z^+$ and incomplete $Z^-$ \\

$L := V \cup W$ & Substantive variables (all non-latent variables other than missingness indicators) \\

$U$ & Latent common causes of members of $L \cup R(L)$ \\

$S \subseteq L$ & Generic subset of substantive variables \\

$S^+,\, S^-$ & Complete and incomplete variables, respectively, in a given set $S$ \\

$R(S) := \{R_{S_j} : S_j \in S^-\}$ & Missingness indicators for incomplete variables in $S$ \\

$\Sobs(r),\, \Smis(r)$ & The subsets of $S$ that are observed and missing, respectively, for pattern $R(S) = r$ \\

\midrule
\multicolumn{2}{@{}l}{\textit{Graphs and models}} \\
$G$ & DAG on $L \cup R(L) \cup U$ \\

$\m$ & A structural causal model compatible with $G$ \\

$\An_G(A)$ & Ancestors of $A$ in $G$, including $A$ itself \\

$G[d]$ & Context-pruned DAG for context value $d$ (Definition \ref{def:pruned_DAG}) \\

$G(A)$ & Latent projection of $G$ onto the nodes $A$, i.e., marginalizing over all other nodes (Section \ref{sec:connections_other_lit}) \\

\midrule
\multicolumn{2}{@{}l}{\textit{Conditions}} \\
$\MARS$ in $\m$ & $1\{ R(S) = r \} \indM \Smis(r) \mid \Sobs(r)$ for every pattern $r$ \\

$\VMARS$ in $G$ & $S^- \ind_G R(S) \mid S^+$ \\

VMAR-violating path & Path between $S^-$ and $R(S)$ that is unblocked conditional on $S^+$ \\

\midrule
\multicolumn{2}{@{}l}{\textit{Functionals and auxiliary sets}} \\
$\pMAR{S}(s)$ & MAR-assuming functional for $p(s)$ (Section \ref{sec:MAR_MNAR}) \\

$\pMAR{V \cup Z}(v)$ & Marginal for $V$ of $\pMAR{V \cup Z}(v, z)$; the target of MAR-assuming analyses using auxiliaries $Z$ \\

$\Z := W^+ \cap \An_G(V \cup R(V))$ & Sufficient auxiliary set, when one exists (Theorem \ref{thm:suff_set_general}) \\
\bottomrule
\end{tabular}
}
\caption{\label{tb:notation} List of key notation.}
\end{table}

\subsection{DAGs representing missing data}
\label{sec:dags}

\newcommand{\SCMFn}{}

A class of DGMs for $L \cup R(L)$ can be represented by a causal DAG denoted $G$. The set of nodes in $G$ comprises $L$, $R(L)$, and any variables $U$ that are latent common causes of members of $L \cup R(L)$ \autocite{nabi2022causal,mohan2021graphical}. Examples of such DAGs appear in Figure \ref{fig:VMAR_VMNAR_examples}, discussed further in Section \ref{sec:equivalent_conditions_unrestricted} below. As usual in causal DAGs, directed edges ($\to$ and $\gets$) represent potential direct causal relationships, which may be deterministic. The \textit{parents} of a variable $A$, denoted $\Pa(A)$, are the variables that have a directed edge into $A$; lowercase $\pa(A)$ denotes a specific value of $\Pa(A)$. The \textit{ancestors} of a variable $A$, denoted $\An_G(A)$, are the variables that have a directed path of any length to $A$; by convention, $\An_G(A)$ includes $A$.

A given data-generating mechanism is represented by a \textit{structural causal model} (SCM).\autocite{pearl2009causality}\SCMFn  In a given SCM $\m$, any given variable $A$ is generated from the structural equation $f_{A}( \pa(A), \epsilon_{A} )$, where $\epsilon_A$ is exogenous random variation; the exogenous errors are mutually independent across variables. The collection of these structural equations induces a joint distribution, $p(l, r, u)$, over $L \cup R(L) \cup U$. We use $\indM$ to denote statistical independence in this induced distribution. We assume throughout that $(L^i, R(L)^i, U^i)$ for $i \in \{ 1, \cdots, n \}$ are independent and identically distributed (iid) samples of $(L, R(L), U)$. 

A given path between two nodes $A$ and $B$ is said to be \textit{blocked} by a set of nodes $C$ (disjoint from $A$ and $B$) if the path includes at least one of the following structures: (1) a chain $N_i \to C_j \to N_k$ or fork $N_i \gets C_j \to N_k$, where $C_j \in C$ and $(N_i, N_k)$ are any other nodes; or (2) a collider  $N_i \to N_j \gets N_k$, where neither $N_j$ nor any of its descendants is in the conditioning set $C$. Given disjoint sets of nodes $A, B, C$, the sets $A$ and $B$ are d-separated by $C$ if and only if for every pair of nodes $A_i, B_j$ such that $A_i \in A$ and $B_j \in B$, the set $C$ blocks every path in $G$ between $A_i$ and $B_j$. %
We denote this as $A \ind_G B \mid C$. (For details, see Section 1.2.3 of reference \cite{pearl2009causality}.) An SCM is \textit{compatible} with $G$ if
a structural equation for each $A$ has the form $f_A(\Pa_G(A), \epsilon_A)$, and furthermore the set of errors $\epsilon_A$ for each $A$ is mutually independent. Compatible SCMs are Markov with respect to $G$, i.e., $A \ind_G B \mid C$ implies that $A \indM B \mid C$.\autocite[Theorem 1.4.1]{pearl2009causality} However, in some compatible SCMs, determinism (e.g., monotone missingness) precludes the occurrence of one or more patterns $r \in \{ 0, 1 \}^{|R(L)|}$.

\subsection{Standard definitions of MAR and MNAR}
\label{sec:MAR_MNAR}

Given a set of variables $S \subseteq L$ and an SCM $\m$ on $L \cup R(L) \cup U$, MAR and MNAR are defined as follows:
\begin{definition}[MAR and MNAR]
\label{def:eMAR}
In $\m$, \textit{missingness at random (MAR)} holds with respect to $S$ if, for every missingness pattern $r$ such that $p( R(S) = r ) > 0$, we have:
$$1\{ R(S) = r \} \indM \Smis(r) \mid \Sobs(r).$$
Otherwise, missing not at random (MNAR) holds with respect to $S$. We use the shorthand $\MARS$ and MNAR($S$). 
\end{definition}

Several definitions of MAR appear in the literature; this one is sometimes called \textit{everywhere-MAR} or \textit{missing always at random} \autocite{seaman2013meant,mealli2015clarifying}. Everywhere-MAR is stronger than Rubin's original definition of MAR, but everywhere-MAR is the relevant condition to justify repeated-sampling properties such as consistency.\autocite{seaman2013meant} Throughout this paper, ``$\text{MAR}$'' refers to everywhere-MAR. If $\MARS$ holds, then the joint distribution of $S$ can be written as:
\begingroup
\addtolength{\jot}{0.5em}
\small
\begin{align}
\label{eq:mar_ipw_true}
p(s) & = \frac{p(s, R(S) = 1)}{p(R(S) = 1 \mid s)} = \frac{p(R(S) = 1) \, p(s \mid R(S) = 1)}{ 1 - \sum_{r \neq 1} p(R(S) = r \mid s) } = \frac{p(R(S) = 1) \, p(s \mid R(S) = 1)}{ 1 - \sum_{r \neq 1} p\left(R(S) = r \mid \sobs(r)\right) }.
\end{align}
\endgroup
The last equality invokes $\MARS$ to drop the conditioning on $\smis(r)$. The numerator is an observed-data quantity; under $\MARS$, the pattern probabilities in the denominator are also identified.\autocite{robins1994estimation,sun2018inverse} 

Under Assumption \ref{ass:cc_positivity}, at most one law of $(S, R(S))$ satisfying $\MARS$ implies the observed-data distribution (i.e., the distribution of $R(S)$ and the observed components of $S$).\autocite{robins1994estimation} This law, called $\pMAR{S}(s, r)$, is guaranteed to exist under mild regularity conditions (E. Cator, personal communication, Oct. 20, 2026).\autocite{cator2004testability} The \textit{MAR-assuming functional for $p(s)$} is defined as the marginal distribution $\pMAR{S}(s)$ that is implied by $\pMAR{S}(s, r)$. When $\MARS$ holds, $\pMAR{S}(s)$ is \textit{correct}, i.e., it is equal to $p(s)$. Regardless of whether $\MARS$ holds, $\pMAR{S}(s)$ is the probability limit of all estimators of $p(s)$ that (possibly incorrectly) assume $\MARS$ and impose no restrictions beyond $\MARS$ (or that impose only restrictions that $\pMAR{S}(s, r)$ satisfies, such as parametric restrictions). That all such estimators share this limit follows from the uniqueness result above: each such estimator converges to a law that satisfies $\MARS$ and implies the observed-data distribution of $(S, R(S))$, and $\pMAR{S}(s, r)$ is the only such law.

\subsection{Context-specific causal irrelevance}

To establish a mechanistic characterization of MAR, we will need the following formalization of context-specific causal irrelevance (CSCI). Examples will be provided at the end of this section. Heuristically, CSCI occurs when an edge in a DAG, say $A \to B$, ceases to operate when certain other variables take particular values. Throughout this section, we treat the DAG $G$ and compatible SCM $\m$ as fixed, and variables are taken to be nodes in $G$. 

\newcommand{\cinactivityFn}{ }

\begin{definition}[c-inactivity]
\label{def:c_inactive}
Let $D$ be a set of variables, called \textit{context variables}, and let $d$ be a possible value of $D$, called a \textit{context value}, with $p(D = d) > 0$. For a node $B \notin \An_G(D)$,\cinactivityFn let $\mathcal A \subseteq \Pa(B)$ be nonempty. The parent subset $\mathcal A$ is \textit{c-inactive in context $d$}, where the prefix ``c'' stands for ``contextually'', if the structural equation $f_B$ is functionally independent of $\mathcal{A}$ over the values of $B$'s parents that are possible in context $d$, holding $B$'s other parents fixed: $f_B(\pa(B), \epsilon_B) = f_B(\pa'(B), \epsilon_B)$ for all $\pa(B), \pa'(B) \in \supp\left( \Pa(B) \mid D = d \right)$ that agree on $\Pa(B) \setminus \mathcal{A}$, and all $\epsilon_B$. We call the edge $A \to B$ \textit{c-inactive in context $d$} if $\{A\}$ is c-inactive in context $d$, and \textit{c-active} otherwise. 
\end{definition}

Note that c-inactivity is a property of the SCM, not its induced distribution: even if $\mathcal{A}$ is c-inactive in context $d$, $B$ may remain statistically dependent on $\mathcal{A}$ given $D = d$, for example if $\mathcal{A}$ is associated with another parent of $B$ or shares a common cause with $B$. More generally, $A \to B$ is c-inactive in context $d$ only if the context confines $A$, $B$'s other parents, or both to a range of values over which $f_B$ does not vary with $A$, either directly, when the context variables are themselves among $B$’s parents, or through a chain of partially deterministic relationships.\autocite{csci} Following the labeled-DAG literature,\autocite{pensar2015labeled,boutilier1996context} we call the set of contexts in which an edge is c-inactive its \textit{label}.
\vspace{0.4cm}
\input{figures/c_inactivity_dags}
\vspace{-0.2cm}

Figure \ref{fig:c_inactivity_dags} shows some mechanisms by which c-inactivity can arise. The structure in panel (a) could arise in a study investigating associations between a terminally ill patient's access to palliative care ($A$) and whether they opt for an assisted suicide ($B$). The study is conducted in the United States, where assisted suicide is only legal in some states (indicated by $D=1$). For simplicity, there are no missing data. In the context $D=0$, assisted suicide is illegal, so $B$ is deterministically 0. Thus $f_B(\cdot)$ is not functionally dependent on $A$ or any other variables. However, for patients with $D=1$, their access to palliative care might affect whether they opt for an assisted suicide, in which case $f_B(\cdot)$ is functionally dependent on $A$. In summary, the edge $A \to B$ is c-inactive in the context $D=0$. 

\newcommand{\boutilierFn}{\footnote{The definitions in this section differ from existing graph-based formalizations of context-specific independence,\autocite{boutilier1996context,pensar2015labeled} which concerned statistical independence conditional on the context variables. In contrast, ours concerns causal irrelevance and does not require context variables to be co-parents with $\mathcal A$. While these generalizations will be needed to establish our main results, in many simple settings our formalization coincides with previous ones.} }

We write $\pd(\cdot) := p(\cdot \mid D = d)$ for the distribution in context $d$ that is implied by $p(\cdot)$. The following pruning operation modifies $G$ into a context-specific DAG to which $\pd$ is Markov in the following sense: for disjoint sets of nodes $X$, $Y$ and $C$ (where $C$ may be empty), $X \ind_{G[d]} Y \mid C \cup D$ implies $X \indM Y \mid (C, D = d)$.\boutilierFn This Markov property is established in Lemma \ref{supp-lm:pruning_soundness} (Supplement).
\begin{definition}[Pruned DAG]
\label{def:pruned_DAG}
Fix a DAG $G$ whose set of nodes is $N$, a nonempty set of context variables $D \subseteq N$, and a context value $d$ with $p(D = d) > 0$. For each node $B \in N \setminus \An_G(D)$, let $\mathcal{A}_B$ be a maximal c-inactive subset of $\Pa(B) \setminus D$ in context $d$, taking $\mathcal{A}_B := \emptyset$ if no such subset is c-inactive; for each $B \in \An_G(D)$, let $\mathcal{A}_B := \emptyset$. The \textit{context-pruned DAG}, denoted $G[d]$, is the subgraph of $G$ obtained by deleting, for every node $B$, all edges from members of $\mathcal{A}_B$ to $B$. Thus $\pa_{G[d]}(B) = \pa_G(B) \setminus \mathcal{A}_B$. When needed for clarity, we write $G[\{D = d\}]$ for $G[d]$. For the empty set of context variables, we define $G[\emptyset] := G$.
\end{definition}

\newcommand{\contextEdgesFn}{\footnote{Thus, edges out of context variables are never deleted. An alternative convention deletes them, since in a point-valued context they are trivially c-inactive. The two conventions imply the same d-separations conditional on any set containing $D$, which includes every conditioning set used here: a path that uses an edge out of a member of $D$ has that member as a non-collider, so conditioning on $D$ already blocks it.} }%
A maximal c-inactive set need not be unique; $G[d]$ denotes the result of any fixed choice used throughout, and Lemma \ref{supp-lm:pruning_soundness} (Supplement) holds for every maximal c-inactive set.\contextEdgesFn The converse of the Markov property is an assumption:
\begin{definition}[c-faithfulness]
\label{def:c_faithfulness}
For every set of context variables $D$, every context value $d$ with $p(D = d) > 0$, every pair of nodes $X \in N$ and $Y \in N$, every set of nodes $C \supseteq D$ disjoint from $\{X, Y\}$, and every $\ct \in \supp(C)$ that agrees with $d$ on $D$, the following implication holds:
$$X \notind_{G[d]} Y \mid C \quad \Rightarrow \quad X \notindM Y \mid (C = \ct).$$
\end{definition}
\newcommand{\faithfulnessFn}{\footnote{C-faithfulness implies faithfulness to $G$: it must hold for every choice of context variables $D$, and choosing $D = \emptyset$ yields faithfulness to $G[\emptyset] = G$. However, faithfulness to $G$ does not imply c-faithfulness. As an example, consider $G$ comprising the paths $C \gets A \to B \to C$ and $D \to B$, and suppose the effects of $A$ and $D$ on $B$ interact such that, when $D=d$, the paths $A \to B \to C$ and $A \to C$ exactly cancel. Then $A \indM C \mid D=d$, violating c-faithfulness because $A \notind_{G[d]} C \mid D$; however, for this pair, faithfulness to $G$ requires only $A \notindM C \mid D$, which holds as long as the cancellation fails for some other value of $D$.} }

This assumption will be invoked in only some results, which will be explicitly stated. C-faithfulness is a form of faithfulness, a standard assumption in the causal graph literature.\autocite{pearl2009causality}\faithfulnessFn Violations of c-faithfulness can occur if, for example, there are perfect cancellations of effects corresponding to multiple c-active paths.

In this paper, the relevant form of CSCI is that involving missingness indicators in causal DAGs. For example, consider a longitudinal study in which the measurements of the outcome, $L_1$, $L_2$, and $L_3$, occur at three time points and may be missing (Figure \ref{fig:monotone_MAR}). Here, $L_1$ and $L_2$ are both marginally associated with one or more of the missingness indicators. However, the missingness pattern is monotone: if $L_1$ is missing, then the participant has permanently dropped out of the study, so $L_2$ is also missing; if instead $L_1$ is observed, then $L_2$ may or may not be missing. That is, in the context $R_{L_1} = 0$, both $R_{L_2}$ and $R_{L_3}$ are forced to 0, so the edges $L_1 \to R_{L_2}$, $L_1 \to R_{L_3}$, $L_2 \to R_{L_3}$, and $R_{L_2} \to R_{L_3}$ are c-inactive. The pruned DAG $G[\{R_{L_1} = 0\}]$ is depicted in Figure \ref{fig:monotone_MAR}(b). Since $L_1$ is d-separated from $\{ R_{L_2}, R_{L_3} \}$ given $R_{L_1}$ in $G[\{R_{L_1} = 0\}]$, the corresponding contextual statistical independence holds by Lemma \ref{supp-lm:pruning_soundness} (Supplement): $L_1 \indM \{ R_{L_2}, R_{L_3} \} \mid (R_{L_1} = 0)$. By similar arguments, $L_2 \indM R_{L_3} \mid (R_{L_2} = 0)$ (Figure \ref{fig:monotone_MAR}(c)). Thus, the form of CSCI in this example produces MAR, a result that will be generalized to arbitrary DGMs in Section \ref{sec:equivalent_conditions_unrestricted}.

\input{figures/tikz_monotone_MAR}

\section{MNAR-equivalent mechanistic conditions}
\label{sec:equivalent_conditions_unrestricted}

We first review VMAR, an established graphical sufficient condition for MAR, and illustrate why it is not necessary. We then generalize VMAR-violating paths to MAR-violating paths, which account for CSCI, and show that under weak intersection and c-faithfulness, MAR holds if and only if there are no MAR-violating paths (Theorem \ref{thm:mnar_equivalent}). Finally, we relate this result to existing work on full-law identification and variable dependence.

\noindent A previously established graphical sufficient condition for MAR is as follows.\autocite{mohan2021graphical,tian2015missing,nabi2022causal}
\begin{definition}%
\label{def:VMAR}
\label{def:vmar_violating_path}
\textit{VMAR with respect to $S$} holds in $G$ if $S^- \ind_G R(S) \mid S^+$; this is denoted $\VMARS$. Otherwise, \textit{VMNAR with respect to $S$} holds in $G$; this is denoted VMNAR($S$). A path between any member of $S^-$ and any member of $R(S)$ that is unblocked conditional on $S^+$ is called a \textit{VMAR-violating path}. Thus, by definition, VMNAR($S$) is equivalent to the existence of a VMAR-violating path.
\end{definition}

\input{figures/tikz_vmar_vmnar}

For example, in Figure \ref{fig:VMAR_VMNAR_examples}(a), if only the solid black edges are present, then VMAR( $\{ L_1, L_2, L_3 \}$ ) holds because a complete member of $S$, namely $L_1$, blocks all paths between incomplete members of $S$ and their missingness indicators. If the red dotted edges are also present, they create VMAR-violating paths, so VMNAR( $\{ L_1, L_2, L_3 \}$ ) holds. In panel (b), the red dotted edges form a path between an incomplete variable ($L_1$) and a missingness indicator ($R_{L_1}$) that is marginally blocked, but is unblocked conditional on the complete variable $L_2$. Thus, VMAR($\{ L_1, L_2 \}$) is violated, yet VMAR($\{ L_1 \}$) holds. More generally, Proposition \ref{supp-prop:minimal_vmar_paths} (Supplement) characterizes the necessary structure of VMAR-violating paths.

Although VMAR implies MAR, the converse is not true in general, even if restricted to structural causal models: under certain forms of CSCI, MAR can hold despite VMNAR, as in the monotone example of Figure \ref{fig:monotone_MAR}. Thus, a graphical equivalent to MAR requires generalizing VMAR-violating paths to account for CSCIs whose independence implications are obtained via d-separation in pruned DAGs. This motivates our definition (see below) of MAR-violating paths. For a path between an incomplete variable $S^*\in S^-$ and its own missingness indicator $R_{S^*}$, the definition refers to the original DAG $G$. However, for a path between an incomplete variable $S^* \in S^-$ and the missingness indicator of another incomplete variable, the definition instead refers to a pruned DAG obtained in the context  $R_{S^*}=0$.
\begin{definition}[MAR-violating paths]
\label{def:mar_violating_path}
For $S^* \in S^-$, let $S^c := S \setminus \{S^*\}$. A \textit{MAR-violating path} is a path of either of the following types:
\begin{itemize}
\item \textit{Own-indicator case:} A path in $G$ between a member $S^* \in S^-$ and $R_{S^*}$ that is unblocked conditional on $S^c$.  

\item \textit{Cross-indicator case:} For $S^* \in S^-$ and $R' \in R(S) \setminus \{ R_{S^*} \}$, a path in $G[\{ R_{S^*} = 0 \}]$ between $S^*$ and $R'$ that is unblocked conditional on $S^c \cup \{ R_{S^*} \}$.
\end{itemize}
\end{definition}

Each type of MAR-violating path has a variable-dependence counterpart: a conditional dependence between $S^*$ and a missingness indicator, given the same conditioning set $S^c$ and, in the cross-indicator case, possibly some other missingness indicators.%

\begin{definition}
\label{def:variable_dependence}
\textit{Own-indicator variable dependence} occurs if there exists $S^* \in S^-$ such that $S^* \notindM R_{S^*} \mid S^c$. \textit{Cross-indicator variable dependence} occurs if there exist $S^* \in S^-$, $R' \in R(S) \setminus \{ R_{S^*} \}$, and $\Psi \subseteq R(S) \setminus \{ R', R_{S^*} \}$ (which may be empty) such that $S^* \notindM R' \mid \left( S^c, \, R_{S^*} = 0, \, \Psi \right)$. \textit{Variable dependence} is defined as the occurrence of either own-indicator or cross-indicator variable dependence.
\end{definition}

The next theorem relates MAR-violating paths and these various dependencies to MNAR($S$), to VMNAR($S$), and to the correctness of the MAR-assuming functional. Only part (3) of the theorem requires the following restriction on CSCI, which essentially states that CSCI cannot occur in contexts defined by any single incomplete variable's being missing:  
\begin{assumption}[No CSCI in missingness contexts with respect to $S$]
\label{ass:no_csci}
For every $S^* \in S^-$, $G[\{ R_{S^*} = 0 \}] = G$; that is, no set of parents of any node, excluding $R_{S^*}$ itself, is c-inactive in the context $R_{S^*} = 0$.%
\end{assumption}
\begin{theorem}[MNAR-equivalent conditions]
\label{thm:mnar_equivalent}
The following implications connect variable dependence, MNAR($S$), MAR-violating paths, and VMAR-violating paths:
\leavevmode
\begin{outline}[enumerate]
\1[(1)] Variable dependence implies MNAR($S$). Conversely, under weak intersection (Assumption \ref{ass:weak_intersection}), MNAR($S$) implies variable dependence.

\1[(2)] Variable dependence implies that a MAR-violating path exists. Conversely, under c-faithfulness (Definition \ref{def:c_faithfulness}), the existence of a MAR-violating path implies variable dependence.

\1[(3)] The existence of a MAR-violating path implies that a VMAR-violating path exists, i.e., that VMNAR($S$) holds. Conversely, if there is no CSCI in missingness contexts (Assumption \ref{ass:no_csci}), the existence of a VMAR-violating path implies that a MAR-violating path exists.

\1[(4)] If $\pMAR{S}(s) \ne p(s)$ for some $s$, then MNAR($S$) holds. Conversely, under no perfect cancellations (Supplement, Assumption \ref{supp-ass:no_mar_law_cancellations}), MNAR($S$) implies $\pMAR{S}(s) \ne p(s)$ for some $s$.
\end{outline}
\end{theorem}

Parts (1)--(3) are proven in the Supplement (Section \ref{supp-sec:thm1_proof}). Part (4) follows from the uniqueness of the MAR law and, for the converse, the assumption of no perfect cancellations (Section \ref{sec:MAR_MNAR}; Supplement, Proposition \ref{supp-prop:mar_fnl_correct}).

Figure \ref{fig:mnar_implication_map} summarizes (1)--(4) in the theorem, along with the standard result that $\VMARS$ implies $\MARS$ in every compatible SCM: by the Markov property, $S^- \ind_G R(S) \mid S^+$ implies $S^- \indM R(S) \mid S^+$, which implies $1\{ R(S) = r \} \indM \Smis(r) \mid \Sobs(r)$ for every pattern $r$. Thus, under c-faithfulness, weak intersection is needed only to establish $\MARS$ in SCMs in which $\VMARS$ fails yet no MAR-violating path exists, that is, in which MAR holds through CSCI. If all four assumptions in Theorem \ref{thm:mnar_equivalent} hold (i.e., weak intersection, c-faithfulness, no CSCI in missingness contexts, and no perfect cancellations), then five conditions are all equivalent: MNAR($S$), variable dependence, existence of a MAR-violating path, existence of a VMAR-violating path, and incorrectness of $\pMAR{S}$. These equivalences hold in every SCM compatible with $G$ that satisfies the four assumptions. MNAR($S$), variable dependence, and incorrectness of $\pMAR{S}$ are properties of a given SCM, whereas the existence of MAR-violating and VMAR-violating paths is a property of $G$; c-faithfulness is what allows the latter to characterize the former in each such SCM. We consider some practical implications in the Discussion. 

As a concrete example, consider again the monotone MAR example in Figure \ref{fig:monotone_MAR}. The VMAR-violating paths are $L_1 \to R_{L_2}$, $L_1 \to R_{L_3}$, $L_2 \to R_{L_3}$, and numerous superpaths of those. However, there are no own-indicator MAR-violating paths because every path between a member of $L$ and its own missingness indicator is blocked conditional on the other members of $L$. Also, there are no cross-indicator MAR-violating paths because: (1) for $S^* = L_1$, the context $R_{L_1} = 0$ forces $R_{L_2} = R_{L_3} = 0$, so $G[\{ R_{L_1} = 0 \}]$ has no paths between $L_1$ and $R(L)$ (Figure \ref{fig:monotone_MAR}(b)); (2) for $S^* = L_2$, the context $R_{L_2} = 0$ forces $R_{L_3} = 0$, so every path between $L_2$ and $R(L)$ in $G[\{ R_{L_2} = 0 \}]$ passes through $L_1$ as a non-collider and is therefore blocked conditional on $\{ L_1, L_3, R_{L_2} \}$ (Figure \ref{fig:monotone_MAR}(c)); and (3) for $S^* = L_3$, $G[\{ R_{L_3} = 0 \}]$ is $G$ itself, in which every path between $L_3$ and $R(L)$ passes through $L_1$ or $L_2$ as a non-collider and is therefore blocked conditional on $\{ L_1, L_2, R_{L_3} \}$ (Figure \ref{fig:monotone_MAR}(d)). Therefore, by Theorem \ref{thm:mnar_equivalent} MAR holds under weak intersection. Monotone MAR is a special case of a more general class of MAR mechanisms called \textit{randomized monotone $\text{MAR}$ (RMM)}, discussed in the Supplement, Section \ref{supp-sec:rmm}.\autocite{robins1997non}

\input{figures/mnar_implication_map}

The following result states that adding incomplete variables to $S$ cannot remove a VMAR violation; it will be useful when considering auxiliary variables (Section \ref{sec:auxiliaries}):

\begin{lemma}[Incomplete variables cannot restore VMAR]
\label{coro:incomplete_vars_cant_fix_mnar}
Let $S \subseteq L$, and let $T \subseteq L^-$ be a set of incomplete variables. If $\text{VMNAR}(S)$ holds, then so does $\text{VMNAR}(S \cup T)$.
\end{lemma}
\begin{proof}
Since every member of $T$ is incomplete, $(S \cup T)^+ = S^+$, whereas $S^- \subseteq (S \cup T)^-$ and $R(S) \subseteq R(S \cup T)$. Thus, any VMAR-violating path with respect to $S$ is also a VMAR-violating path with respect to $S \cup T$, so $\text{VMNAR}(S)$ implies $\text{VMNAR}(S \cup T)$. \qedhere
\end{proof}

\subsection{Connections to existing literature}
\label{sec:connections_other_lit}

Theorem \ref{thm:mnar_equivalent} has interesting connections to recent work on identification of the \textit{full law} $p(l, r(L))$ rather than $p(s)$. This work uses \textit{latent projection}, an operation that marginalizes a DAG over some of its nodes while preserving the independences it implies among the remaining nodes.\autocite{richardson2003markov} For a set of nodes $A$, the latent projection of $G$ onto $A$ is an acyclic directed mixed graph (ADMG) over $A$, denoted $G(A)$, which may contain bidirected edges ($\leftrightarrow$) representing associations arising from the marginalized nodes (Supplement, Section \ref{supp-sec:DAGs}). For example, Figure \ref{fig:VMAR_VMNAR_examples}(c) is the latent projection of panel (b) onto $L \cup R(L)$. In $G(S \cup R(S))$, VMAR-violating paths take a simple form: VMNAR($S$) holds if and only if $G(S \cup R(S))$ contains a path between a member of $S^-$ and a member of $R(S)$ whose nonterminal nodes, if any, are all colliders in $S^+$ (Supplement, Proposition \ref{supp-prop:minimal_vmar_paths}), such as $L_1 \leftrightarrow L_2 \leftrightarrow R_{L_1}$ in Figure \ref{fig:VMAR_VMNAR_examples}(c). In $G(L \cup R(L))$, Nabi et al. (2020)\autocite{nabi2020full} defined a \textit{colluding path} as a path between an incomplete variable $L^* \in L$ and its own missingness indicator $R_{L^*}$, such that all nonterminal nodes (if any) on the path are colliders. In the missing-data model of $G$ (encompassing all compatible SCMs), $p(l, r(L))$ is identified if and only if $G(L \cup R(L))$ contains no colluding paths.\autocite{nabi2020full} A colluding path is similar to a VMAR-violating path in that, in the respective latent projections, both imply the existence of an all-collider path between an incomplete variable and a missingness indicator (Supplement, Proposition \ref{supp-prop:minimal_vmar_paths}). However, as we discuss in the Supplement (Section \ref{supp-sec:vmar_vs_colluding}), neither is a generalization of the other. 

Theorem \ref{thm:mnar_equivalent} also relates to the results of Mealli \& Rubin (2015)\autocite{mealli2015clarifying} and its correction.\autocite{mealli2016correction} They provided an alternative variable dependence condition under the assumption that $R_i \indM (R(S) \setminus R_i) \mid S$ holds for all $R_i \in R(S)$: if there exist $R^* \in R(S)$ and $S^* \in S^-$ such that $S^* \notindM R^* \mid S^c$, and $p(R_{S^*} = 0 \mid s) > 0$ wherever this dependence occurs, then MNAR($S$) holds. Their dependence condition implies $S^- \notindM R(S) \mid S^+$, and the converse implication holds under weak intersection together with their independence assumption that $R_i \indM (R(S) \setminus R_i) \mid S$. However, that assumption is strong\autocite{little2021missing} and is violated in canonical cases such as RMM, including the basic monotone case, whereas Theorem \ref{thm:mnar_equivalent}(1) does not require this assumption. Figure \ref{supp-fig:mnar_implication_map_with_mealli} (Supplement) summarizes connections between our results and those of Mealli \& Rubin (2015).

\section{Identification using auxiliary variables}
\label{sec:auxiliaries}

\noindent We now consider MAR-assuming identification of the target law $p(v)$ when the analysis also includes complete auxiliary variables $Z \subseteq W^+$. The standard approach is to fit a model assuming $\text{MAR}(V \cup Z)$, for example by multiple imputation, and then marginalize over $Z$. This estimates $\pMAR{V \cup Z}(v)$, the marginal for $V$ of $\pMAR{V \cup Z}(v, z)$ (Section \ref{sec:MAR_MNAR}). Taking $S := V \cup Z$, we then have $S^- = V^-$, $S^+ = V^+ \cup Z$, and $R(S) = R(V)$. Recovering $p(v)$ is, in general, less demanding than recovering the joint law $p(v, z)$. However, for MAR-assuming identification with complete auxiliaries, the two coincide barring certain perfect cancellations:

\begin{lemma}[Correctness for the target law]
\label{lm:margin_correct}
Let $Z \subseteq W^+$. Barring perfect cancellations when integrating over $Z$ (Supplement, Assumption \ref{supp-ass:no_margin_cancellations}), $\pMAR{V \cup Z}(v) = p(v)$ for all $v$ if and only if $\pMAR{V \cup Z}(v, z) = p(v, z)$ for all $(v, z)$.
\end{lemma}

Heuristically, any discrepancy between $p(v,z)$ and $\pMAR{V \cup Z}(v, z)$ arises in $\pMAR{V \cup Z}(v^- \mid v^+, z)$ because the rest of the distribution is always observed. Then, Assumption \ref{supp-ass:no_margin_cancellations} rules out errors in this conditional distribution that exactly cancel upon marginalizing over $Z$ (Supplement, Section \ref{supp-sec:aux_results}). Combining Lemma \ref{lm:margin_correct} with Theorem \ref{thm:mnar_equivalent}, applied with $S := V \cup Z$, gives:

\begin{corollary}[MAR-assuming identification with complete auxiliaries]
\label{coro:aux_equivalent}
Let $Z \subseteq W^+$, and take $S := V \cup Z$. Suppose that c-faithfulness, no CSCI in missingness contexts (Assumption \ref{ass:no_csci}), and no perfect cancellations in the MAR law or over the auxiliaries (Supplement, Assumptions \ref{supp-ass:no_mar_law_cancellations} and \ref{supp-ass:no_margin_cancellations}) hold. Then the following are equivalent: (1) $\pMAR{V \cup Z}(v) = p(v)$ for all $v$; (2) $\text{MAR}(V \cup Z)$; (3) $\text{VMAR}(V \cup Z)$; and (4) there is no VMAR-violating path with respect to $V \cup Z$.
\end{corollary}
\begin{proof}
By Lemma \ref{lm:margin_correct}, (1) holds if and only if $\pMAR{V \cup Z}(v, z) = p(v, z)$ for all $(v, z)$, which by Theorem \ref{thm:mnar_equivalent}(4) holds if and only if (2) holds. Next, (3) implies (2) by the Markov property (Section \ref{sec:equivalent_conditions_unrestricted}). Conversely, if (3) fails, a VMAR-violating path exists; by the converse of Theorem \ref{thm:mnar_equivalent}(3), which uses Assumption \ref{ass:no_csci}, a MAR-violating path exists; by the converse of Theorem \ref{thm:mnar_equivalent}(2), which uses c-faithfulness, variable dependence holds; and by Theorem \ref{thm:mnar_equivalent}(1), MNAR($V \cup Z$) holds, so (2) fails. Finally, (3) and (4) are equivalent by definition. \qedhere
\end{proof}

If there is CSCI violating Assumption \ref{ass:no_csci}, (1) and (2) remain equivalent, and under weak intersection both hold whenever there is no MAR-violating path with respect to $V \cup Z$. $\text{VMAR}(V \cup Z)$ is then sufficient but not necessary. Corollary \ref{coro:aux_equivalent} also clarifies the relationship between MAR-assuming analyses and inverse-probability weighting of the complete cases with weights $1 / p\big( R(V) = 1 \mid v^+, z \big)$. The latter corresponds to the functional $\sum_{z} p(v^+, z) \, p\big( v^- \mid v^+, z, R(V) = 1 \big)$, which equals $p(v)$ if $\text{VMAR}(V \cup Z)$ holds. Thus, under the assumptions of Corollary \ref{coro:aux_equivalent}, complete-case inverse-probability weighting is correct in every SCM in which the MAR-assuming analysis is correct. If Assumption \ref{ass:no_csci} is violated, the two can differ: in the monotone MAR example of Figure \ref{fig:monotone_MAR}, the MAR-assuming analysis is correct, whereas complete-case analysis is generally not, because $\text{VMAR}$ fails.

\subsection{Constructing sufficient auxiliary sets}
\label{sec:constructing_suff_aux_set}

Corollary \ref{coro:aux_equivalent} has an immediate implication for choosing auxiliary variables: including additional complete auxiliaries can introduce MAR violations that would not otherwise occur, because conditioning on a collider, or a descendant of one, on a path between $V^-$ and $R(V)$ can unblock that path.\autocite{curnow2023multiple} For example, in Figure \ref{fig:VMAR_VMNAR_examples}(b), $\text{VMAR}(\{ L_1 \})$ holds, but $\text{VMAR}(\{ L_1, L_2 \})$ does not because $L_2$ unblocks the VMAR-violating path $L_1 \gets U_1 \to L_2 \gets U_2 \to R_{L_1}$. Further examples and simulations appear in the Supplement, Section \ref{supp-sec:examples_sims}. Given this pitfall, we seek auxiliary sets that are sufficient in the following sense:
\begin{definition}[Sufficient auxiliary set]
\label{def:suff_aux_set}
\noindent Fix $G$. A set $Z \subseteq W^+$ is a \textit{sufficient auxiliary set with respect to $V$} if $\text{VMAR}(V \cup Z)$ holds in $G$, i.e., if $V^- \ind_G R(V) \mid (V^+ \cup Z)$.
\end{definition}

By Corollary \ref{coro:aux_equivalent}, under its assumptions, $\pMAR{V \cup Z}(v)$ is correct in an SCM compatible with $G$ if and only if $Z$ is a sufficient auxiliary set. If $G$ is fully known, one can check directly whether a given set is sufficient. In practice, such detailed knowledge of the DGM is rarely available. The next result instead requires knowing only certain ancestral relationships. Recall that $\An_G(A)$ denotes the ancestors of $A$ in $G$.

\begin{theorem}[A sufficient auxiliary set for partially unknown DAGs]
\label{thm:suff_set_general}
Fix $G$. If $\text{VMAR}(V \cup Z)$ holds for some $Z \subseteq W$, which may include incomplete variables, then $\Z := W^+ \cap \An_G( V \cup R(V) )$ is a sufficient auxiliary set with respect to $V$.
\end{theorem}

That is, $\Z$ comprises the complete candidates that causally affect a target variable or its missingness. Theorem \ref{thm:suff_set_general} is somewhat analogous to the disjunctive cause criterion for confounding, which likewise yields a sufficient adjustment set whenever one exists (among pre-treatment variables), using only knowledge of ancestral relationships.\autocite{vanderweele2011new} As a caveat, if no sufficient auxiliary set exists, then including variables that affect missingness but not the incomplete target variables can \emph{amplify} bias rather than reduce it.\autocite{curnow2024multiple,pearl2010class}

For example, in Figure \ref{fig:suff_aux_set}, $V := \{ V_1, V_2 \}$ and $W := \{ W_1, W_2 \}$. Of the complete candidates, $W_2$ is an ancestor of $V \cup R(V)$ but $W_1$ is not, so $\Z = \{ W_2 \}$. Thus, Theorem \ref{thm:suff_set_general} indicates that either $\{ W_2 \}$ is a sufficient auxiliary set or $\text{VMAR}(V \cup Z)$ fails for every $Z \subseteq W$. Since the full DAG is known here, it is easy to see that $\{ W_2 \}$ is indeed sufficient. No other subset of $W$ is sufficient: excluding $W_2$ leaves the path $V_2 \gets W_2 \to R_{V_2}$ unblocked, and including $W_1$ unblocks the path $V_2 \gets U_1 \to W_1 \gets U_2 \to R_{V_2}$.

\input{figures/tikz_admgs_aux_sets}

Combined with Corollary \ref{coro:aux_equivalent}, Theorem \ref{thm:suff_set_general} implies that, in any SCM satisfying the corollary's assumptions, if the MAR-assuming analysis using some set of complete auxiliaries is correct, then so is the analysis using $\Z$. The same holds for auxiliary sets $Z$ that include incomplete variables: if c-faithfulness and no CSCI in missingness contexts (Assumption \ref{ass:no_csci}) hold with $S := V \cup Z$, then $\text{MAR}(V \cup Z)$ implies $\text{VMAR}(V \cup Z)$ by Theorem \ref{thm:mnar_equivalent}, so $\Z$ is a sufficient auxiliary set by Theorem \ref{thm:suff_set_general}, and hence $\text{MAR}(V \cup \Z)$ holds. In this sense, $\Z$ is never worse than any other choice of auxiliaries, regardless of whether that choice includes incomplete variables. Without Assumption \ref{ass:no_csci}, this implication can fail.\footnote{As an example of $\text{MAR}(V \cup Z)$ holding due to CSCI while $\text{MAR}(V \cup \Z)$ is violated, take $V := \{ L_3 \}$ and $W = Z := \{ L_1, L_2 \}$ in Figure \ref{fig:monotone_MAR}, so that $\Z = \emptyset$.} (For a detailed treatment of incomplete auxiliaries, see Mathur et al. (2026),\autocite{iai} which also shows, under different assumptions, that incomplete auxiliaries cannot restore MAR.)%

\subsection{Practical implications}
\label{sec:practical_implications_aux}

These results regarding auxiliary variables expose problems with widespread recommendations to 
use all candidate auxiliary variables 
\autocite{little2012prevention,perkins2018principled,collins2001comparison}, or all candidates 
associated with missingness \autocite{carpenter2021missing}. Even complete auxiliary variables can introduce MAR violations that would not have arisen with an appropriately chosen auxiliary set. Theorem~\ref{thm:suff_set_general} may inform the selection of auxiliary variables, though 
it should not be taken to imply that $\Z$ is always an optimal choice overall (see the Discussion). Although incomplete auxiliaries fall outside the scope of Corollary \ref{coro:aux_equivalent}, Theorem \ref{thm:mnar_equivalent} and Lemma \ref{coro:incomplete_vars_cant_fix_mnar} indicate that under c-faithfulness and no CSCI in missingness contexts (both with $S := V \cup Z$), incomplete auxiliaries cannot restore MAR: $\text{MAR}(V \cup Z)$ implies $\text{MAR}(V \cup Z^+)$. These considerations suggest the following rules of thumb for cases in which the data-generating DAG is uncertain:
\begin{enumerate}

    \item \textit{By default, use only complete auxiliary variables.} Including incomplete auxiliary variables is warranted only when both: (1) the investigator has sufficient knowledge of the DAG to be confident that inclusion will not introduce VMAR-violating paths with respect to $V \cup Z$; and (2) there is a compelling reason to include them, such as a substantial gain in precision (barring CSCI, restoring MAR cannot be such a reason).

    \item \textit{Restrict complete auxiliary variables to those affecting missingness indicators or target variables (complete or incomplete).} 
    Complete variables that affect neither missingness indicators nor target variables may introduce VMAR-violating paths and 
    hence MNAR \autocite{curnow2024multiple}. 

    \item \textit{Perform estimation under at least two auxiliary sets.} As noted in Section \ref{sec:constructing_suff_aux_set}, auxiliaries that affect missingness indicators but not target variables could amplify bias if the auxiliary set is not, in fact, sufficient. Therefore, we recommend comparing results using $Z_{V,W}$ to results using $Z := W^+ \cap \An_G(V)$, i.e., only candidates believed to affect target variables.
\end{enumerate}

\subsection{Applied example}
\label{sec:applied}

Following work led by author WZ \autocite{zhang2024population}, we investigated the research question: Are individuals who identify as non-heterosexual more or less likely to report changes in sexual identity over time, compared to those who initially identify as heterosexual? We used data from the Stockholm Public Health Cohort (2010–2021); details are provided in the Supplement, Section \ref{supp-sec:applied}. Figure \ref{fig:applied_example_dag} shows a plausible causal structure, where a complete exposure variable --- baseline sexual identity (ID$_1$) --- influences identity at follow-up (ID$_2$), partially mediated by structural stigmatization of non-heterosexual identities ($U_1$). Missingness in ID$_2$ primarily reflected attrition ($R_{\text{ID}_2}$), and ID$_1$ affects $R_{\text{ID}_2}$. Parental leave ($U_2$) is a mediator: sexual identity may affect probability of parenthood or distribution of parental responsibilities between partners; in turn, parental leave might affect a person's availability for survey research. Variables $U_1$ and $U_2$ are latent. The target variables are $V := \{ \text{ID}_1, \text{ID}_2 \}$.

\input{figures/applied_example_dags}

Income is a candidate auxiliary variable that is a collider between $U_1$ and $U_2$ as follows. Experimental evidence suggests that non-heterosexual applicants face hiring discrimination in the Swedish labor market, suggesting the relationship $U_1 \to \text{Income}$ \autocite{ahmed2013gay}. Additionally, in Sweden and elsewhere, parental leave ($U_2$) appears to affect the size of the ``child penalty'' in income,\autocite{kleven2019child} suggesting the relationship $U_2 \to \text{Income}$. Sex (female vs. male) is also a common cause of $U_1$ and $U_2$. Other presumed common causes that were treated as candidate auxiliary variables included age, country of birth, antidepressant use, and anxiolytic use (not depicted in the figure for simplicity). Income and the other candidates are obtained from registers; income is $>99\%$ complete so is treated as complete, and the others are complete (Supplement, Section \ref{supp-sec:applied}). In the depicted DAG, $\Z = \{ \text{Sex} \}$: sex is an ancestor of $\text{ID}_2$ and of $R_{\text{ID}_2}$, whereas income is an ancestor of neither. 

Of course, in reality, the data-generating mechanism probably does not conform exactly to the mechanism in Figure \ref{fig:applied_example_dag}. However, to reiterate a key point, only partial knowledge of the DAG is needed to assess whether $\text{VMAR}(V \cup Z)$ holds and hence whether income should be used as an auxiliary variable. Specifically, to determine that income should not be included in the imputation model, we only need to know that income shares latent causes with ID$_2$ and with $R_{\text{ID}_2}$. This is enough to determine that the resulting path ID$_2 \leftrightarrow \text{Income} \leftrightarrow R_{\text{ID}_2}$ is a VMAR-violating path with respect to $V \cup Z$ if $Z$ includes income. The depicted DAG is only one of many possible DAGs that include this collider structure while also obeying $\text{VMAR}(V \cup Z)$; that is, many additional variables and edges could be added to Figure \ref{fig:applied_example_dag} without changing the identification results.

For our analyses, we used multiple imputation by chained equations (MICE) \autocite{van2011mice} to impute missing data separately by sex, comparing five auxiliary-variable sets. Critically, some models did not include income, while others did (which we expect would lead to bias). We fit log-binomial models regressing change in sexual identity (yes vs. no) on baseline identity. Across specifications, non-heterosexual identity at baseline was associated with an increased probability of reporting changes in sexual identity. However, the choice of auxiliary variables mattered: including income consistently attenuated risk-ratio estimates toward the null. This attenuation is consistent with the hypothesized collider structure (Supplement, Section \ref{supp-sec:applied_bias}): if parental leave reduces income but increases response, nonrespondents tend to have higher income than respondents with the same baseline identity. In the imputation model, fit among respondents, higher income is associated with less stigmatization. If stigmatization makes reporting a heterosexual identity at follow-up more likely, the imputations then underestimate identity change among baseline non-heterosexual participants but overestimate identity change among baseline heterosexual participants, thus biasing the risk ratio downward. Full results are provided in the Supplement, Section \ref{supp-sec:applied}, including forest plots for the most comprehensive imputation models (with and without income) for the two subcohorts and three follow-up periods (Figure \ref{supp-fig:forest_plots}).

\section{Discussion}
\label{sec:discussion}

This paper has made two primary contributions. First, we characterized MAR in terms of the data-generating mechanism, showing that, under weak intersection and c-faithfulness, MAR holds if and only if there are no MAR-violating paths. When no VMAR-violating paths exist, MAR holds in every compatible SCM. When such paths do exist, then under c-faithfulness, MAR can hold only through CSCI; that is, some edges must cease to operate in a context in which an incomplete variable is missing. Second, we addressed the choice of complete auxiliary variables in MAR-assuming analyses, showing that, barring perfect cancellations, such an analysis recovers the target law if and only if MAR holds with respect to the target and auxiliary variables jointly. When a sufficient auxiliary set exists, one such set is the set of complete candidate auxiliaries that are ancestors of the target variables or their missingness indicators. Using this set is never worse than using any other choice of complete auxiliaries; moreover, barring CSCI in missingness contexts, if MAR holds with respect to the target variables and any other auxiliary set, including one with incomplete variables, then it also holds with this set.

Our work has limitations related to our exclusive focus on identification versus non-identification of $p(v)$ by $\pMAR{V \cup Z}(v)$. First, we have not addressed the magnitude of bias arising under non-identification. That is, Theorem \ref{thm:suff_set_general} and Corollary \ref{coro:aux_equivalent} establish that MAR-assuming identification using the auxiliary set $\Z$ is never worse than using any other choice of auxiliaries (under the assumed restrictions on CSCI); however, this does not imply that $\Z$ minimizes bias when $\text{MAR}(V \cup \Z)$ is violated. As noted above, in the latter case, auxiliary variables that only affect missingness indicators may amplify bias. Second, we have not provided theoretical results regarding efficiency. It is possible that certain auxiliary sets that induce MNAR could nevertheless provide a better bias-variance tradeoff than $\Z$ or another sufficient auxiliary set, although this did not occur in our simulations (Supplement, Section \ref{supp-sec:sim_study}). Identifying auxiliary sets that optimize, for example, mean-squared error would be a natural extension, analogous to recent results for optimal adjustment sets in causal inference.\autocite{henckel2022graphical,rotnitzky2020efficient} Third, our focus on MAR-assuming methods is not intended to undercut the importance of other identification approaches. Rather, our results reveal exactly what is being assumed, mechanistically, when MAR methods are used. Arguably, those stringent mechanistic conditions render the assumption implausible in many substantive contexts. If the data-generating DAG is fully known and $p(v)$ is identified, then the correct identifying expression can be derived from the DAG using existing identification algorithms.\autocite{nabi2020full,bhattacharya2020identification} Indeed, there are numerous cases in which MNAR holds for every choice of auxiliaries but $p(v)$ --- or even all of $p(l)$ --- is identified by a functional other than MAR. Finally, our results pertain to the target law $p(v)$ as a whole, but a specific estimand may be identifiable under weaker conditions by the corresponding MAR functional. Developing graphical criteria for MAR-assuming identification of specific estimands could yield more permissive sufficient conditions for valid MAR-assuming analyses, along the lines of past work on estimand-specific identification.\autocite{correa2018generalized,correa2019identification,robins1995analysis,nmar2,van2023missing} Similarly, we restricted attention to complete auxiliaries. Lemma \ref{lm:margin_correct} relies on this restriction: with incomplete auxiliaries, $\pMAR{V \cup Z}(v)$ can in principle be correct even when $\text{MAR}(V \cup Z)$ fails, and characterizing when it is correct remains an open problem. Our ongoing work makes some progress on both topics.\autocite{iai}

Section \ref{sec:practical_implications_aux} discussed immediate practical implications of our results. More broadly, these results help address the longstanding critique that MAR is too abstract and divorced from concrete data-generating mechanisms to be assessed with confidence in practice.\autocite{seaman2013meant,mealli2015clarifying,chen2022pattern,doretti2018missing,ccvmi} When investigators address missing data at all, MAR-assuming methods are typically applied with little or no justification,\autocite{mainzer2024gap} and the assumption's abstract nature perhaps enables its de facto treatment as a reasonable default. This practice is concerning. Our results clarify precisely what every MAR-assuming analysis assumes, shifting the burden on investigators from passively invoking MAR to actively justifying, on substantive grounds, why no VMAR-violating paths are plausible or, if they are, why the specific form of CSCI needed to preserve MAR would hold, that is, why the relevant edges would cease to operate when the relevant incomplete variable is missing. Our results similarly challenge the default inclusion of all candidate auxiliary variables: incomplete candidates should generally be excluded, and even complete candidates should be selected with attention to whether they unblock VMAR-violating paths through collider stratification. We believe that careful engagement with these considerations would, in many applied settings, reduce rather than increase investigators' confidence that MAR holds and underscore the need for sensitivity analyses, including those accommodating various forms of MNAR.\autocite{carpenter2015,nmar2,mvalue,nabi2020full,carpenter2021missing,white2011strategy,little2012prevention,bell2014handling,phung2025recursive}

\pagebreak
\singlespacing

\small
\paragraph{Data availability.}\label{sec:data-availability} Public repositories contain all code required to reproduce the applied example (\maybeRedact{\url{https://github.com/willizhang/applied-example-collider-bias/tree/main}}; DOI \url{10.5281/zenodo.23040910}
). The dataset on sexual identity can be accessed after approved ethics application to the Steering Committee of the Stockholm Public Health Cohort (\maybeRedact{\url{https://www.ces.regionstockholm.se/projekt-och-uppdrag/halsa-stockholm/SPHC-data}}).

\vspace{-0.7cm}

\paragraph{Funding.} \maybeRedact{MBM was supported by NIH grants UL1TR003142, P30CA124435, and P30DK116074. IS was supported by NIH 1-R56-AI-0191526-01 and ONR N00014-24-1-2701. RB was supported by NSF CRII 2348287 and EPSRC grant EP/Z000580/1. WZ was supported by the Swedish Research Council for Health, Working Life and Welfare (Grant nr 2022-00159 to Dr. Kyriaki Kosidou) and Region Stockholm.} The funders had no role in the design, conduct, or reporting of this research.

\vspace{-0.7cm}

\paragraph{Ethics approval.} The analyses for the applied example were conducted under existing ethical approvals \maybeRedact{(DNR: 2022-02626-01, 2023-06610-02, 2024-02155-02)}.

\vspace{-0.7cm}

\paragraph{Acknowledgments.} \maybeRedact{Dr. Kyriaki Kosidou} coordinated and acquired funding for the analyses of the Stockholm Public Health Cohort. RB and IS would like to thank the
Isaac Newton Institute for Mathematical Sciences for the
support and hospitality during the program \emph{Causal Inference: From Theory to Practice and Back Again} when some
of the work on this paper was undertaken. 

\vspace{-0.7cm}

\paragraph{Conflicts of interest.} None.

\vspace{-0.7cm}

\paragraph{Generative AI.} Claude Opus assisted with searching for related articles, drafting visuals and R code, and endless proofreading. All AI-drafted code was reviewed line-by-line and tested by one or more authors.

\pagebreak
\printbibliography

\wordCountSection

\end{document}


\thispagestyle{fancy} %

\singlespacing

\maketitle
\begingroup
\hypersetup{linkcolor=black}
\tableofcontents
\endgroup

\pagebreak

\input{figures/results_index_table}

\input{figures/proof_dependency_flowchart}

\pagebreak
\section{Theoretical results}

Supporting definitions and results appearing only here in the Supplement have numbers prefaced with ``S'' (e.g., Lemma S1).

\subsection{Additional preliminaries}
\label{sec:addl_prelims}

\noindent Throughout this Supplement, we fix a set of substantive variables $S \subseteq L$, where $S^-$ are the \textit{incomplete} variables and $S^+ := S \setminus S^-$ are the complete ones. Let $G$ be a DAG whose node set comprises $L$, $R(L)$, and the latent common causes $U$ of $L \cup R(L)$; in general, $G$ contains nodes outside $S \cup R(S)$. Let $\m$ be a compatible SCM. A \textit{pattern} is a value $r \in \{0, 1\}^{|R(S)|}$ of $R(S)$. For a pattern $r$, let $\Smis(r) := \{ S_j \in S^- : R_{S_j} = 0 \text{ in } r \}$ and $\Sobs(r) := S \setminus \Smis(r)$, so that $S^+ \subseteq \Sobs(r)$ for every $r$. We write $R(S) = 1$ for the complete-case pattern. As in the main text, we write the argument of $R(\cdot)$ explicitly outside of proofs; within proofs, we abbreviate $R := R(S)$ where stated. Lowercase letters with tildes (e.g., $\rt$) are values chosen to witness some property; bare lowercase letters are arbitrary values.

\begin{definition}[Minimal VMAR-violating paths]
\label{def:minimal_vmar_path}
A VMAR-violating path (Definition \ref{main-def:VMAR} in the main text) is \textit{minimal} if, additionally, no member of $S^- \cup R(S)$ is a nonterminal node of the path.
\end{definition}

\subsubsection{Causal graphs with missing data}
\label{sec:DAGs}

Here we briefly summarize a standard formalization of DAGs; further details are provided in Section 1.2 of reference \cite{pearl2009causality}. In an arbitrary DAG $G_L$ comprising nodes $L$, a \textit{path} between two members $L_i$ and $L_j$ of $L$ is an unbroken sequence of nodes and edges of any orientation between $L_i$ and $L_j$. A given path between two disjoint sets $A \subset L$ and $B \subset L$ is said to be \textit{d-separated} or \textit{blocked} by a set $C \subseteq ( L \setminus (A \cup B) )$ if, for some set of variables $\left( L_i, L_j, L_k \right)$: (i) the path includes a chain $L_i \rightarrow L_j \rightarrow L_k$ or a fork $L_i \leftarrow L_j \rightarrow L_k$, where $L_j \in C$; or (ii) the path includes a collider $L_i \rightarrow L_j \leftarrow L_k$, where neither $L_j$ nor any of its descendants is in $C$. A set $C$ is said to \textit{d-separate} $A$ from $B$ in a graph $G_L$ if and only if $C$ blocks every path in $G_L$ from $A$ to $B$. D-separation between $A$ and $B$ conditional on $C$ is denoted $A \ind_{G_L} B \mid C$.  

A distribution $p$ over $L$ is \textit{Markov wrt $G_L$} if, for all disjoint sets of nodes $A$, $B$, and $C$ such that $C$ d-separates $A$ from $B$, the corresponding statistical independence $A \ind B \mid C$ holds in $p$. Conversely, $p$ is \textit{faithful} to $G_L$ if, for all disjoint sets of nodes $A$, $B$, and $C$ such that $A \ind B \mid C$ holds in $p$, the corresponding d-separation holds in $G_L$. Without restrictions on $p$, trivially, there exists DAG(s) such that $p$ is Markov wrt the DAG(s). However, if $p$ exhibits (for example) certain forms of determinism or perfect cancellations of association, then there may not exist a DAG to which $p$ is faithful.

Informally, a DAG is \textit{causal} if each edge $A \to B$ represents a potential causal relationship of $A$ on $B$.\autocite{pearl2009causality} While our theoretical results do not require a causal interpretation of the DAG, it is difficult to envision DGMs that do not reflect an underlying causal process.\autocite{robins1997non} We therefore use causal language when discussing the practical implications of our results, since doing so facilitates assessing the assumptions' plausibility (see Section \ref{main-sec:practical_implications_aux} in the main text). 

In our context, as above, $G$ has node set $L \cup R(L) \cup U$. Every compatible SCM is Markov with respect to $G$, and, by Lemma \ref{lm:pruning_soundness}, $p(\cdot \mid D = d)$ is Markov with respect to each context-pruned DAG $G[d]$ for conditioning sets that contain $D$. The converse, c-faithfulness (Definition \ref{main-def:c_faithfulness}), is not assumed in general; results that use it say so explicitly. Recall that each member $L_j \in L$ denotes the underlying value of that variable; the analyst instead observes a proxy that equals $L_j$ when $L_j$ is observed and is missing otherwise. In some existing formalizations of DAGs involving missingness indicators,\autocite{nabi2022causal} both underlying variables and their proxies are depicted in the DAG. We do not show the proxies for compactness. Following existing formalizations, members of $R(L)$ may be ancestors of proxies but not of the underlying variables; this is a notational convention rather than a restriction on the DGM. The notational convention essentially reflects the fact that each underlying variable represents a counterfactual under \textit{fixed} intervention setting its missingness indicator to 1, so it is not affected by the intervention variable itself.%

\textit{Acyclic directed mixed graphs (ADMG)} are a generalization of DAGs that may include not only directed edges but also bidirected edges ($\leftrightarrow$).\autocite{richardson2003markov} Bidirected edges represent associations originating from variables that have been marginalized out, per the next definition. (We use ADMGs in Proposition \ref{prop:minimal_vmar_paths} and in Section \ref{sec:vmar_vs_colluding}.)%

\begin{definition}[Latent projection]
\label{def:projection}
Let $N := L \cup R(L) \cup U$ be the set of nodes in $G$, and let $N_S \subseteq N$. The ADMG obtained by marginalizing $G$ over $N_S$, denoted $G(N \setminus N_S)$, is the ADMG with node set $N \setminus N_S$ whose edges are defined as follows. For distinct $N_i, N_j \in (N \setminus N_S)$:
\begin{itemize}
    \item $N_i \to N_j$ in $G(N \setminus N_S)$ if and only if there exists a directed path from $N_i$ to $N_j$ in $G$ such that any nonterminal nodes are in $N_S$;
    \item $N_i \leftrightarrow N_j$ in $G(N \setminus N_S)$ if and only if there exists in $G$ a path of the form $N_i \gets \cdots \to N_j$ such that every nonterminal node is in $N_S$ and is a non-collider on the path.
\end{itemize}
\end{definition}

Independence in ADMGs is characterized by m-separation, which is the same as d-separation but with the definition of a collider generalized to allow either edge into the collider to be bidirected (i.e., $\leftrightarrow N_i \gets$, $\to N_i \leftrightarrow$, or $\leftrightarrow N_i \leftrightarrow$).\autocite{richardson2003markov} Examples appear in Figure \ref{fig:vmar_vs_colluding}(b)--(c), which will be discussed in Section \ref{sec:vmar_vs_colluding}.

\subsubsection{Review of assumptions}
\label{sec:review_assumptions}
\label{sec:regularity_assumptions}

\noindent For brevity, we will not restate the regularity assumptions in each result, and invoke them only when they play a specific technical role. Weak intersection (Assumption \ref{ass:weak_intersection}) and c-faithfulness (Definition \ref{main-def:c_faithfulness}) are not regularity assumptions; every result that uses either one says so explicitly.

\setcounter{assumption}{-1} %
\begin{assumption}[Regularity assumptions] \label{ass:boring_assumptions}
\hfill\\ %
\vspace{-1cm}

\begingroup
\renewcommand{\theenumi}{(\roman{enumi})}
\renewcommand{\labelenumi}{\theenumi}
\makeatletter
\renewcommand{\p@enumi}{\theassumption}
\makeatother
\begin{enumerate}

\item \label{ass:no_margin_cancellations} \textit{No perfect cancellations over the auxiliaries.} For every $Z \subseteq W^+$, writing $q := \pMAR{V \cup Z}$, if
$$ \sum_{z} p(v^+, z) \Big[ q\big(v^- \mid v^+, z\big) - p\big(v^- \mid v^+, z\big) \Big] = 0 \quad \text{for all } v \in \supp(V),$$
then $q(v^- \mid v^+, z) = p(v^- \mid v^+, z)$ for all $(v, z)$ with $p(v^+, z) > 0$ (for almost every such $(v, z)$ when $Z$ is continuous), with the sum replaced by an integral for continuous $Z$. (When $Z = \emptyset$, the sum has a single term and this holds trivially.)%

\item \label{ass:no_mar_law_cancellations} \textit{No perfect cancellations in the MAR law.} For every $S \subseteq L$, if MNAR($S$) holds, then
$$ \sum_{r} p\big( R(S) = r \mid \sobs(r) \big) \ne 1 \quad \text{for some } s \in \supp(S).$$

\end{enumerate}
\endgroup
\end{assumption}

The next two assumptions are reproduced from the main text.

\fromMain{ass:cc_positivity}
\begin{assumption}[Complete-case positivity]
\label{ass:cc_positivity}
The complete-case pattern occurs with positive probability conditional on any value of the substantive variables, i.e., $p( R(L) = 1 \mid l ) > 0$ for all $l \in \supp(L)$.
\end{assumption}

\fromMain{ass:weak_intersection}
\begin{assumption}[Weak intersection with respect to $S$]
\label{ass:weak_intersection}
Let $S \subseteq L$. For all disjoint sets $A \subseteq (S \cup R(L))$, $B \subseteq S$, and $C \subseteq S$, and every value $a$ of $A$, the following holds. For any $B_i \in B$, let $B_{-i} := B \setminus \{B_i\}$. If, for each $B_i \in B$, $p(a \mid b, c) = p(a \mid b_{-i}, c)$ for all values $(b, c)$ of $(B, C)$, then $p(a \mid b, c) = p(a \mid c)$ for all $(b, c)$. 
\end{assumption}

\subsection{General framework for context-specific causal irrelevance}
\label{sec:csci}

We give four remarks on our definition of c-inactivity (Definition \ref{main-def:c_inactive}), prove that the context-specific distribution is Markov to the pruned DAG for conditioning sets that contain the context variables (Lemma \ref{lm:pruning_soundness}), and discuss RMM as a worked example (Section \ref{sec:rmm_example}). This material applies to any DAG, not only those involving missingness indicators.%

Remarks on the definition of c-inactivity:
\begin{outline}[itemize]
\1 \textbf{Ways that c-inactivity can arise.} The context can restrict $B$'s other parents, $\pa_G(B) \setminus \mathcal{A}$, to values at which $f_B$ does not depend on $\mathcal{A}$, as in RMM (Section \ref{sec:rmm_example}). Alternatively, the context can restrict $\mathcal{A}$ itself to a region within which $f_B$ does not depend on it. A special case of the latter is a context that renders $\mathcal{A}$ degenerate; then c-inactivity holds trivially, because $a = a'$ for all $a, a' \in \supp(\mathcal{A} \mid D = d)$.

\1 \textbf{Context variables need not be parents.} The definition requires only that the context restrict the conditional support of $\Pa(B)$; the variables in $D$ need not be parents of $B$, nor even adjacent to it. This generality is needed for RMM, in which the context variables are missingness indicators but the parent that deactivates the edge is the latent $U$.

\1 \textbf{c-inactivity is a property of a set of parents, not of each parent separately.} Parents that are individually c-inactive can be jointly c-active. For example, suppose that $A_1$, $A_2$, and $A_3$ are binary, that $A_1 = A_2 = A_3$ whenever $D = 1$, and that $B = \text{maj}(A_1, A_2, A_3)$. In the context $D = 1$, the joint support of the parents is $\{ (0,0,0), (1,1,1) \}$, so fixing any one or two parents fixes the rest; thus, each singleton and each pair is c-inactive. However, the three together are not, since $\text{maj}(0,0,0) \ne \text{maj}(1,1,1)$. For this reason, the pruned DAG (Definition \ref{main-def:pruned_DAG} in the main text) deletes the edges from a single c-inactive \textit{set} of parents of each node: deleting several edges into $B$ at once is legitimate only if $f_B$ ignores the corresponding parents jointly.

\1 \textbf{c-inactivity is defined only at nodes outside $\An_G(D)$.} Thus, edges are not pruned from members of $\An_G(D)$. If they were, then the Markov property (Lemma \ref{lm:pruning_soundness}) would not hold.\footnote{For example, let $X$, $Y$, and $\varepsilon$ be iid $N(0,1)$, $V = X + Y + \varepsilon$, and $T = 1\{V \ge 0\}$, with context $T = 1$. Then $f_T$ is constant on $\supp(V \mid T = 1) = [0, \infty)$, so without the restriction, $V \to T$ would be pruned and $X \ind_{G[d]} Y \mid T$, yet $X \notindM Y \mid T = 1$ because $T = 1$ selects on the collider $V$.}

\end{outline}

\begin{lemma}[Markov property with c-inactivity]
\label{lm:pruning_soundness}
    Fix a DAG $G$ with node set $N$ and a compatible SCM ${\cal M}$ that may display c-inactivity. 
    For all disjoint sets of nodes $X, Y, C, D$ and every context value $d$ with $p(D = d) > 0$, we have:
    \begin{align*}
        X \ind_{G[d]} Y \mid C \cup D \quad \implies \quad X \ind_p Y \mid (C, D=d).
    \end{align*}
\end{lemma}
\begin{proof}
Let $\mathcal{B}_d$ be the set of nodes that have at least one incoming edge pruned in $G[d]$. Let $\m^*$ be the SCM obtained from $\m$ by replacing, for each node $B \in \mathcal{B}_d$, the structural equation $f_B$ with
$$ f^*_B\big( \pa_{G[d]}(B), \epsilon_B \big) := f_B\Big( a^*_B\big( \pa_{G[d]}(B) \big), \pa_{G[d]}(B), \epsilon_B \Big), $$
\newcommand{\SCMFn}{\footnote{The SCM $\m^*$ is merely a mathematical construction used for the proof; it need not correspond to any data-generating mechanism. As we will see, $\m^*$ is constructed to have two useful properties: (1) it agrees with $\m$ on independence statements that condition on $D=d$, which is what we need to establish the Markov property; and (2) it will be easy to show that $\m^*$ obeys the Markov property with respect to $G[d]$.} }
where, for each value $w$ of $\Pa_{G[d]}(B)$, $a^*_B(w)$ is a fixed value of $\mathcal{A}_B$ such that $\big( a^*_B(w), w \big) \in \supp\big( \Pa_G(B) \mid D = d \big)$ if such a value exists, and an arbitrary fixed value otherwise.\SCMFn Since $\m^*$ has the same mutually independent errors as $\m$ and each of its structural equations depends only on parents in $G[d]$, its induced distribution $p^*$ is Markov wrt $G[d]$. For the remainder of the proof, fix a realization of the errors in all variables' structural equations, and use this realization for both $\m$ and $\m^*$.

\noindent \textbf{Step 1: The context event coincides.} By the restriction that c-inactivity occurs only at nodes outside $\An_G(D)$, no structural equation of a member of $\An_G(D)$ was changed. Since the value of $D$ is determined by the errors and structural equations of members of $\An_G(D)$, $D$ takes the same value in $\m$ and $\m^*$ for every realization of the errors. Thus the event $\{ D = d \}$ is the same in both, and $p^*(D = d) = p(D = d) > 0$. 

\noindent \textbf{Step 2: All variables coincide on the context event.} Consider the nodes in a topological order of $G$, and suppose all nodes preceding $B$ take the same values in $\m$ and $\m^*$ on $\{ D = d \}$. If $B \notin \mathcal{B}_d$, its structural equation is unchanged, so $B$ also takes the same value in the two SCMs. If $B \in \mathcal{B}_d$, then on $\{ D = d \}$, the value of $\Pa_G(B) = \mathcal{A}_B \cup \Pa_{G[d]}(B)$ lies in $\supp(\Pa_G(B) \mid D = d)$; hence, writing $w$ for the value of $\Pa_{G[d]}(B)$, so does $\big( a^*_B(w), w \big)$. These two values of $\Pa_G(B)$ agree on $\Pa_G(B) \setminus \mathcal{A}_B$, so by c-inactivity (Definition \ref{main-def:c_inactive}), $f_B(\mathcal{A}_B, w, \epsilon_B) = f^*_B(w, \epsilon_B)$. Hence, almost surely on $\{ D = d \}$, every variable takes the same value in $\m$ and $\m^*$, so $p(\cdot \mid D = d) = p^*(\cdot \mid D = d)$.

\noindent \textbf{Step 3: Conclusion.} If $X \ind_{G[d]} Y \mid C \cup D$, then since $p^*$ is Markov wrt $G[d]$, $X \ind_{p^*} Y \mid (C, D = d)$, and by Step 2 the same holds under $p$.
\end{proof}

\subsubsection{Example: RMM}
\label{sec:rmm_example}

To illustrate the definitions, consider 2-variable RMM as defined in Section \ref{sec:rmm}. The structural equations are:
\begingroup
\addtolength{\jot}{0.5em}
\begin{align*}
R_{L_1} \; &= \; f_{R_{L_1}}(L_2, U, \epsilon_{R_{L_1}}) \; = \;
  \begin{cases}
    0 & U = 0 \\
    1 & U = 1 \\
    \mathbf 1\{ \epsilon_{R_{L_1}} > \lambda_1(L_2) \} & U = 2,
  \end{cases}
\\
R_{L_2} \; &= \; f_{R_{L_2}}(L_1, U, \epsilon_{R_{L_2}}) \; = \;
  \begin{cases}
    0 & U = 0 \\
    \mathbf 1\{ \epsilon_{R_{L_2}} > \lambda_2(L_1) \} & U = 1 \\
    1 & U = 2,
  \end{cases}
\end{align*}
\endgroup
with $\epsilon_{R_{L_1}}, \epsilon_{R_{L_2}} \sim \operatorname{Unif}(0,1)$ the error terms of the two indicators, mutually independent and independent of $(L_1, L_2, U)$, and with $\lambda_1$ and $\lambda_2$ arbitrary functions taking values in $[0, 1]$. Thus $\lambda_1(l_2) = p(R_{L_1} = 0 \mid U = 2, L_2 = l_2)$ and $\lambda_2(l_1) = p(R_{L_2} = 0 \mid U = 1, L_1 = l_1)$. 

Now take the context variables $D := \{ R_{L_1} \}$ with context value $d := 0$. Since $U = 1$ forces $R_{L_1} = 1$, this context implies $U \in \{0, 2\}$. At both of those values, $f_{R_{L_2}}$ evaluates to $0$ and $1$ respectively, so is functionally independent of $L_1$. Hence $\{ L_1 \}$ is c-inactive at $R_{L_2}$ in context $d$, and the edge $L_1 \to R_{L_2}$ is deleted in $G[d]$; see Figure \ref{fig:ex_rmm}. The context $\{ R_{L_2} = 0 \}$ is symmetric, deleting $L_2 \to R_{L_1}$ instead.

\begin{figure}[H]
\centering
\begin{subfigure}[t]{0.48\textwidth}
\centering
\begin{tikzpicture}[scale=0.8, transform shape]
\tikzset{
  ell/.style={draw, fill=white, inner sep=2pt, line width=1.5pt},
  every path/.style={>=Stealth, line width=1.2pt, >={Stealth[scale=\arrowheadScale]}},
  csi/.style={->, line width=1.2pt, dotted, color=blue, >=stealth},
  determ/.style={draw=red, line width=3pt, -{Stealth[length=10pt,width=10pt]}}
}
\node[name=L1, ell, shape=ellipse, fill={rgb:black,1;white,2}]{$L_1$};
\node[name=L2, ell, shape=ellipse, right=45mm of L1, fill={rgb:black,1;white,2}]{$L_2$};
\node[name=RL1, ell, shape=ellipse, below=20mm of L1]{$R_{L_1}$};
\node[name=RL2, ell, shape=ellipse, below=20mm of L2]{$R_{L_2}$};
\node[name=U, ell, shape=ellipse] at ($(RL1)!0.5!(RL2)$) {$U$};
\draw[->, line width=1.2pt, >=stealth] (L1) edge (L2);
\draw[csi] (L1) edge node[pos=0.25, sloped, color=blue, below, font=\scriptsize] {$R_{L_1} = 0$} (RL2);
\draw[csi] (L2) edge node[pos=0.25, sloped, color=blue, below, font=\scriptsize] {$R_{L_2} = 0$} (RL1);
\draw[determ] (U) -- (RL1);
\node[below=3pt of RL1, text=red, font=\scriptsize, align=center] {$U = 0 \Rightarrow R_{L_1} = 0$ \\ $U = 1 \Rightarrow R_{L_1} = 1$};
\draw[determ] (U) -- (RL2);
\node[below=3pt of RL2, text=red, font=\scriptsize, align=center] {$U = 0 \Rightarrow R_{L_2} = 0$ \\ $U = 2 \Rightarrow R_{L_2} = 1$};
\end{tikzpicture}
\caption{$G$}
\end{subfigure}
\hfill
\begin{subfigure}[t]{0.48\textwidth}
\centering
\begin{tikzpicture}[scale=0.8, transform shape]
\tikzset{
  ell/.style={draw, fill=white, inner sep=2pt, line width=1.5pt},
  every path/.style={>=Stealth, line width=1.2pt, >={Stealth[scale=\arrowheadScale]}},
  csi/.style={->, line width=1.2pt, dotted, color=blue, >=stealth},
  determ/.style={draw=red, line width=3pt, -{Stealth[length=10pt,width=10pt]}}
}
\node[name=L1, ell, shape=ellipse, fill={rgb:black,1;white,2}]{$L_1$};
\node[name=L2, ell, shape=ellipse, right=45mm of L1, fill={rgb:black,1;white,2}]{$L_2$};
\node[name=RL1, ell, shape=ellipse, below=20mm of L1]{$R_{L_1}$};
\node[name=RL2, ell, shape=ellipse, below=20mm of L2]{$R_{L_2}$};
\node[name=U, ell, shape=ellipse] at ($(RL1)!0.5!(RL2)$) {$U$};
\draw[->, line width=1.2pt, >=stealth] (L1) edge (L2);
\draw[csi] (L2) edge node[pos=0.25, sloped, color=blue, below, font=\scriptsize] {$R_{L_2} = 0$} (RL1);
\draw[determ] (U) -- (RL1);
\node[below=3pt of RL1, text=red, font=\scriptsize, align=center] {$U = 0 \Rightarrow R_{L_1} = 0$ \\ $U = 1 \Rightarrow R_{L_1} = 1$};
\draw[determ] (U) -- (RL2);
\node[below=3pt of RL2, text=red, font=\scriptsize, align=center] {$U = 0 \Rightarrow R_{L_2} = 0$ \\ $U = 2 \Rightarrow R_{L_2} = 1$};
\end{tikzpicture}
\caption{$G[\{R_{L_1} = 0\}]$}
\end{subfigure}
\caption{\label{fig:ex_rmm}Randomized monotone MAR. Gray shading indicates incomplete variables. Bold red edges: determinism. Blue dotted edges are annotated with their labels, i.e., contexts in which they are c-inactive; each is also c-inactive in any context that forces the annotated value. In the context $R_{L_1} = 0$, the edge $L_1 \to R_{L_2}$ is c-inactive, so it is deleted in panel (b).}
\end{figure}

\subsection{Theorem \ref{main-thm:mnar_equivalent}: MNAR-equivalent conditions}
\label{sec:thm1_proof}

This section restates and proves parts (1)--(3) of Theorem \ref{main-thm:mnar_equivalent} in the main text. Part (4) follows from Proposition \ref{prop:mar_fnl_correct} below together with Assumption \ref{ass:no_mar_law_cancellations}. The proofs in this section rely on supporting lemmas that are relegated to Section \ref{sec:boring}. We first give MAR-equivalent conditions that describe variable-level statistical independences, then we parlay those conditions into graphical criteria.

Throughout this section, fix $S$ as in Section \ref{sec:addl_prelims}, and for $S^* \in S^-$, let $S^c := S \setminus \{S^*\}$. Own-indicator, cross-indicator, and variable-level dependence are as defined in Definition \ref{main-def:variable_dependence} of the main text.

\fromMainPart{thm:mnar_equivalent}{1}
\begin{theorem}[MNAR-equivalent conditions: variable-level independences]
\label{thm:mnar_variable_level}
Variable-level dependence implies MNAR($S$). Conversely, under weak intersection (Assumption \ref{ass:weak_intersection}), MNAR($S$) implies variable-level dependence.

\begin{proof}
Throughout, write $R := R(S)$. We prove the contrapositive of each direction, i.e., we work with $\MARS$ and with the absence of variable-level dependence.

\noindent \textbf{Claim 1: If $\MARS$ holds, then variable-level dependence does not occur.}

\noindent Suppose that $\MARS$ holds, so for every pattern $r$ (trivially so if $p(R = r) = 0$, since then $p(r \mid s) = 0$):
$$p(r \mid s) = p(r \mid \sobs(r)).$$ 
Note that whenever $r_{S^*} = 0$, we have $S^* \in \Smis(r)$, so $p(r \mid s)$ is functionally independent of $s^*$; and hence so is any sum of such terms.

\noindent \textbf{Own-indicator case.} Fix any $S^* \in S^-$. By the law of total probability, for any $\st^c$ we can write:
$$ p\left( R_{S^*} = 0 \mid s^*, \st^c \right) \; = \; \sum_{r \, : \, r_{S^*} = 0} p(r \mid s^*, \st^c).$$
Every pattern in the sum has $S^* \in \Smis(r)$, so every term is functionally independent of $s^*$, and so is the sum. Since $R_{S^*}$ is binary, the same holds at $R_{S^*} = 1$. Hence $S^* \indM R_{S^*} \mid S^c$, and since $S^*$ was arbitrary, own-indicator variable dependence does not occur.

\noindent \textbf{Cross-indicator case.} Fix any $S^* \in S^-$, any $R' \in R \setminus \{ R_{S^*} \}$, and any $\Psi \subseteq R \setminus \{R', R_{S^*}\}$. By the law of total probability, for any supported $(\st^c, \tilde{\psi})$ we can write:
$$ p\left( r' \mid s^*, \st^c, R_{S^*} = 0, \tilde{\psi} \right) \; = \; \frac{ p(r', \tilde{\psi}, R_{S^*} = 0 \mid s^*, \st^c) }{ p(\tilde{\psi}, R_{S^*} = 0 \mid s^*, \st^c) } \; = \; \frac{ \sum_{r \, : \, r_{S^*} = 0, \, r_{R'} = r', \, r_\Psi = \tilde{\psi}} p(r \mid s^*, \st^c) }{ \sum_{r \, : \, r_{S^*} = 0, \, r_\Psi = \tilde{\psi}} p(r \mid s^*, \st^c) }.$$
Every pattern contributing to either sum has $R_{S^*} = 0$ and thus $S^* \in \Smis(\cdot)$. By $\MARS$, each term $p(r \mid s^*, \st^c)$ is functionally independent of $s^*$, and so is the ratio. Hence $S^* \indM R' \mid \left( S^c, R_{S^*} = 0, \Psi \right)$, and since the triple was arbitrary, cross-indicator variable dependence does not occur.

\noindent \textbf{Claim 2: Under weak intersection, if variable-level dependence does not occur, then $\MARS$ holds.}

\noindent The absence of variable-level dependence means that $S^* \indM R_{S^*} \mid S^c$ for every $S^* \in S^-$, and $S^* \indM R' \mid \left( S^c, R_{S^*} = 0, \Psi \right)$ for every triple $(S^*, R', \Psi)$ as in Definition \ref{main-def:variable_dependence}.

\noindent For any pattern $r$ with $\Smis(r) = \emptyset$, the MAR condition holds trivially. The only such pattern is the complete-case pattern. Fix a pattern $r$ with $\Smis(r) \ne \emptyset$. Choose any $S_k \in \Smis(r)$, so $r_k = 0$, and fix a value $\st_{-k}$ of $S \setminus \{S_k\}$. We will show that $p(r \mid s)$ is functionally independent of $s_k$. Since $S_k \in S^-$, we can set $S^* := S_k$ in Definition \ref{main-def:variable_dependence}, so that $S^c = S \setminus \{S_k\}$ and $\st^c := \st_{-k}$. Write $R_k := R_{S_k}$, $R_{-k} := R \setminus \{R_k\}$, and $r_k$ and $r_{-k}$ for the corresponding entries of $r$.

\noindent Enumerate the members of $R_{-k}$ in any order as $R_{(1)}, \ldots, R_{(J)}$. For $t = 1, \ldots, J$, the absence of cross-indicator variable dependence, applied with $R_{S^*} := R_k$, $R' := R_{(t)}$, and $\Psi := \{ R_{(1)}, \ldots, R_{(t-1)} \}$, gives
$$ S_k \indM R_{(t)} \mid \left( S^c, \, R_k = 0, \, R_{(1)}, \ldots, R_{(t-1)} \right).$$
By the contraction property of conditional independence, which holds for every distribution, these $J$ statements together give $S_k \indM R_{-k} \mid \left( S^c, R_k = 0 \right)$. (Thus the argument uses the absence of cross-indicator variable dependence only with $\Psi$ ranging over the initial segments of a single ordering of $R_{-k}$.) Now write
$$ p(r \mid s_k, \st^c) \; = \; p\left( R_k = 0 \mid s_k, \st^c \right) \, p\left( R_{-k} = r_{-k} \mid s_k, \st^c, R_k = 0 \right),$$
where the second factor is needed only where the first is positive. By the absence of own-indicator variable dependence for $S^* := S_k$, the first factor is constant in $s_k$ over $\supp(S_k \mid \st^c)$. If this constant is zero, then $p(r \mid s_k, \st^c) = 0$ for every $s_k$. Otherwise, the first factor is positive for every $s_k \in \supp(S_k \mid \st^c)$, so conditioning on $R_k = 0$ does not change the conditional support of $S_k$, and by $S_k \indM R_{-k} \mid \left( S^c, R_k = 0 \right)$ the second factor is also constant in $s_k$. Either way, $p(r \mid s_k, \st^c)$ is constant in $s_k$.

\noindent Since $S_k \in \Smis(r)$ was arbitrary, for each $S_j \in \Smis(r)$, $p(r \mid s)$ does not depend on $s_j$ given the remaining components of $s$, i.e., $\mathbb{1}\{R = r\} \indM S_j \mid S_{-j}$. By Assumption \ref{ass:weak_intersection} (weak intersection), $\mathbb{1}\{R = r\} \indM \Smis(r) \mid \Sobs(r)$; that is, $p(r \mid s)$ is a function of $\sobs(r)$ alone. Since $r$ was arbitrary, $\MARS$ holds.
\qedhere
\end{proof}
\end{theorem}

MAR-violating paths, including their own-indicator and cross-indicator cases, are defined in Definition \ref{main-def:mar_violating_path} of the main text.

\fromMainPart{thm:mnar_equivalent}{2}
\begin{theorem}[MNAR-equivalent mechanistic conditions]
\label{thm:mnar_mechanistic}
Variable-level dependence implies that a MAR-violating path exists. Conversely, under c-faithfulness (Definition \ref{main-def:c_faithfulness} in the main text), the existence of a MAR-violating path implies variable-level dependence.

\begin{proof}
Throughout, write $R := R(S)$, and for $S^* \in S^-$, write $G_{S^*} := G[\{ R_{S^*} = 0 \}]$.

\noindent \textbf{Direction 1: Path existence implies variable dependence.} This is c-faithfulness (Definition \ref{main-def:c_faithfulness}), applied with the context and conditioning set of the relevant case. In the own-indicator case, take the context variables to be empty, so that the pruned DAG is $G[\emptyset] = G$ (Definition \ref{main-def:pruned_DAG} in the main text), and take $C := S^c$. An own-indicator MAR-violating path gives $S^* \notind_G R_{S^*} \mid S^c$, so c-faithfulness gives $S^* \notindM R_{S^*} \mid S^c$, which is own-indicator variable dependence. (In this case, c-faithfulness implies ordinary faithfulness.) In the cross-indicator case, take the context $D := \{ R_{S^*} \}$ with $d := 0$, which has positive probability because $S^*$ is incomplete, and take $C := S^c \cup \{ R_{S^*} \}$, so that $D \subseteq C$. A cross-indicator MAR-violating path to $R'$ gives $S^* \notind_{G_{S^*}} R' \mid S^c \cup \{ R_{S^*} \}$, so c-faithfulness gives $S^* \notindM R' \mid \left( S^c, R_{S^*} = 0 \right)$, which is cross-indicator variable dependence with $\Psi := \emptyset$.

\noindent \textbf{Direction 2: Variable dependence implies path existence.} We prove the contrapositive for each case. In the own-indicator case, if there is no own-indicator MAR-violating path for $S^*$, then $S^* \ind_G R_{S^*} \mid S^c$, so by the Markov property, $S^* \indM R_{S^*} \mid S^c$, and own-indicator variable dependence does not occur for $S^*$. In the cross-indicator case, suppose that there is no cross-indicator MAR-violating path for $S^*$, i.e., $S^* \ind_{G_{S^*}} R'' \mid S^c \cup \{ R_{S^*} \}$ for every $R'' \in R \setminus \{ R_{S^*} \}$. By the composition property of d-separation,\autocite{pearl1988probabilistic} $S^* \ind_{G_{S^*}} \left( R \setminus \{ R_{S^*} \} \right) \mid S^c \cup \{ R_{S^*} \}$, and by weak union and decomposition, $S^* \ind_{G_{S^*}} R' \mid S^c \cup \{ R_{S^*} \} \cup \Psi$ for every $R' \in R \setminus \{ R_{S^*} \}$ and every $\Psi \subseteq R \setminus \{ R', R_{S^*} \}$. Lemma \ref{lm:pruning_soundness}, applied with $C := S^c \cup \Psi$ and context $D := \{ R_{S^*} \}$ with $d := 0$, then gives $S^* \indM R' \mid \left( S^c, R_{S^*} = 0, \Psi \right)$ for every such $(R', \Psi)$, so cross-indicator variable dependence does not occur for $S^*$. Lemma \ref{lm:pruning_soundness} relies on c-inactivity being defined only at nodes outside $\An_G(D)$ (Definition \ref{main-def:c_inactive} in the main text). 
\end{proof}
\end{theorem}

Chaining parts (1) and (2): under weak intersection, MNAR($S$) implies that a MAR-violating path exists; and under c-faithfulness, the existence of a MAR-violating path implies MNAR($S$). Weak intersection enters only through the converse direction of part (1). It is not needed when $\VMARS$ holds: then the Markov property gives $S^- \indM R(S) \mid S^+$, which directly implies $\MARS$. Under c-faithfulness, it is needed only for SCMs in which $\VMARS$ fails yet no MAR-violating path exists. In such SCMs, the d-separations that rule out MAR-violating paths hold in different graphs, namely $G$ for the own-indicator case and $G\big[ \{ R_{S^*} = 0 \} \big]$ for each $S^* \in S^-$ in the cross-indicator case, so they cannot be combined into a single d-separation. Instead, the proof of part (1) combines the corresponding independences in the distribution, where intersection can fail.

The next part establishes VMAR-violating paths as the special case of MAR-violating paths that arises when there is no CSCI in missingness contexts.

\fromMainPart{thm:mnar_equivalent}{3}
\begin{theorem}[MAR-violating and VMAR-violating paths]
\label{thm:mnar_paths}
The existence of a MAR-violating path implies that a VMAR-violating path exists, i.e., that VMNAR($S$) holds. Conversely, under no CSCI in missingness contexts (Assumption \ref{main-ass:no_csci} in the main text), the existence of a VMAR-violating path implies that a MAR-violating path exists.

\begin{proof}
Throughout, write $R := R(S)$. We use the following properties of d-separation, which hold in every DAG without any positivity condition: decomposition, weak union, contraction, composition, and intersection.\autocite{pearl1988probabilistic} We also use the fact that, for any $S^* \in S^-$, every d-separation that holds in $G$ also holds in $G\big[ \{ R_{S^*} = 0 \} \big]$. This is because $G\big[ \{ R_{S^*} = 0 \} \big]$ is a subgraph of $G$ on the same nodes: every path in $G\big[ \{ R_{S^*} = 0 \} \big]$ is also a path in $G$ with the same colliders, and every node's descendants in $G\big[ \{ R_{S^*} = 0 \} \big]$ are also its descendants in $G$, so a path that is unblocked conditional on a given set in $G\big[ \{ R_{S^*} = 0 \} \big]$ is also unblocked conditional on that set in $G$. We prove the contrapositive of each direction.

\noindent \textbf{MAR-violating path exists $\Rightarrow$ VMAR-violating path exists.} Suppose that $\VMARS$ holds, i.e., $S^- \ind_G R \mid S^+$, and fix any $S^* \in S^-$. Since $S^c = \left( S^- \setminus \{ S^* \} \right) \cup S^+$, weak union gives $S^* \ind_G R \mid S^c$, and decomposition then gives $S^* \ind_G R_{S^*} \mid S^c$, so there is no own-indicator MAR-violating path for $S^*$. Applying weak union and then decomposition to $S^* \ind_G R \mid S^c$ gives $S^* \ind_G R' \mid S^c \cup \{ R_{S^*} \}$ for every $R' \in R \setminus \{ R_{S^*} \}$, and this d-separation also holds in $G\big[ \{ R_{S^*} = 0 \} \big]$, so there is no cross-indicator MAR-violating path for $S^*$. Since $S^*$ was arbitrary, no MAR-violating path exists. This direction does not use Assumption \ref{main-ass:no_csci}.

\noindent \textbf{VMAR-violating path exists $\Rightarrow$ MAR-violating path exists.} Suppose that no CSCI in missingness contexts holds and that no MAR-violating path exists, and fix any $S^* \in S^-$. Since $G\big[ \{ R_{S^*} = 0 \} \big] = G$, the absence of cross-indicator MAR-violating paths gives $S^* \ind_G R' \mid S^c \cup \{ R_{S^*} \}$ for every $R' \in R \setminus \{ R_{S^*} \}$, so by composition, $S^* \ind_G \left( R \setminus \{ R_{S^*} \} \right) \mid S^c \cup \{ R_{S^*} \}$. The absence of own-indicator MAR-violating paths gives $S^* \ind_G R_{S^*} \mid S^c$. By contraction, $S^* \ind_G R \mid S^c$. Since this holds for every $S^* \in S^-$, that is, $R \ind_G S^* \mid \left( S^- \setminus \{ S^* \} \right) \cup S^+$ for every $S^* \in S^-$, repeated application of intersection gives $R \ind_G S^- \mid S^+$. Thus $\VMARS$ holds, so no VMAR-violating path exists. \qedhere
\end{proof}
\end{theorem}

\begin{proposition}[Correctness of the MAR-assuming functional]
\label{prop:mar_fnl_correct}
Suppose, as in the main text, that a law satisfying $\MARS$ and implying the observed-data distribution exists. Under Assumption \ref{ass:cc_positivity} (complete-case positivity), $\pMAR{S}(s) = p(s)$ for all $s$ if and only if
$$ \sum_{r} p\big( R(S) = r \mid \sobs(r) \big) = 1 \quad \text{for all } s \in \supp(S).$$
In particular, $\MARS$ implies $\pMAR{S}(s) = p(s)$ for all $s$; and under Assumption \ref{ass:no_mar_law_cancellations} (no perfect cancellations in the MAR law), MNAR($S$) implies $\pMAR{S}(s) \ne p(s)$ for some $s$. Thus Assumption \ref{ass:no_mar_law_cancellations} excludes exactly the MNAR mechanisms under which the MAR-assuming functional is nonetheless correct.

\begin{proof}
Throughout, write $R := R(S)$. By Assumption \ref{ass:cc_positivity}, at most one law satisfying $\MARS$ implies the observed-data distribution;\autocite{robins1994estimation} write $q := \pMAR{S}$ for this law.

\noindent ($\Rightarrow$) Suppose that $q(s) = p(s)$ for all $s$. Then $q(\sobs(r)) = p(\sobs(r))$ for every $r$. Since $q$ and $p$ imply the same observed-data distribution, $q\big(R = r \mid \sobs(r)\big) = p\big(R = r \mid \sobs(r)\big)$ for every $r$. Since $q$ satisfies $\MARS$, for every $s$:
$$\sum_r p\big(R = r \mid \sobs(r)\big) \; = \; \sum_r q\big(R = r \mid \sobs(r)\big) \; = \; \sum_r q(R = r \mid s) \; = \; 1.$$

\noindent ($\Leftarrow$) Suppose that the displayed sum equals 1 for all $s \in \supp(S)$, and define $q'(s, r) := p(s) \, p\big( R = r \mid \sobs(r) \big)$. By the hypothesis, $q'$ is a law whose marginal for $S$ is $p(s)$. It satisfies $\MARS$, since its pattern probabilities depend on $s$ only through $\sobs(r)$. It implies the observed-data distribution, since summing over the missing components gives $q'(\sobs(r), r) = p(\sobs(r)) \, p\big( R = r \mid \sobs(r) \big) = p(\sobs(r), r)$ for every $r$. By uniqueness, $q' = q$, so $q(s) = p(s)$ for all $s$.

\noindent For the two particular statements: under $\MARS$, $p(R = r \mid \sobs(r)) = p(R = r \mid s)$ for every $r$, so the sum is 1 for every $s$; and Assumption \ref{ass:no_mar_law_cancellations} states that under MNAR($S$), the sum differs from 1 for some $s$. \qedhere
\end{proof}
\end{proposition}

\begin{proposition}[Minimal VMAR-violating paths in the latent projection]
\label{prop:minimal_vmar_paths}
Let $G' := G(S \cup R(S))$ be the ADMG obtained by marginalizing $G$ over all nodes outside $S \cup R(S)$. Then VMNAR($S$) holds if and only if $G'$ contains a path $\pi'$ between some $S^* \in S^-$ and some $R^* \in R(S)$ such that every nonterminal node of $\pi'$ is a collider in $S^+$. If such a path $\pi'$ has no nonterminal nodes, then it takes the form $S^* \to R^*$ or $S^* \leftrightarrow R^*$. Otherwise, it takes the form
$$ S^* \rightBiarrow S^+_{i} \leftrightarrow S^+_{j} \leftrightarrow \cdots \leftrightarrow S^+_{n} \leftrightarrow R^*,$$
where $\{ S^+_{i}, \cdots, S^+_{n} \} \subseteq S^+$ and the first edge is either $S^* \to S^+_i$ or $S^* \leftrightarrow S^+_i$. In particular, with $S := V \cup Z$ for complete auxiliaries $Z \subseteq W^+$, we have $G' = G(V \cup Z \cup R(V))$, which marginalizes over $(W \setminus Z) \cup R(W) \cup U$.%

\begin{proof}
Throughout, write $R := R(S)$. Since $S^-$, $R$, and $S^+$ are all sets of nodes of $G'$, and (conditional) d-separation in a DAG is equivalent to the corresponding (conditional) m-separation in the latent projection,\autocite{richardson2003markov} VMNAR($S$) holds if and only if $G'$ contains a path between a member of $S^-$ and a member of $R$ that is unblocked conditional on $S^+$.

\noindent ($\Leftarrow$) A path $\pi'$ as in the statement is unblocked conditional on $S^+$, since each of its nonterminal nodes is a collider in the conditioning set. Hence VMNAR($S$) holds.

\noindent ($\Rightarrow$) Suppose that VMNAR($S$) holds, so that $G'$ contains a path between a member of $S^-$ and a member of $R$ that is unblocked conditional on $S^+$. Applying Lemma \ref{lm:general_path_truncation_helper} in $H := G'$ with $A := S^-$, $B := R$, and $\Sigma := S^+$ yields a path $\pi'$ between some $S^* \in S^-$ and some $R^* \in R$ that is unblocked conditional on $S^+$ and has no member of $S^- \cup R$ as a nonterminal node. Since the nodes of $G'$ are $S^- \cup S^+ \cup R$, every nonterminal node of $\pi'$ is in $S^+$; since $\pi'$ is unblocked conditional on $S^+$, every such node is a collider.

\noindent For the form of $\pi'$, note that an edge $R^* \to N_i$ in $G'$ with $N_i \in S$ would require a directed path in $G$ from $R^*$ to $N_i$, contradicting the standing convention that no member of $R(L)$ is an ancestor of a member of $L$ (Section \ref{sec:DAGs}). If $\pi'$ has no nonterminal nodes, then it is a single edge between $S^*$ and $R^*$, which therefore is $S^* \to R^*$ or $S^* \leftrightarrow R^*$. Otherwise, every edge between two consecutive nonterminal nodes has arrowheads at both of its endpoints, since both are colliders, and so is bidirected. The edge between $S^+_n$ and $R^*$ has an arrowhead at $S^+_n$ and is not $R^* \to S^+_n$, so it is $S^+_n \leftrightarrow R^*$. The edge between $S^*$ and $S^+_i$ has an arrowhead at $S^+_i$, so it is $S^* \to S^+_i$ or $S^* \leftrightarrow S^+_i$. \qedhere
\end{proof}
\end{proposition}

Figure \ref{main-fig:mnar_implication_map} in the main text summarizes the implications established by the results in this section.

\subsubsection{Supporting lemmas}
\label{sec:boring}

\begin{lemma}[Selecting a single member]
\label{lm:product_support_implies_weak_intersection}
Let $A$, $B$, and $C$ be disjoint sets of variables, and fix a value $a$ of $A$. Suppose that, for every $c \in \supp(C)$,
$$\supp(B \mid C = c) = \prod_{B_i \in B} \supp(B_i \mid C = c).$$
For any $B_i \in B$, let $B_{-i} := B \setminus \{B_i\}$. If, for each $B_i \in B$,
$$p(a \mid b, c) = p(a \mid b_{-i}, c) \quad \text{for all supported } (b, c),$$
then $p(a \mid b, c) = p(a \mid c)$ for all supported $(b, c)$.

\begin{proof}
Fix $c \in \supp(C)$, and write $B = \{B_1, \ldots, B_n\}$. By the premise for $B_i$, $p(a \mid b, c) = p(a \mid b_{-i}, c)$, so $p(a \mid b, c)$ is unchanged when only the $i$-th coordinate of $b$ changes within $\supp(B \mid C = c)$.

Fix any $b' \in \supp(B \mid C = c)$. For a given $b \in \supp(B \mid C = c)$ and $k = 0, \ldots, n$, let
$$b^{(k)} := \big( b_1, \ldots, b_k, b'_{k+1}, \ldots, b'_n \big),$$
so that $b^{(0)} = b'$ and $b^{(n)} = b$. Each coordinate of $b^{(k)}$ lies in the corresponding marginal support, so $b^{(k)}$ is supported because $\supp(B \mid C = c)$ is a product set. Consecutive points $b^{(k-1)}$ and $b^{(k)}$ differ only in the $k$-th coordinate, so
$$p\big(a \mid b^{(k)}, c\big) = p\big(a \mid b^{(k-1)}, c\big) \quad \text{for each } k.$$
Hence $p(a \mid b, c) = p(a \mid b', c)$ for every $b \in \supp(B \mid C = c)$. By the law of total probability,
$$p(a \mid c) = \sum_{b} p(b \mid c) \, p(a \mid b, c) = p(a \mid b', c),$$
so $p(a \mid b, c) = p(a \mid c)$ for every supported $b$. (For continuous $B$, replace the sum with an integral.)  \qedhere
\end{proof}
\end{lemma}

The lemma's conclusion, with its $(A, B, C)$ taken to be the corresponding sets in Assumption \ref{ass:weak_intersection}, is exactly that assumption, and it requires product support only of $B$. Hence product support of $S$ implies weak intersection: product support of $S$ gives the lemma's conditional product-support hypothesis for any disjoint $B, C \subseteq S$, and no support condition on $A$, and hence on $R(L)$, is needed.

\begin{lemma}[Path truncation helper]
\label{lm:general_path_truncation_helper}
Let $A$, $B$, and $\Sigma$ be disjoint sets of nodes in an ADMG, $H$. If there exists a path in $H$ between a member of $A$ and a member of $B$ that is unblocked conditional on $\Sigma$, then there exists a path $\pi^*$ between some $A^* \in A$ and some $B^* \in B$ such that: (i) $\pi^*$ is unblocked conditional on $\Sigma$; and (ii) no member of $A \cup B$ is a nonterminal node of $\pi^*$.

\begin{proof}
Let $\pi$ be a path between some $A_i \in A$ and some $B_j \in B$ that is unblocked conditional on $\Sigma$. Suppose some nonterminal node $J$ of $\pi$ is in $A \cup B$. If $J \in B$, replace $\pi$ by its $A_i$-to-$J$ subpath; if $J \in A$, replace $\pi$ by its $J$-to-$B_j$ subpath. Either way, the result is a strictly shorter path between a member of $A$ and a member of $B$, and it is still unblocked conditional on $\Sigma$, since each of its nonterminal nodes is a nonterminal node of $\pi$ with the same collider status. Iterating until no nonterminal node is in $A \cup B$ yields $\pi^*$. \qedhere
\end{proof}
\end{lemma}

\subsection{Lemma \ref{lm:margin_correct}: Correctness for the target law}
\label{sec:aux_results}

\fromMain{lm:margin_correct}
\begin{lemma}[Correctness for the target law]
\label{lm:margin_correct}
Let $Z \subseteq W^+$. Under Assumption \ref{ass:no_margin_cancellations} (no perfect cancellations over the auxiliaries) for this $Z$, $\pMAR{V \cup Z}(v) = p(v)$ for all $v$ if and only if $\pMAR{V \cup Z}(v, z) = p(v, z)$ for all $(v, z)$.

\begin{proof}
\noindent \textbf{Sufficiency.} This direction is trivial: if $\pMAR{V \cup Z}(v, z) = p(v, z)$ for all $(v, z)$, then marginalizing over $Z$ gives $\pMAR{V \cup Z}(v) = p(v)$ for all $v$.

\noindent \textbf{Necessity.} Suppose $\pMAR{V \cup Z}(v) = p(v)$ for all $v$. 

Since $\pMAR{V \cup Z}$ implies the observed-data distribution and $V^+ \cup Z$ is observed for all patterns, $\pMAR{V \cup Z}(v^+, z, R(V) = r) = p(v^+, z, R(V) = r)$ for every pattern $r$, and summing over $r$ gives $\pMAR{V \cup Z}(v^+, z) = p(v^+, z)$. Hence $\pMAR{V \cup Z}(v^- \mid v^+, z)$ is well defined wherever $p(v^+, z) > 0$, and for every $v$:
$$0 \; = \; \pMAR{V \cup Z}(v) - p(v) \; = \; \sum_{z} p(v^+, z) \Big[ \pMAR{V \cup Z}\big(v^- \mid v^+, z\big) - p\big(v^- \mid v^+, z\big) \Big].$$
Since this holds for all $v$, Assumption \ref{ass:no_margin_cancellations} gives $\pMAR{V \cup Z}(v^- \mid v^+, z) = p(v^- \mid v^+, z)$ for all $(v, z)$ with $p(v^+, z) > 0$ (for almost every such $(v, z)$ when $Z$ is continuous). Therefore $\pMAR{V \cup Z}(v, z) = \pMAR{V \cup Z}(v^+, z)\, \pMAR{V \cup Z}(v^- \mid v^+, z) = p(v, z)$ wherever $p(v^+, z) > 0$, and both sides are zero wherever $p(v^+, z) = 0$. \qedhere
\end{proof}
\end{lemma}

\subsection{Theorem \ref{thm:suff_set_general}: A sufficient auxiliary set for partially unknown DAGs}

\begin{lemma}[Conditional d-separating set among a subset of nodes]
\label{lm:constructing_a_dsep_set}
Let $G$ be an arbitrary DAG. Let $A$, $B$, $C$, and $T$ be disjoint subsets of the nodes in $G$. A set $Q$ disjoint from $A \cup B$ is called a \textit{$C$-conditional d-separating set} if $A \ind_{G} B \mid (C \cup Q)$. If there exists a subset $T^* \subseteq T$ that is a $C$-conditional d-separating set, then the
set $Q^* := T \cap \An_{G}(A \cup B \cup C)$ is also a $C$-conditional d-separating set.%

\begin{proof}
This result is closely related to known results on separators restricted to ancestral sets;\autocite{vanderzander2019separators} we give a self-contained proof. For any subset $N$ of nodes in $G$, let $G_N$ be the subgraph of $G$ induced by $N$. Let $M_{T^*}$ be the moralization of $G_{\An(A \cup B \cup C \cup T^*)}$. By a canonical result (Proposition 3 in Lauritzen et al. (1990)\autocite{lauritzen1990independence}), the d-separation $A \ind_{G} B \mid (C \cup T^*)$ is equivalent to the separation $A \ind_{M_{T^*}} B \mid (C \cup T^*)$. (``Separation'' here refers to separation in an undirected graph.)

To prove the contrapositive, suppose $Q^*$ is not a $C$-conditional d-separating set in $G$. Let $M_{Q^*}$ be the moralization of $G_{\An(A \cup B \cup C)}$. Note that by the definition of $Q^*$, $\An(A \cup B \cup C) = \An(A \cup B \cup C \cup Q^*)$, so $G_{\An(A \cup B \cup C)}$ is the same object as $G_{\An(A \cup B \cup C \cup Q^*)}$. Since $M_{Q^*}$ is the moralization of this object, we can apply the Lauritzen et al. result again, as follows. Since $A \notind_{G} B \mid (C \cup Q^*)$, we also have $A \notind_{M_{Q^*}} B \mid (C \cup Q^*)$. Let $\pi$ be a witnessing path in $M_{Q^*}$, i.e., a path between $A$ and $B$ such that no nonterminal nodes are in $C \cup Q^*$. 

We now claim that every edge in $M_{Q^*}$ is also an edge of $M_{T^*}$. Since $\An(A \cup B \cup C) \subseteq \An(A \cup B \cup C \cup T^*)$, the smaller induced subgraph $G_{\An(A \cup B \cup C)}$ is contained in the larger one, $G_{\An(A \cup B \cup C \cup T^*)}$. Also, every 3-node structure in the smaller subgraph that comprises two parents and their shared child is also in the larger subgraph. This establishes that every edge in $M_{Q^*}$ is also an edge of $M_{T^*}$.

Therefore, the exact node-and-edge sequence that comprises $\pi$ is also in $M_{T^*}$. We claim that $\pi$ in $M_{T^*}$ is unblocked conditional on $C \cup T^*$. Arguing by contradiction, suppose a member $T_j$ of $T^*$ is a nonterminal node on $\pi$ (hence blocking the path conditional on $C \cup T^*$, since the graph is undirected). Since $\pi$ also exists in $M_{Q^*}$ and that graph's nodes are exactly $\An(A \cup B \cup C)$, it follows that $T_j \in T^* \cap \An(A \cup B \cup C)$. But since $T^* \subseteq T$ and $Q^* := T \cap \An(A \cup B \cup C)$, this implies $T_j \in Q^*$. This is a contradiction, so we have established that $A \notind_{M_{T^*}} B \mid (C \cup T^*)$.

In turn, this implies $A \notind_{G} B \mid (C \cup T^*)$, so $T^*$ is not a $C$-conditional d-separating set. This establishes the contrapositive. \qedhere
\end{proof}
\end{lemma}

\fromMain{thm:suff_set_general}
\begin{theorem}[A sufficient auxiliary set for partially unknown DAGs]
\label{thm:suff_set_general}
Fix $G$. If $\text{VMAR}(V \cup Z)$ holds for some $Z \subseteq W$, which may include incomplete variables, then $\Z := W^+ \cap \An_G( V \cup R(V) )$ is a sufficient auxiliary set with respect to $V$.

\begin{proof}
Suppose $\text{VMAR}(V \cup Z)$ holds for some $Z \subseteq W$. By the contrapositive of Lemma \ref{main-coro:incomplete_vars_cant_fix_mnar} in the main text, applied with $S := V \cup Z^+$ and $T := Z^-$, $\text{VMAR}(V \cup Z^+)$ also holds; that is, $V^- \ind_G R(V) \mid (V^+ \cup Z^+)$, so by Definition \ref{main-def:suff_aux_set} in the main text, $Z^+$ is a sufficient auxiliary set. Applying Lemma \ref{lm:constructing_a_dsep_set} with $A := V^-$, $B := R(V)$, $C := V^+$, and $T := W^+$ yields the result. \qedhere
\end{proof}
\end{theorem}

\pagebreak
\section{Connections to existing results}

\subsection{Mealli \& Rubin (2015)}

Figure \ref{fig:mnar_implication_map_with_mealli} extends Figure \ref{main-fig:mnar_implication_map} of the main text with the condition of Mealli \& Rubin (2015)\autocite{mealli2015clarifying} and the implications discussed in Section \ref{main-sec:connections_other_lit} of the main text.

\input{figures/mnar_implication_map_with_mealli}

\subsection{VMAR-violating paths versus colluding paths}
\label{sec:vmar_vs_colluding}

Colluding paths and minimal VMAR-violating paths are similar in that, in the respective latent projections, both are all-collider paths between incomplete variables and missingness indicators, but there are three key distinctions. Consider VMAR-violating paths with respect to $V \cup Z$ for complete auxiliaries $Z \subseteq W^+$. First, the definition of a VMAR-violating path imposes a weaker condition on the terminal nodes: the terminal missingness indicator need not specifically be $R_{L^*}$. Second, minimal VMAR-violating paths in $G(V \cup Z \cup R(V))$ impose a stronger condition on the nonterminal nodes: they must be in $V^+ \cup Z$, whereas colluding paths permit any member of $L \cup R(L)$. Third, minimal VMAR-violating paths are all-collider paths in the ADMG $G(V \cup Z \cup R(V))$ (Proposition \ref{prop:minimal_vmar_paths}), whereas colluding paths are all-collider paths in $G(L \cup R(L))$. The latter marginalizes out only $U$; the former additionally marginalizes out $(W \setminus Z) \cup R(W)$.

Figure~\ref{fig:vmar_vs_colluding} illustrates these distinctions. In panel~(a), the DAG $G$ contains target variables $V := \{V_1, V_2, V_3\}$, candidate auxiliary variables $W := \{W_1\}$, missingness indicators $R(V \cup W)$, and a latent variable $U_1$. Panel~(b) shows the corresponding ADMG $G(S \cup R(S))$ with $S := V \cup W$, in which each all-collider path between a variable $S_i \in S^-$ and its own missingness indicator is a colluding path (dotted red and solid green edges). Panel~(c) shows the ADMG $G(V \cup R(V))$, i.e., with $Z := \emptyset$, in which every all-collider path between a member of $V^-$ and any missingness indicator is a VMAR-violating path (dashed blue and solid green paths). The green path is both a VMAR-violating path with respect to $V$ and a colluding path with respect to $S$ because it is a bidirected edge between $V_3$ and its own missingness indicator $R_{V_3}$ in both $G(S \cup R(S))$ and $G(V \cup R(V))$. The blue path $V_2 \gets W_1 \to R_{V_2}$ is a VMAR-violating path with respect to $V$ because marginalizing out $W_1$ yields a bidirected edge. However, when considering colluding paths with respect to $S$, $W_1$ is not marginalized out, so $V_2 \gets W_1 \to R_{V_2}$ is not a colluding path.

\definecolor{bothColor}{HTML}{1B7A3E}
\setlength{\hnodesep}{8mm}
\setlength{\vnodesep}{12mm}

\begin{figure}[h]
\centering
\begin{subfigure}[t]{0.42\textwidth}
\centering
\begin{tikzpicture}[baseline=(current bounding box.center),
                    node distance=12mm and 12mm]
    \sansmath
    \tikzset{
        line width=1.2pt,
        ell/.style={draw, fill=white, inner sep=2pt, line width=1.5pt},
        ellInc/.style={draw, fill={rgb:black,1;white,2}, inner sep=2pt, line width=1.5pt},
        every path/.style={>=Stealth, line width=1.2pt, >={Stealth[scale=\arrowheadScale]}},
        bidirected/.style={Latex-Latex},
        swig vsplit={gap=3pt, line color right=\bidirectedEdgeColor}
    }
    \node[name=V1, ellInc, shape=ellipse] {$V_1$};
    \node[name=V2, ellInc, shape=ellipse, right=\hnodesep of V1] {$V_2$};
    \node[name=W1, ellInc, shape=ellipse, right=\hnodesep of V2] {$W_1$};
    \node[name=V3, ellInc, shape=ellipse, right=\hnodesep of W1] {$V_3$};
    \node[name=U1, ell, shape=ellipse, right=\hnodesep of V3] {$U_1$};
    \node[name=RV1, ell, shape=ellipse, below=\vnodesep of V1] {$R_{V_1}$};
    \node[name=RV2, ell, shape=ellipse, below=\vnodesep of V2] {$R_{V_2}$};
    \node[name=RW1, ell, shape=ellipse, below=\vnodesep of W1] {$R_{W_1}$};
    \node[name=RV3, ell, shape=ellipse, below=\vnodesep of V3] {$R_{V_3}$};
    \draw[->, line width=1.2pt, >=stealth, color=blue, dashed]
        (V1) edge (RV2)
        (W1) edge (V2)
        (W1) edge (RV2);
    \draw[->, line width=1.2pt, >=stealth, color=red, dotted]
        (W1) edge (RW1);
    \draw[->, line width=1.2pt, >=stealth, color=bothColor]
        (U1) edge (V3)
        (U1) edge (RV3);
\end{tikzpicture}
\caption{$G$}
\label{fig:vmar_vs_colluding_G}
\end{subfigure}
\hfill
\begin{subfigure}[t]{0.42\textwidth}
\centering
\begin{tikzpicture}[baseline=(current bounding box.center),
                    node distance=12mm and 12mm]
    \sansmath
    \tikzset{
        line width=1.2pt,
        ell/.style={draw, fill=white, inner sep=2pt, line width=1.5pt},
        ellInc/.style={draw, fill={rgb:black,1;white,2}, inner sep=2pt, line width=1.5pt},
        every path/.style={>=Stealth, line width=1.2pt, >={Stealth[scale=\arrowheadScale]}},
        bidirected/.style={Latex-Latex},
        swig vsplit={gap=3pt, line color right=\bidirectedEdgeColor}
    }
    \node[name=V1, ellInc, shape=ellipse] {$V_1$};
    \node[name=V2, ellInc, shape=ellipse, right=\hnodesep of V1] {$V_2$};
    \node[name=W1, ellInc, shape=ellipse, right=\hnodesep of V2] {$W_1$};
    \node[name=V3, ellInc, shape=ellipse, right=\hnodesep of W1] {$V_3$};
    \node[name=RV1, ell, shape=ellipse, below=\vnodesep of V1] {$R_{V_1}$};
    \node[name=RV2, ell, shape=ellipse, below=\vnodesep of V2] {$R_{V_2}$};
    \node[name=RW1, ell, shape=ellipse, below=\vnodesep of W1] {$R_{W_1}$};
    \node[name=RV3, ell, shape=ellipse, below=\vnodesep of V3] {$R_{V_3}$};
    \draw[->, line width=1.2pt, >=stealth, color=blue, dashed]
        (V1) edge (RV2)
        (W1) edge (V2)
        (W1) edge (RV2);
    \draw[->, line width=1.2pt, >=stealth, color=red, dotted]
        (W1) edge (RW1);
    \draw[bidirected, color=bothColor]
        (V3) -- (RV3);
\end{tikzpicture}
\caption{$G(S \cup R(S))$ with $S := V \cup W$}
\label{fig:vmar_vs_colluding_full}
\end{subfigure}

\vspace{6mm}

\begin{subfigure}[t]{0.42\textwidth}
\centering
\begin{tikzpicture}[baseline=(current bounding box.center),
                    node distance=12mm and 12mm]
    \sansmath
    \tikzset{
        line width=1.2pt,
        ell/.style={draw, fill=white, inner sep=2pt, line width=1.5pt},
        ellInc/.style={draw, fill={rgb:black,1;white,2}, inner sep=2pt, line width=1.5pt},
        every path/.style={>=Stealth, line width=1.2pt, >={Stealth[scale=\arrowheadScale]}},
        bidirected/.style={Latex-Latex},
        swig vsplit={gap=3pt, line color right=\bidirectedEdgeColor}
    }
    \node[name=V1, ellInc, shape=ellipse] {$V_1$};
    \node[name=V2, ellInc, shape=ellipse, right=\hnodesep of V1] {$V_2$};
    \node[name=V3, ellInc, shape=ellipse, right=\hnodesep of V2] {$V_3$};
    \node[name=RV1, ell, shape=ellipse, below=\vnodesep of V1] {$R_{V_1}$};
    \node[name=RV2, ell, shape=ellipse, below=\vnodesep of V2] {$R_{V_2}$};
    \node[name=RV3, ell, shape=ellipse, below=\vnodesep of V3] {$R_{V_3}$};
    \draw[->, line width=1.2pt, >=stealth, color=blue, dashed]
        (V1) edge (RV2);
    \draw[bidirected, color=blue, dashed]
        (V2) -- (RV2);
    \draw[bidirected, color=bothColor]
        (V3) -- (RV3);
\end{tikzpicture}
\caption{$G(V \cup R(V))$}
\label{fig:vmar_vs_colluding_empty}
\end{subfigure}

\caption{\label{fig:vmar_vs_colluding}Gray shading indicates incomplete variables. Edge types: \textcolor{blue}{dashed blue} = VMAR-violating path with respect to $V$ but not colluding path with respect to $S := V \cup W$; \textcolor{red}{dotted red} = colluding path with respect to $S$ but not VMAR-violating path with respect to $V$; \textcolor{bothColor}{solid green} = both. (a) The DAG $G$. (b) The ADMG $G(S \cup R(S))$ with $S := V \cup W$, where $U_1$ has been marginalized out producing the bidirected edge $V_3 \leftrightarrow R_{V_3}$. (c) The ADMG $G(V \cup R(V))$, where $W_1$ is also marginalized out, producing $V_2 \leftrightarrow R_{V_2}$.}
\end{figure}

\section{Randomized monotone MAR}
\label{sec:rmm}

Monotone MAR is a special case of a more general class of MAR mechanisms called ``randomized monotone $\text{MAR}$'' (RMM).\autocite{robins1997non} In monotone MAR, the missingness indicators are generated in a fixed chronological ordering, and if any indicator $R_{L_i} = 0$, then $R_{L_{i+1}} = 0$ deterministically. In RMM, the chronological ordering is random rather than fixed. In the 2-variable case shown in Figure \ref{fig:ex_rmm}(a), data generation proceeds as follows. A latent variable $U$, which will control the chronological ordering of the missingness indicators, is generated first; here $U \sim \operatorname{Unif}(\{ 0, 1, 2 \} )$. If $U=1$, then the first missingness indicator generated is $R_{L_1}$, and $R_{L_1} = 1$ deterministically. The one remaining missingness indicator, $R_{L_2}$, is generated next from a distribution that may depend on the value of $L_1$. If $U=2$, the same holds except with the roles of $R_{L_1}$ and $R_{L_2}$ reversed. Finally, if $U=0$, then $R_{L_1} = R_{L_2} = 0$. This setup ensures that MAR holds. The only minimal VMAR-violating paths in Figure \ref{fig:ex_rmm}(a) are $L_1 \to R_{L_2}$ and $L_2 \to R_{L_1}$, but there are no MAR-violating paths. For $S^* = L_1$, the context $R_{L_1} = 0$ deletes $L_1 \to R_{L_2}$ (Figure \ref{fig:ex_rmm}(b)), and every other path from $L_1$ to $R_{L_1}$ or $R_{L_2}$ is blocked by $L_2$, which is conditioned on, or passes through an unconditioned collider. The case $S^* = L_2$ is symmetric.

\pagebreak
\pagebreak
\section{Example DAGs and simulation study}
\label{sec:sim_study}

\subsection{Example DAGs}
\label{sec:examples_sims}

\input{figures/table_simulation_dags}

We now use the causal graphs in Table \ref{tb:simulation_dags} to illustrate our results regarding auxiliary variables (Section \ref{main-sec:auxiliaries} in the main text). In these graphs, all auxiliary variables are complete. For compactness, associations arising from latent common causes are depicted as bidirected edges; hence each graph is the latent projection of the DAG onto the depicted variables, with $U$ marginalized out. The theoretical implications described in this subsection are also illustrated in the simulations below.

In graph (a), $A$ is marginally d-separated from $R_A$, but is d-connected to $R_A$ conditional on $B$. Hence, $\text{VMAR}( \{ A \} )$ holds, but $\text{VMAR}(\{ A, B \})$ is violated. Thus, if $V := \{ A \}$ (e.g., if the estimand of interest is $E[A]$), then $Z := \emptyset$ is a sufficient auxiliary set because there is no path between $R(V)$ and $V^-$ that is marginally unblocked. However, $Z := \{ B \}$ is not a sufficient auxiliary set because the path $A \leftrightarrow B \leftrightarrow R_A$ is unblocked conditional on $B$ and is thus a VMAR-violating path. Critically, the standard recommendation is indeed to use this insufficient choice of auxiliary variables.\autocite{little2012prevention,perkins2018principled,collins2001comparison} If instead the target variables are $V := \{ A, B \}$ (e.g., if the estimand of interest is $E[B \mid A]$), such that $W = \emptyset$, then no sufficient auxiliary set exists because the path $A \leftrightarrow B \leftrightarrow R_A$ is unblocked conditional on $V^+ = \{ B \}$, and $W = \emptyset$. That is, even though a sufficient auxiliary set exists with respect to $V := \{ A \}$, none exists with respect to $V := \{ A, B \}$.

Graph (b) differs from graph (a) in that $B$ is also incomplete, though its own missingness is independent of all other variables. As in graph (a), if $V := \{ A \}$, then $Z := \emptyset$ is a sufficient auxiliary set. If instead $V := \{ A, B \}$, then $B \in V^-$ and $V^+ = \emptyset$. Again, no sufficient auxiliary set exists, but for a different reason: $B$ and $R_A$ are adjacent via the bidirected edge $B \leftrightarrow R_A$, which no choice of $Z$ can block.

In graph (c), if $V := \{ B \}$, then $Z := \emptyset$ is not a sufficient auxiliary set because the path $B \gets A \to R_B$ is unblocked marginally (note that $V^+ \cup Z = \emptyset$ here). Likewise, the standard auxiliary set $\{ A, C \}$ is not sufficient because $B \leftrightarrow C \leftrightarrow R_B$ is unblocked conditional on $V^+ \cup Z = \{ A, C \}$. However, $Z := \{ A \}$ is a sufficient auxiliary set. If instead $V := \{ A, B \}$, then the empty set is a sufficient auxiliary set because $A \in V^+$ itself blocks the path $B \gets A \to R_B$. However, $Z := \{ C \}$ is not a sufficient auxiliary set because $B \leftrightarrow C \leftrightarrow R_B$ is unblocked conditional on $V^+ \cup Z = \{ A, C \}$.

\subsection{Data generation}

For each graph in Table \ref{tb:simulation_dags}, we generated 5,000 datasets each with $n = 1{,}000$ individuals. All variables other than missingness indicators were normally distributed. The primary incomplete variable ($A$ in graphs (a) and (b); $B$ in graph (c)) was missing for 50\% of individuals; in graph (b), $B$ was additionally missing completely at random for 10\% of individuals. For each graph, the data-generating parameters and empirical correlation matrices are listed below. Variables called $U_i$ are latent variables, represented by bidirected edges in Table \ref{tb:simulation_dags}.

\singlespacing
\begin{table}[H]
\centering
\resizebox{0.97\columnwidth}{!}{%
\renewcommand{\arraystretch}{1.3}
\begin{tabular}{@{}c c l l@{}}
\toprule
 & \textbf{Graph} & \textbf{Data-generating mechanism} & \textbf{Correlations} \\
\midrule
(a) & \input{figures/table1_dags/table1_dag_c}
    & $\begin{aligned}[c]
        &U_1, U_2, C \sim N(0,1) \\
        &A \sim N(2U_1, 1) \\
        &B \sim N(3U_1 + 3U_2, 1) \\
        &R_A \sim \text{Bern}(\text{expit}(3U_2))
      \end{aligned}$
    & $\begin{array}{l|ccc}
          & B   & C   & R_A \\ \hline
        A & .62 &     &     \\
        B &     &     & .47 \\
        C &     &     &
      \end{array}$ \\
\addlinespace[0.5em]
\midrule
\addlinespace[0.5em]
(b) & \input{figures/table1_dags/table1_dag_d}
    & $\begin{aligned}[c]
        &U_1, U_2, C \sim N(0,1) \\
        &A \sim N(2U_1, 1) \\
        &B \sim N(2U_1 + 2U_2, 1) \\
        &R_A \sim \text{Bern}(\text{expit}(3U_2)) \\
        &R_B \sim \text{Bern}(0.9)
      \end{aligned}$
    & $\begin{array}{l|cccc}
          & B   & C   & R_A & R_B \\ \hline
        A & .60 &     &     &     \\
        B &     &     & .46 &     \\
        C &     &     &     &     \\
        R_A &   &     &     &
      \end{array}$ \\
\addlinespace[0.5em]
\midrule
\addlinespace[0.5em]
(c) & \input{figures/table1_dags/table1_dag_e}
    & $\begin{aligned}[c]
        &U_1, U_2, A \sim N(0,1) \\
        &C \sim N(1.5U_1 + 1.5U_2, 1) \\
        &B \sim N(1.5A + 1.5U_1, 1) \\
        &R_B \sim \text{Bern}(\text{expit}(A + U_2))
      \end{aligned}$
    & $\begin{array}{l|ccc}
          & B   & C   & R_B \\ \hline
        A & .64 &     & .36 \\
        B &     & .41 & .23 \\
        C &     &     & .23
      \end{array}$ \\
\bottomrule
\end{tabular}
}
\caption{\label{tb:dgp_correlations}Data-generating mechanisms and empirical marginal correlations for simulation scenarios (rounded to two decimals; blank cells indicate $|\rho| < 0.02$).}
\end{table}
\singlespacing

For each graph, we considered two possible choices of target variables $V$: (1) a single variable, in which case the estimand of interest was the marginal mean of this variable; or (2) two variables, in which case the estimand of interest was the coefficient of $A$ in the conditional expectation $E[B \mid A]$, that is, the slope of the linear regression of $B$ on $A$. For each missing-data method described below, we estimated these quantities using a correctly-specified ordinary least squares regression.

\subsection{Missing-data methods}

Our simulations assessed three forms of MAR-based imputation using different choices of auxiliary variables, along with several comparison methods. The first two forms of imputation were Amelia and multiple imputation by chained equations (MICE). These are two widely-used implementations of, respectively, joint modeling under a multivariate normal model \autocite{honaker2011amelia,van2018flexible} and fully conditional imputation by chained equations \autocite{van2007multiple,van2011mice,van2018flexible}. 
When implementing MICE, we generated imputations using semiparametric predictive mean matching, which is the default approach in the R package \texttt{mice} (van Buuren (2018) \autocite{van2018flexible}, p. 77). The last form of imputation was a MAR-based model in which imputation models were fit using only the complete cases. For this method, termed ``CC MVN'', we used a custom R implementation of multivariate normal modeling \autocite{honaker2011amelia,van2018flexible}. For all imputation methods, we generated 50 imputations (using 200 iterations for MICE), and we pooled estimates and variances using Rubin's Rules (Rubin (2004) \autocite{rubin2004multiple}, p. 76).

Overall, our simulations used eight missing-data methods. (1) The \textbf{benchmark} method proceeded as if all variables had been complete; that is, it analyzed the full data, including the values that were missing for the other methods. (2) The \textbf{complete-case} method involved no imputation and instead simply excluded all individuals with missing data on any variable. (3) For the imputation models \textbf{Amelia MVN (standard)}, (4) \textbf{MICE (standard)}, and (5) \textbf{CC MVN (standard)}, we included all available covariates in the imputation model, per standard recommendations.\footnote{Because the software for Amelia and MICE will not run if there are no covariates in the imputation model, we generated a covariate $C$ that was independent of all others for graphs (a) and (b). In graph (c), $C$ instead denotes the auxiliary variable shown in Table \ref{tb:simulation_dags}.} For (6) \textbf{Amelia MVN (updated)}, (7) \textbf{MICE (updated)}, and (8) \textbf{CC MVN (updated)}, we included a sufficient auxiliary set in the imputation model, as listed in Table \ref{tb:simulation_dags}. When a sufficient auxiliary set did not exist, the ``updated'' methods did not apply. For graph (b) with $V := \{ A \}$, we omitted the ``standard'' methods because the standard auxiliary set would include the incomplete variable $B$.

\subsection{Results}

Table \ref{tb:sim_results} shows the results in terms of bias, root mean-square error (RMSE), and the coverage of 95\% confidence intervals. The findings regarding bias agreed with the implications from theory discussed in Section \ref{sec:examples_sims}. In graphs (a) and (b), complete-case analysis was nonetheless unbiased for the slope of $E[B \mid A]$: because $R_A$ depends only on $U_2$, which is independent of $A$ and enters $B$ additively, restricting to complete cases shifts only the intercept.

\singlespacing
\renewcommand{\arraystretch}{1}

\begin{longtable}{lllllll}
\hline
\textbf{Graph} & \textbf{Estimand} & \textbf{$Z$} & \textbf{Method} & \textbf{Bias} & \textbf{RMSE} & \textbf{Coverage} \\ \hline
\endhead

(a) & \multirow{8}{*}{$E[A]$}
    & --          & Benchmark                & 0.00 & 0.07 & 0.95 \\
    &             & --          & Complete-case            & 0.00 & 0.10 & 0.95 \\
    &             & $\{B, C\}$  & {\color{red} CC MVN (standard)}        & {\color{red} -0.84} & {\color{red} 0.85} & {\color{red} 0.00} \\
    &             & $\{B, C\}$  & {\color{red} Amelia MVN (standard)}    & {\color{red} -0.84} & {\color{red} 0.85} & {\color{red} 0.00} \\
    &             & $\{B, C\}$  & {\color{red} MICE (standard)}          & {\color{red} -0.84} & {\color{red} 0.85} & {\color{red} 0.00} \\
    &             & $C$         & CC MVN (updated)      & 0.00 & 0.10 & 0.95 \\
    &             & $C$         & Amelia MVN (updated)  & 0.00 & 0.10 & 0.95 \\
    &             & $C$         & MICE (updated)        & 0.00 & 0.10 & 0.95 \\
\cline{2-7}
    & \multirow{8}{*}{$E[B \mid A]$}
    & --          & Benchmark                & 0.00 & 0.05 & 0.95 \\
    &             & --          & Complete-case            & 0.00 & 0.05 & 0.95 \\
    &             & $C$         & {\color{red} CC MVN (standard)}        & {\color{red} 0.16} & {\color{red} 0.16} & {\color{red} 0.09} \\
    &             & $C$         & {\color{red} Amelia MVN (standard)}    & {\color{red} 0.16} & {\color{red} 0.16} & {\color{red} 0.09} \\
    &             & $C$         & {\color{red} MICE (standard)}          & {\color{red} 0.15} & {\color{red} 0.16} & {\color{red} 0.10} \\
    &             & --          & CC MVN (updated)      & -- & -- & -- \\
    &             & --          & Amelia MVN (updated)  & -- & -- & -- \\
    &             & --          & MICE (updated)        & -- & -- & -- \\
\hline

(b) & \multirow{5}{*}{$E[A]$}
    & --          & Benchmark                & 0.00 & 0.07 & 0.95 \\
    &             & --          & Complete-case            & 0.00 & 0.11 & 0.95 \\
    &             & $C$         & CC MVN (updated)      & 0.00 & 0.10 & 0.95 \\
    &             & $C$         & Amelia MVN (updated)  & 0.00 & 0.10 & 0.95 \\
    &             & $C$         & MICE (updated)        & 0.00 & 0.10 & 0.95 \\
\cline{2-7}
    & \multirow{8}{*}{$E[B \mid A]$}
    & --          & Benchmark                & 0.00 & 0.03 & 0.95 \\
    &             & --          & Complete-case            & 0.00 & 0.04 & 0.95 \\
    &             & $C$         & {\color{red} CC MVN (standard)}        & {\color{red} 0.11} & {\color{red} 0.11} & {\color{red} 0.17} \\
    &             & $C$         & {\color{red} Amelia MVN (standard)}    & {\color{red} 0.11} & {\color{red} 0.11} & {\color{red} 0.18} \\
    &             & $C$         & {\color{red} MICE (standard)}          & {\color{red} 0.10} & {\color{red} 0.11} & {\color{red} 0.21} \\
    &             & --          & CC MVN (updated)      & -- & -- & -- \\
    &             & --          & Amelia MVN (updated)  & -- & -- & -- \\
    &             & --          & MICE (updated)        & -- & -- & -- \\
\hline

(c) & \multirow{8}{*}{$E[B]$}
    & --          & Benchmark                & 0.00 & 0.07 & 0.96 \\
    &             & --          & {\color{red} Complete-case}            & {\color{red} 0.54} & {\color{red} 0.55} & {\color{red} 0.00} \\
    &             & $\{A, C\}$  & {\color{red} CC MVN (standard)}        & {\color{red} -0.27} & {\color{red} 0.29} & {\color{red} 0.18} \\
    &             & $\{A, C\}$  & {\color{red} Amelia MVN (standard)}    & {\color{red} -0.27} & {\color{red} 0.29} & {\color{red} 0.18} \\
    &             & $\{A, C\}$  & {\color{red} MICE (standard)}          & {\color{red} -0.27} & {\color{red} 0.29} & {\color{red} 0.18} \\
    &             & $A$         & CC MVN (updated)      & 0.00 & 0.10 & 0.95 \\
    &             & $A$         & Amelia MVN (updated)  & 0.00 & 0.10 & 0.95 \\
    &             & $A$         & MICE (updated)        & 0.00 & 0.10 & 0.95 \\
\cline{2-7}
    & \multirow{8}{*}{$E[B \mid A]$}
    & --          & Benchmark                & 0.00 & 0.06 & 0.95 \\
    &             & --          & Complete-case            & 0.00 & 0.09 & 0.95 \\
    &             & $C$         & {\color{red} CC MVN (standard)}        & {\color{red} 0.10} & {\color{red} 0.13} & {\color{red} 0.77} \\
    &             & $C$         & {\color{red} Amelia MVN (standard)}    & {\color{red} 0.10} & {\color{red} 0.13} & {\color{red} 0.77} \\
    &             & $C$         & {\color{red} MICE (standard)}          & {\color{red} 0.10} & {\color{red} 0.13} & {\color{red} 0.77} \\
    &             & $\emptyset$ & CC MVN (updated)      & 0.00 & 0.09 & 0.95 \\
    &             & $\emptyset$ & Amelia MVN (updated)  & 0.00    & 0.09    & 0.95    \\
    &             & $\emptyset$ & MICE (updated)        & 0.00    & 0.09    & 0.95    \\
\hline

\caption{\label{tb:sim_results}$Z$: Auxiliary variables in imputation model. ``CC MVN'': Custom implementation of multivariate joint modeling with the imputation model fit using only the complete cases. ``Standard'': Imputation model included all available covariates. ``Updated'': Imputation model included a sufficient auxiliary set; when no such set existed, this is indicated by ``--''. Standard methods are omitted for graph (b), $E[A]$, because the standard auxiliary set would include the incomplete variable $B$. ``RMSE'': Root mean-square error. ``Coverage'': Coverage probability of 95\% confidence intervals. Methods shown in {\color{red} red} were biased.}
\end{longtable}

\renewcommand{\arraystretch}{1}
\singlespacing
\section{Applied example: Changes in sexual identity}
\label{sec:applied}

\subsection{Datasets}

The Stockholm Public Health Cohort (SPHC; \url{https://www.ces.regionstockholm.se/projekt-och-uppdrag/halsa-stockholm/SPHC-data/}) is a prospective, population-based cohort initiated in Stockholm County, Sweden. To date, five subcohorts have been established in the years 2002, 2006, 2010, 2014, and 2021. For this study, we used data from the 2010 and 2014 subcohorts (i.e., SPHC 2010 and 2014).

At baseline, area-stratified random samples were drawn from the Stockholm County population: individuals aged 18 years and older in SPHC 2010, and 16 years and older in SPHC 2014. Sampling frame was selected from the Total Population Register on 2009-12-31 in SPHC 2010 and 2014-08-31 in SPHC 2014. The approximate sampling frames were 1.6 million in 2010 and 1.8 million in 2014. Participants from SPHC 2010 were followed up via surveys in 2014 and 2021, while those from SPHC 2014 were followed up in 2021. From 2010 onward, sexual identity was assessed using the question: “How do you define your sexual orientation?” with four response options: “heterosexual,” “homosexual,” “bisexual,” and “uncertain” (in 2010), or “none of the above” (in 2014 and 2021). For simplicity, participants were categorized as heterosexual or non-heterosexual. In SPHC 2010, T1 refers to 2010 and T2 to 2014 or 2021; in SPHC 2014, T1 refers to 2014 and T2 to 2021.

In Sweden, each individual registered as a resident is assigned a unique personal identity number, which enables linkage to the national and regional registers. In this study, we linked participants to the National Prescribed Drug Register (\url{https://www.socialstyrelsen.se/en/statistics-and-data/registers/national-prescribed-drug-register/}) to obtain data on dispensed prescriptions of antidepressants and anxiolytics during the first six months following their selection into the sampling frame. This relatively short time frame minimizes the likelihood of participants having temporarily or permanently emigrated after their selection. The definitions of self-reported and registry-based variables are available in Table \ref{tb:applied_var_defns}.

In total, 55,340 individuals were sampled in SPHC 2010 and 53,037 in SPHC 2014. Of these, 30,767 (55.6\%) in SPHC 2010 and 22,250 (42.0\%) in SPHC 2014 responded at baseline. Among participants, in SPHC 2010, 29,607 (96.2\%) reported sexual identity in 2010, 18,087 (58.8\%) in 2014, and 12,930 (42.0\%) in 2021; in SPHC 2014, 20,249 (91.0\%) reported sexual identity in 2014 and 13,885 (62.4\%) in 2021. Meanwhile, 18,518 individuals from SPHC 2010 took part in the 2014 follow-up, and 13,613 in the 2021 follow-up; for SPHC 2014, 14,704 individuals were followed up in 2021. Thus, missingness in sexual identity at follow-up surveys was primarily attributable to loss to follow-up over time. In this study, the analysis was restricted to individuals aged 25–45 years with observed sexual identity at baseline, as this age group is generally considered to be relatively socioeconomically stable. The characteristics of participants are summarized in Tables \ref{tb:baseline_characteristics_1} and \ref{tb:baseline_characteristics_2}.

\singlespacing
\begin{table}[H]
\centering
\resizebox{0.85\columnwidth}{!}{%
\begin{tabular}{@{}lll@{}}
\toprule
\textbf{Variable} & \textbf{Definition} & \textbf{Source} \\
\midrule

\makecell[l]{Sexual identity \\ (T1, T2)} 
& Heterosexual or non-heterosexual 
& Self-report \\ \hline

Age (y) 
& \makecell[l]{SPHC 2010: age in 2010;\\ SPHC 2014: age in 2014} 
& TPR \\ \hline

Sex 
& \makecell[l]{Female or male,\\ as assigned at birth or legally registered} 
& TPR \\ \hline

Country of birth 
& \makecell[l]{Sweden, Europe (excl. Sweden), or\\ Outside Europe} 
& TPR \\ \hline 

Disposable income 
& \makecell[l]{
$\leq$2,500, (2,500, 3,500], (3,500, 4,500],\\ or $>$4,500 (in 100 SEK/y),\\CPI-adjusted to 2024;\\
SPHC 2010: annual income in 2010;\\
SPHC 2014: annual income in 2014} 
& LISA \\ \hline  

Antidepressant use 
& \makecell[l]{Yes or no;\\
SPHC 2010: any dispensed prescription\\ between 2009-12-31 and 2010-06-30;\\
SPHC 2014: between 2014-08-31 and\\ 2015-02-28} 
& \makecell[l]{PDR} \\ \hline 

Anxiolytic use 
& Yes or no; same as above 
& \makecell[l]{PDR} \\

\bottomrule
\end{tabular}
}
\vspace{0.3em}

\begin{minipage}{0.85\columnwidth}
\caption{\label{tb:applied_var_defns}Variable definitions and sources. Sampling frame was selected from the TPR on 2009-12-31 in SPHC 2010 cohort, and 2014-08-31 in SPHC 2014 cohort. In SPHC 2010, T1 refers to 2010 and T2 to 2014 or 2021; in SPHC 2014, T1 refers to 2014 and T2 to 2021. Abbreviations: CPI, Consumer Price Index; LISA, the Swedish Longitudinal Integrated Database for Health Insurance and Labor Market Studies; PDR, Prescribed Drug Register; SEK, Swedish Krona; SPHC, Stockholm Public Health Cohort; TPR, Total Population Register.}
\end{minipage}
\end{table}
\singlespacing

\begin{table}[H]
\centering
\small
\resizebox{1\textwidth}{!}{%
\begin{tabular}{l c c c c}
\toprule
& \multicolumn{2}{c}{\textbf{Sexual Identity in 2010}} & \\
\cmidrule(lr){2-3}
\textbf{Characteristics} & \textbf{Heterosexual (\%)} & \textbf{Non-heterosexual (\%)} & \textbf{Total N (\%)} \\
\midrule

Total N & 9,398 (100) & 434 (100) & 9,832 (100) \\

Age (y), mean (SD) & 36.2 (5.8) & 34.4 (6.1) & 36.1 (5.8) \\

Sex & & & & \\
\quad Male & 4,031 (42.9) & 177 (40.8) & 4,208 (42.8) \\
\quad Female & 5,367 (57.1) & 257 (59.2) & 5,624 (57.2) \\

Country of birth & & & & \\
\quad Sweden & 7,717 (82.1) & 311 (71.7) & 8,028 (81.7) \\
\quad Europe (excl. Sweden) & 789 (8.4) & 40 (9.2) & 829 (8.4) \\
\quad Outside Europe & 892 (9.5) & 83 (19.1) & 975 (9.9) \\

Income (100 SEK/y) & & & & \\
\quad $\leq$2,500 & 2,463 (26.2) & 195 (44.9) & 2,658 (27.0) \\
\quad (2,500, 3,500] & 3,002 (31.9) & 121 (27.9) & 3,123 (31.8) \\
\quad (3,500, 4,500] & 2,050 (21.8) & 57 (13.1) & 2,107 (21.4) \\
\quad $>$4,500 & 1,883 (20.0) & 61 (14.1) & 1,944 (19.8) \\

Use of antidepressants & & & & \\
\quad Yes & 461 (4.9) & 45 (10.4) & 506 (5.1) \\
\quad No & 8,937 (95.1) & 389 (89.6) & 9,326 (94.9) \\

Use of anxiolytics & & & & \\
\quad Yes & 176 (1.9) & 20 (4.6) & 196 (2.0) \\
\quad No & 9,222 (98.1) & 414 (95.4) & 9,636 (98.0) \\

Sexual identity in 2014 & & & & \\
\quad Heterosexual & 4,972 (52.9) & 55 (12.7) & 5,027 (51.1) \\
\quad Non-heterosexual & 193 (2.1) & 170 (39.2) & 363 (3.7) \\
\quad Missing & 4,233 (45.0) & 209 (48.2) & 4,442 (45.2) \\

Sexual identity in 2021 & & & & \\
\quad Heterosexual & 3,759 (40.0) & 37 (8.5) & 3,796 (38.6) \\
\quad Non-heterosexual & 131 (1.4) & 128 (29.5) & 259 (2.6) \\
\quad Missing & 5,508 (58.6) & 269 (62.0) & 5,777 (58.8) \\

\bottomrule
\end{tabular}
}
\caption{\label{tb:baseline_characteristics_1}Characteristics of participants aged 25–45 years with observed sexual identity at baseline in SPHC 2010 cohort. Variable definitions are available in Table \ref{tb:applied_var_defns}. Abbreviations: SD, standard deviation; SEK, Swedish Krona; SPHC, Stockholm Public Health Cohort.}
\end{table}

\begin{table}[H]
\centering
\small
\resizebox{1\textwidth}{!}{%
\begin{tabular}{l c c c c}
\toprule
& \multicolumn{2}{c}{\textbf{Sexual Identity in 2014}} & \\
\cmidrule(lr){2-3}
\textbf{Characteristics} & \textbf{Heterosexual (\%)} & \textbf{Non-heterosexual (\%)} & \textbf{Total N (\%)} \\
\midrule

Total N & 5,480 (100) & 510 (100) & 5,990 (100) \\

Age (y), mean (SD) & 36.2 (5.9) & 35.0 (5.8) & 36.1 (5.9) \\

Sex & & & & \\
\quad Male & 2,393 (43.7) & 180 (35.3) & 2,573 (43.0) \\
\quad Female & 3,087 (56.3) & 330 (64.7) & 3,417 (57.0) \\

Country of birth & & & & \\
\quad Sweden & 4,525 (82.6) & 276 (54.1) & 4,801 (80.2) \\
\quad Europe (excl. Sweden) & 454 (8.3) & 66 (12.9) & 520 (8.7) \\
\quad Outside Europe & 501 (9.1) & 168 (32.9) & 669 (11.2) \\

Income (100 SEK/y) & & & & \\
\quad $\leq$2,500 & 947 (17.3) & 245 (48.0) & 1,192 (19.9) \\
\quad (2,500, 3,500] & 1,452 (26.5) & 115 (22.5) & 1,567 (26.2) \\
\quad (3,500, 4,500] & 1,363 (24.9) & 70 (13.7) & 1,433 (23.9) \\
\quad $>$4,500 & 1,710 (31.2) & 79 (15.5) & 1,789 (29.9) \\
\quad Missing & 8 (0.1) & 1 (0.2) & 9 (0.2) \\

Use of antidepressants & & & & \\
\quad Yes & 395 (7.2) & 65 (12.7) & 460 (7.7) \\
\quad No & 5,085 (92.8) & 445 (87.3) & 5,530 (92.3) \\

Use of anxiolytics & & & & \\
\quad Yes & 156 (2.8) & 30 (5.9) & 186 (3.1) \\
\quad No & 5,324 (97.2) & 480 (94.1) & 5,804 (96.9) \\

Sexual identity in 2021 & & & & \\
\quad Heterosexual & 3,464 (63.2) & 79 (15.5) & 3,543 (59.1) \\
\quad Non-heterosexual & 118 (2.2) & 198 (38.8) & 316 (5.3) \\
\quad Missing & 1,898 (34.6) & 233 (45.7) & 2,131 (35.6) \\

\bottomrule
\end{tabular}
}
\caption{\label{tb:baseline_characteristics_2}Characteristics of participants aged 25–45 years with observed sexual identity at baseline in SPHC 2014 cohort. Variable definitions are available in Table \ref{tb:applied_var_defns}. Abbreviations: SD, standard deviation; SEK, Swedish Krona; SPHC, Stockholm Public Health Cohort.}
\end{table}

\subsection{Methods}

Similarly to previous work led by author WZ \autocite{zhang2024population}, 
we investigated the research question: Are individuals who identify as non-heterosexual more or less likely to report changes in sexual identity over time, compared to those who initially identify as heterosexual? The current analysis falls under existing ethical approvals (DNR: 2022-02626-01, 2023-06610-02, 2024-02155-02) and informs imputation strategies for our ongoing research on sexual identity and health.

To accommodate the mixed-type variables, we used MICE \autocite{van2011mice} to impute missing data separately for females and males, using 5 different sets of auxiliary variables. Because follow-up sexual identity (ID$_2$) was the only incomplete variable, chained-equations imputation reduced to a single imputation model, which is trivially compatible with the joint distribution $p(\text{id}_1, z)\, p(\text{id}_2 \mid \text{id}_1, z)$ and requires no iteration.\autocite[Section 4.3]{van2018flexible} Critically, some models did not include income, while others did (which we expect would lead to bias). We fitted log-binomial regression models of change in sexual identity between T1 and T2 (yes vs. no) on sexual identity at T1, stratified by sex.

\subsection{Results}

All results are presented in Table \ref{tab:imputation_model_summary} and are presented as risk ratios (RR) with 95\% confidence intervals. Across both cohorts, including income in the imputation models consistently attenuated the estimated RRs toward the null. Figure \ref{fig:forest_plots} shows results for the most comprehensive imputation models (with and without income) for the two subcohorts and three follow-up periods.

\begin{table}[H]
\resizebox{\textwidth}{!}{%
\begin{tabular}{p{7cm} c c c c}
\toprule

& \multicolumn{2}{c}{\textbf{Female}} & \multicolumn{2}{c}{\textbf{Male}} \\
\cmidrule(lr){2-3} \cmidrule(lr){4-5}
\textbf{Imputation Model} & \textbf{Sample Size} & \textbf{RR [95\% CI]} & \textbf{Sample Size} & \textbf{RR [95\% CI]} \\
    \midrule
    \multicolumn{5}{l}{
    \parbox{\linewidth}{
        \textbf{SPHC 2010 (followed up until 2014), \( m = 50 \)}
        }
        } \\

  Complete-case analysis & 3,231 & 5.88 [4.20, 7.98] & 2,159 & 8.24 [4.89, 13.18] \\
  Model 1: Sexual identity (T1, T2) & 5,624 & 5.91 [4.38, 7.98] & 4,208 & 8.01 [4.91, 13.07] \\
  Model 2: Sexual identity (T1, T2), age, country of birth & 5,624 & 4.80 [3.50, 6.57] & 4,208 & 5.99 [3.56, 10.06] \\
  Model 3: Sexual identity (T1, T2), age, country of birth, \textbf{income} & 5,624 & 4.51 [3.29, 6.17] & 4,208 & 5.75 [3.60, 9.20] \\
  Model 4: Sexual identity (T1, T2), age, country of birth, antidepressant use, anxiolytic use & 5,624 & 4.83 [3.47, 6.72] & 4,208 & 5.95 [3.73, 9.50] \\
  Model 5: Sexual identity (T1, T2), age, country of birth, \textbf{income}, antidepressant use, anxiolytic use & 5,624 & 4.65 [3.34, 6.46] & 4,208 & 5.53 [3.42, 8.95] \\

  \midrule
  \multicolumn{5}{l}{
    \parbox{\linewidth}{
        \textbf{SPHC 2010 (followed up until 2021), \( m = 70 \)}
        }
        } \\

  Complete-case analysis & 2,463 & 6.17 [3.98, 9.12] & 1,592 & 7.71 [4.26, 13.01] \\
  Model 1: Sexual identity (T1, T2) & 5,624 & 6.07 [4.00, 9.20] & 4,208 & 7.69 [4.43, 13.36] \\
  Model 2: Sexual identity (T1, T2), age, country of birth & 5,624 & 4.72 [2.97, 7.49] & 4,208 & 6.49 [3.39, 12.45] \\
  Model 3: Sexual identity (T1, T2), age, country of birth, \textbf{income} & 5,624 & 4.30 [2.85, 6.49] & 4,208 & 5.65 [3.15, 10.15] \\
  Model 4: Sexual identity (T1, T2), age, country of birth, antidepressant use, anxiolytic use & 5,624 & 4.57 [2.91, 7.16] & 4,208 & 6.16 [3.49, 10.88] \\
  Model 5: Sexual identity (T1, T2), age, country of birth, \textbf{income}, antidepressant use, anxiolytic use & 5,624 & 4.26 [2.70, 6.72] & 4,208 & 5.19 [3.01, 8.94] \\

   \midrule
  \multicolumn{5}{l}{
    \parbox{\linewidth}{
        \textbf{SPHC 2014 (followed up until 2021), \( m = 50 \)}
        }
        } \\

  Complete-case analysis & 2,285 & 8.62 [6.36, 11.62] & 1,574 & 7.85 [4.62, 12.83] \\
  Model 1: Sexual identity (T1, T2) & 3,417 & 8.45 [6.26, 11.41] & 2,573 & 8.07 [4.84, 13.46] \\
  Model 2: Sexual identity (T1, T2), age, country of birth & 3,417 & 7.80 [5.71, 10.65] & 2,573 & 6.46 [3.74, 11.17] \\
  Model 3: Sexual identity (T1, T2), age, country of birth, \textbf{income} & 3,417 & 7.48 [5.53, 10.14] & 2,573 & 6.34 [3.98, 10.11] \\
  Model 4: Sexual identity (T1, T2), age, country of birth, antidepressant use, anxiolytic use & 3,417 & 7.68 [5.54, 10.63] & 2,573 & 6.59 [4.04, 10.75] \\
  Model 5: Sexual identity (T1, T2), age, country of birth, \textbf{income}, antidepressant use, anxiolytic use & 3,417 & 7.34 [5.52, 9.78] & 2,573 & 6.10 [3.65, 10.20] \\
  
  \bottomrule
  \end{tabular}
}

\caption{Effect estimates by imputation model, stratified by sex. In SPHC 2010 cohort, T1 refers to 2010 and T2 to 2014 or 2021; in SPHC 2014 cohort, T1 refers to 2014 and T2 to 2021. The study samples included participants aged 25 to 45 with observed sexual identity at T1. The substantive analysis model used a log-binomial regression, with sexual identity change (yes vs. no) as the outcome and sexual identity at T1 as the exposure. Missingness in sexual identity at T2 was primarily due to loss to follow-up. \( m \) denotes the number of imputed datasets. Bold indicates the assumed collider variable. Abbreviations: CI, confidence interval; RR, risk ratio; SPHC, Stockholm Public Health Cohort.}
\label{tab:imputation_model_summary}
\end{table}

\input{figures/forest_plot_applied_example}

\subsection{Direction of bias from including income}
\label{sec:applied_bias}

Here we show why including income as an auxiliary would plausibly attenuate estimated risk ratios. Write $X := \text{ID}_1$, $Y := \text{ID}_2$, $R := R_{\text{ID}_2}$, and $W_i := \text{Income}$, and let $W_{-i} := W^+ \setminus \{ W_i \}$ denote the other complete auxiliaries, including sex. Suppose that $\text{VMAR}(V \cup W_{-i})$ holds, as in Figure \ref{main-fig:applied_example_dag} with $W_{-i} = \{ \text{Sex} \}$, so that $Y \indM R \mid (X, W_{-i})$, whereas including $W_i$ unblocks the VMAR-violating path $Y \gets U_1 \to W_i \gets U_2 \to R$. Because $Y$ is the only incomplete variable, $\text{MAR}(V \cup W_{-i} \cup \{ W_i \})$ is equivalent to $p(y \mid x, w_{-i}, w_i, R = 0) = p(y \mid x, w_{-i}, w_i, R = 1)$. Thus, with $Z := W_{-i} \cup \{ W_i \}$, the MAR-assuming functional when including income is:
\begingroup
\addtolength{\jot}{0.5em}
\begin{align*}
\pMAR{V \cup Z}(y \mid x) &= \sum_{w_{-i}, w_i} p(y \mid x, w_{-i}, w_i, R = 1) \, p(w_{-i} \mid x) \, p(w_i \mid x, w_{-i}),
\end{align*}
\endgroup
whereas, because $\text{VMAR}(V \cup W_{-i})$ implies $\text{MAR}(V \cup W_{-i})$, the true conditional is
\begingroup
\addtolength{\jot}{0.5em}
{\footnotesize
\begin{align*}
p(y \mid x) &= \sum_{w_{-i}} p(y \mid x, w_{-i}, R = 1) \, p(w_{-i} \mid x) = \sum_{w_{-i}, w_i} p(y \mid x, w_{-i}, w_i, R = 1) \, p(w_{-i} \mid x) \, p(w_i \mid x, w_{-i}, R = 1).
\end{align*}%
}%
\endgroup
Sums are replaced by integrals for continuous components, such as age. Subtracting, and using $p(w_{-i} \mid x) \, p(R = 0 \mid x, w_{-i}) = p(w_{-i}, R = 0 \mid x)$ and
\begingroup
\addtolength{\jot}{0.5em}
{\footnotesize
\begin{align*}
p(w_i \mid x, w_{-i}) - p(w_i \mid x, w_{-i}, R = 1) &= p(R = 0 \mid x, w_{-i}) \big[ p(w_i \mid x, w_{-i}, R = 0) - p(w_i \mid x, w_{-i}, R = 1) \big],
\end{align*}%
}%
\endgroup
gives
\begingroup
\addtolength{\jot}{0.5em}
\begin{align*}
\pMAR{V \cup Z}(y \mid x) - p(y \mid x) &= \sum_{w_{-i}} p(w_{-i}, R = 0 \mid x) \sum_{w_i} p(y \mid x, w_{-i}, w_i, R = 1) \\
&\qquad \times \big[ p(w_i \mid x, w_{-i}, R = 0) - p(w_i \mid x, w_{-i}, R = 1) \big].
\end{align*}
\endgroup

Within each stratum of $(X, W_{-i})$, the bias is thus the average of the respondents' regression $p(y \mid x, w_{-i}, w_i, R = 1)$ over the income distribution of nonrespondents, minus its average over the income distribution of respondents. Each factor corresponds to one subpath of the VMAR-violating path: income is distributed differently in nonrespondents and respondents only through $W_i \gets U_2 \to R$, and the respondents' regression varies with income only through $W_i \gets U_1 \to Y$. If either subpath were absent, the bias would be zero.

The estimand is the probability of identity change, $p(Y \ne x \mid x)$, whose bias is obtained by summing the expression above over $y \ne x$. To sign it heuristically, suppose the relevant effects are monotone within strata of $W_{-i}$. Parental leave reduces income;\autocite{kleven2019child} suppose it also increases response. Then nonrespondents tend to have higher income than respondents. Stigmatization reduces income;\autocite{ahmed2013gay} suppose it also makes reporting a heterosexual identity at follow-up more likely. Then, among respondents, higher income indicates less stigmatization and hence a lower risk of change for baseline non-heterosexual participants (for whom change is to a heterosexual identity), but a higher risk of change for baseline heterosexual participants (for whom change is to a non-heterosexual identity). The bias in $p(Y \ne x \mid x)$ is therefore negative for non-heterosexual $x$ and positive for heterosexual $x$, so the risk ratio is biased downward, which attenuates risk ratios above 1, as in our estimates.

\pagebreak
\scriptsize
\printbibliography

%% file: toggles.tex
\newcommand{\annotationMode}{cameraReady}

\newcommand{\blindedReview}{no}

\newcommand{\lineNumbers}{no}

\newcommand{\abstractWordLimit}{300}

\newcommand{\bodyStretch}{1.655}

\newcommand{\mainFileName}{main}

%% file: preamble_files/preamble_main.tex
\input{preamble_files/preamble_shared.tex}

\usepackage{abstract}

\makeatletter
\NewCommandCopy\mm@abstractbegin\abstract
\NewCommandCopy\mm@abstractend\endabstract
\RenewDocumentEnvironment{abstract}{+b}
  {\ifshowAnnotations\mm@countAbstract{#1}\fi
   \mm@abstractbegin#1\mm@abstractend}
  {}
\makeatother

\newcommand{\wordCountSection}{%
  \ifshowAnnotations
    \clearpage
    {\color{red}%
     \section*{Word count}%
     \detailtexcount{\mainFileName}}%
  \fi}

\input{xref_supp_labels}  %

\title{\vspace{-15mm}\fontsize{15pt}{15pt}\selectfont\textbf{A mechanistic characterization of missingness at random}}
\date{\vspace{-5ex}}

\usepackage{authblk}

\ifredact
  \author{[Authors redacted for blinded review]}
\else
  \author[1]{\normalsize Maya B. Mathur\thanks{Correspondence to: Maya B. Mathur (\href{mailto:mmathur@stanford.edu}{mmathur@stanford.edu}), Quantitative Sciences Unit, 3180 Porter Drive, Palo Alto, CA, 94304.}}
  \author[2,3]{Willi Zhang}
  \author[4]{Shaun Seaman}
  \author[5]{Rohit Bhattacharya}
  \author[6]{Ilya Shpitser}

  \affil[1]{Quantitative Sciences Unit and Department of Pediatrics, Stanford University, Palo Alto, CA, USA}
  \affil[2]{Centre for Epidemiology and Community Medicine, Region Stockholm, Stockholm, Sweden}
  \affil[3]{Department of Global Public Health, Karolinska Institutet, Stockholm, Sweden}
  \affil[4]{MRC Biostatistics Unit, University of Cambridge, Cambridge, UK}
  \affil[5]{Department of Computer Science, Williams College, Williamstown, MA, USA}
  \affil[6]{Department of Computer Science, Whiting School of Engineering, Johns Hopkins University, Baltimore, MD, USA}
\fi

%% file: preamble_files/preamble_shared.tex

\providecommand{\annotationMode}{draft}
\providecommand{\blindedReview}{no}
\providecommand{\lineNumbers}{no}
\providecommand{\abstractWordLimit}{200}
\providecommand{\bodyStretch}{1.655}
\providecommand{\mainFileName}{main}

\newif\ifshowAnnotations %
\newif\ifshowSecrets   %
\newif\ifshowhlDef     %
\newif\ifredact        %
\newif\ifshowLineNumbers
\newif\ifcameraReady     %

\makeatletter
\newcommand{\mm@lower}[2]{%
  \edef#1{#2}%
  \lowercase\expandafter{\expandafter\def\expandafter#1\expandafter{#1}}}
\newcommand{\mm@yesno}[4]{%
  \mm@lower\mm@tmpa{#1}%
  \def\mm@tmpb{yes}\def\mm@tmpc{true}%
  \ifx\mm@tmpa\mm@tmpb #2\else
  \ifx\mm@tmpa\mm@tmpc #2\else
  \def\mm@tmpb{no}\def\mm@tmpc{false}%
  \ifx\mm@tmpa\mm@tmpb #3\else
  \ifx\mm@tmpa\mm@tmpc #3\else
    \PackageError{toggles}{#4 must be 'yes' or 'no' (got '#1')}{Fix it in toggles.tex.}%
  \fi\fi\fi\fi}

\mm@lower\mm@mode{\annotationMode}
\def\mm@tmpb{draft}%
\ifx\mm@mode\mm@tmpb
  \showAnnotationstrue \showSecretstrue \showhlDeftrue
  \def\mm@modeName{draft}%
\else\def\mm@tmpb{hidesecrets}%
\ifx\mm@mode\mm@tmpb
  \showAnnotationstrue \showSecretsfalse \showhlDeftrue
  \def\mm@modeName{hideSecrets}%
\else\def\mm@tmpb{justhldef}%
\ifx\mm@mode\mm@tmpb
  \showAnnotationsfalse \showSecretsfalse \showhlDeftrue
  \def\mm@modeName{justhlDef}%
\else\def\mm@tmpb{cameraready}%
\ifx\mm@mode\mm@tmpb
  \showAnnotationsfalse \showSecretsfalse \showhlDeffalse \cameraReadytrue
  \def\mm@modeName{cameraReady}%
\else
  \PackageError{toggles}{annotationMode must be draft, hideSecrets, justhlDef, or cameraReady (got '\annotationMode')}{Fix it in toggles.tex.}%
  \showAnnotationstrue \showSecretstrue \showhlDeftrue \def\mm@modeName{draft}%
\fi\fi\fi\fi

\mm@yesno{\blindedReview}{\redacttrue}{\redactfalse}{blindedReview}
\mm@yesno{\lineNumbers}{\showLineNumberstrue}{\showLineNumbersfalse}{lineNumbers}
\makeatother

\usepackage[T1]{fontenc}
\usepackage{setspace}
\usepackage{microtype}
\usepackage[margin=1in,headheight=15pt,marginparwidth=2cm]{geometry}
\usepackage[hang,small,labelfont={bf,up},textfont={it,up}]{caption}
\usepackage{subcaption}
\usepackage{booktabs, makecell, multirow, array}
\usepackage{siunitx}
\usepackage{float}
\usepackage{graphicx}
\usepackage{paralist}
\usepackage{outlines}
\usepackage[svgnames]{xcolor}
\usepackage{comment}
\usepackage{verbatim}
\usepackage[bottom]{footmisc}   %
\usepackage{amsmath, amsfonts, amssymb, amsthm, textcomp, bm, bbm, dsfont}
\usepackage{sansmath}           %
\usepackage{soul}
\soulregister\ref7              %
\usepackage[normalem]{ulem}     %
\usepackage{xspace}
\usepackage{refcount}           %
\usepackage{catchfile}          %
\usepackage{pdftexcmds}         %
\usepackage{algorithm, algorithmic}

\usepackage{lineno}

\newcommand{\bodyspacing}{\setstretch{\bodyStretch}}

\usepackage{titlesec}
\titleformat{\section}[block]{\large\bfseries}{\thesection.}{1em}{}
\titleformat{\subsection}[block]{\large}{\thesubsection.}{1em}{}

\renewcommand{\thefootnote}{\textit{\alph{footnote}}}

\makeatletter
\renewcommand*\env@matrix[1][\arraystretch]{%
  \edef\arraystretch{#1}%
  \hskip -\arraycolsep
  \let\@ifnextchar\new@ifnextchar
  \array{*\c@MaxMatrixCols c}}
\renewcommand{\boxed}[1]{\text{\fboxsep=.2em\fbox{\m@th$\displaystyle#1$}}}
\makeatother

\newcolumntype{P}[1]{>{\centering\arraybackslash}p{#1}}

\NewCommandCopy\oldtextbf\textbf
\RenewDocumentCommand{\textbf}{m}{{\boldmath\oldtextbf{#1}}}

\usepackage{listings}
\usepackage{fancyhdr}
\ifshowLineNumbers
  \newcommand*\patchAmsMathEnvironmentForLineno[1]{%
    \expandafter\let\csname old#1\expandafter\endcsname\csname #1\endcsname
    \expandafter\let\csname oldend#1\expandafter\endcsname\csname end#1\endcsname
    \renewenvironment{#1}%
      {\linenomath\csname old#1\endcsname}%
      {\csname oldend#1\endcsname\endlinenomath}}%
  \newcommand*\patchBothAmsMathEnvironmentsForLineno[1]{%
    \patchAmsMathEnvironmentForLineno{#1}%
    \patchAmsMathEnvironmentForLineno{#1*}}%
  \AtBeginDocument{%
    \patchBothAmsMathEnvironmentsForLineno{equation}%
    \patchBothAmsMathEnvironmentsForLineno{align}%
    \patchBothAmsMathEnvironmentsForLineno{flalign}%
    \patchBothAmsMathEnvironmentsForLineno{alignat}%
    \patchBothAmsMathEnvironmentsForLineno{gather}%
    \patchBothAmsMathEnvironmentsForLineno{multline}%
    \linenumbers}
\fi

\definecolor{junglegreen}{rgb}{0.16, 0.67, 0.53}
\definecolor{lav}{rgb}{0.71, 0.49, 0.86}

\makeatletter
\newcommand{\mm@annotcolor}[2]{\ifshowAnnotations\textcolor{#1}{#2}\else#2\fi}
\newcommand{\red}{\mm@annotcolor{red}}
\newcommand{\orange}{\mm@annotcolor{orange}}
\newcommand{\blue}{\mm@annotcolor{blue}}
\newcommand{\pink}{\mm@annotcolor{pink}}
\newcommand{\lav}{\mm@annotcolor{lav}}
\newcommand{\green}{\mm@annotcolor{junglegreen}}
\newcommand{\ilya}{\mm@annotcolor{red!70!blue}}

\ifshowAnnotations\else\fi

\newcommand{\mm@colorfootnote}[2]{%
  \ifshowAnnotations
    \begingroup
    \renewcommand{\thefootnote}{\textcolor{#1}{\alph{footnote}}}%
    \footnote{\textcolor{#1}{#2}}%
    \endgroup
  \fi}
\newcommand{\redf}{\mm@colorfootnote{red}}
\newcommand{\orangef}{\mm@colorfootnote{orange}}
\newcommand{\bluef}{\mm@colorfootnote{blue}}
\newcommand{\greenf}{\mm@colorfootnote{junglegreen}}
\makeatother

\ifshowSecrets
  \newcommand{\secretf}[1]{\footnote{\textcolor{junglegreen}{#1}}}
  \newcommand{\secreti}[1]{\textcolor{junglegreen}{#1}}

\else
  \newcommand{\secretf}[1]{}
  \newcommand{\secreti}[1]{\unskip}   %
  \NewDocumentEnvironment{secret}{+b}{}{\ignorespacesafterend}
  
\fi

\ifshowAnnotations
  
  \newcommand{\rc}[1][]{\textsuperscript{\orange{(cite)}}\xspace}

  \usepackage{todonotes}
\else
  
  \newcommand{\rc}[1][]{\unskip}

  \usepackage[disable]{todonotes}
\fi

\ifshowhlDef

\else

\fi

\makeatletter
\newcommand{\annotationBanner}{%
  \ifshowhlDef
    {\scriptsize\sffamily\color{gray}%
     mode: \mm@modeName\ifredact\ \textbullet\ BLINDED\fi\ \textbullet\ toggles.tex}%
  \fi}
\makeatother
\ifcameraReady
  \AtBeginDocument{\fancyhead{}}
\fi

\usepackage{pgfplotstable}
\pgfplotsset{compat=1.18}
\usepackage{csvsimple}

\usepackage{tikz}
\usetikzlibrary{shapes,decorations,arrows,calc,arrows.meta,fit,positioning,
  decorations.pathreplacing,shapes.arrows,shapes.geometric,shapes.multipart,
  decorations.pathmorphing,swigs}
\tikzset{
    -Latex,auto,node distance =1 cm and 1 cm,semithick,
    state/.style ={ellipse, draw, minimum width = 0.7 cm},
    point/.style = {circle, draw, inner sep=0.04cm,fill,node contents={}},
    bidirected/.style={Latex-Latex,dashed},
    el/.style = {inner sep=2pt, align=left, sloped}
}

\newtheoremstyle{spaceythm}
  {2\topsep} %
  {2\topsep} %
  {}         %
  {}         %
  {\bfseries}%
  {.}        %
  {.5em}     %
  {}         %
\theoremstyle{spaceythm}
\newtheorem{theorem}{Theorem}
\newtheorem{assumption}{Assumption}

\newtheorem{lemma}{Lemma}
\newtheorem{corollary}{Corollary}
\newtheorem{definition}{Definition}

\newcommand{\thistheoremname}{}
\newtheorem{spaceythmgeneric}[theorem]{\thistheoremname}

\newtheorem*{spaceythmgeneric*}{\thistheoremname}
\makeatletter
\newcounter{namedthmctr}
\newenvironment{namedthm*}[1]
  {\renewcommand{\thistheoremname}{#1}%
   \def\thm@name{#1}%
   \refstepcounter{namedthmctr}%
   \let\thm@oldlabel\label
   \renewcommand{\label}[1]{%
     \begingroup\protected@edef\@currentlabel{\thm@name}\thm@oldlabel{##1}\endgroup}%
   \begin{spaceythmgeneric*}}
  {\end{spaceythmgeneric*}\let\label\thm@oldlabel}
\makeatother

\newtheoremstyle{named}{}{}{\itshape}{}{\bfseries}{.}{.5em}{\thmnote{#3}}
\theoremstyle{named}

\theoremstyle{spaceythm}

\usepackage[american]{babel}
\usepackage[style=american]{csquotes}
\usepackage[backend=biber,style=vancouver,autocite=superscript,sorting=none,
            sortcites=true,citestyle=numeric-comp]{biblatex}
\renewbibmacro{in:}{}   %

\newcommand{\redactedText}{[REDACTED FOR BLINDED REVIEW]}
\newcommand{\maybeRedact}[1]{\ifredact\redactedText\else#1\fi}

\ifredact
  \DeclareSourcemap{
    \maps[datatype=bibtex]{
      \map[overwrite=true]{
        \step[fieldsource=keywords, match=\regexp{redactme}, final]
        \step[fieldset=author, fieldvalue={{[REDACTED-FOR-BLINDED-REVIEW]}}]
        \step[fieldset=url,  null]
        \step[fieldset=doi,  null]
        \step[fieldset=note, null]
      }
    }
  }
\fi

\makeatletter
\newcommand{\mm@texcount}[2]{%
  \ifnum\pdf@shellescape=\@ne
    \immediate\write18{texcount #1 > #2}%
  \fi}

\newcommand{\detailtexcount}[1]{%
  \ifnum\pdf@shellescape=\@ne
    \mm@texcount{-merge -sum -q #1.tex \jobname.bbl}{\jobname.wcdetail}%
    \verbatiminput{\jobname.wcdetail}%
  \else
    (Word count needs shell escape, which is off for this compile.)%
  \fi}

\newcommand{\quickwordcount}[1]{%
  \mm@texcount{-1 -sum -merge -q #1.tex \jobname.bbl}{\jobname-words.sum}%
  \input{\jobname-words.sum} words}

\newcommand{\quickcharcount}[1]{%
  \mm@texcount{-1 -sum -merge -char -q #1.tex \jobname.bbl}{\jobname-chars.sum}%
  \input{\jobname-chars.sum} characters (not including spaces)}

\newcommand{\abstractWordCount}{?}
\newwrite\mm@abswrite
\newif\ifabstractOverLimit
\ExplSyntaxOn
\cs_generate_variant:Nn \regex_match:nnTF { nV }
\cs_new_protected:Npn \mm_validate_count:N #1
  {
    \tl_gset:Nx #1 { \tl_trim_spaces:o {#1} }
    \regex_match:nVTF { \A \d+ \Z } #1
      {
        \int_compare:nNnTF {#1} > { \abstractWordLimit }
          { \global\abstractOverLimittrue } { \global\abstractOverLimitfalse }
      }
      { \tl_gset:Nn #1 { ? } \global\abstractOverLimitfalse }
  }
\NewDocumentCommand{\mmValidateCount}{m}{ \mm_validate_count:N #1 }
\ExplSyntaxOff
\newcommand{\mm@countAbstract}[1]{%
  \ifnum\pdf@shellescape=\@ne
    \immediate\openout\mm@abswrite=\jobname-abstract.wctmp\relax
    \immediate\write\mm@abswrite{\string\input{preamble_files/texcount_rules}}%
    \immediate\write\mm@abswrite{\unexpanded{#1}}%
    \immediate\closeout\mm@abswrite
    \mm@texcount{-1 -sum -merge -q -utf8 \jobname-abstract.wctmp}{\jobname-abstract.wc}%
    \IfFileExists{\jobname-abstract.wc}%
      {\newcommand{\abstractWordCount}{\input{\jobname-abstract.wc}}}{}%
  \fi
  \mmValidateCount\abstractWordCount}
\makeatother

\usepackage{xr-hyper}
\usepackage{hyperref}
\hypersetup{colorlinks=true, linkcolor=black, urlcolor=cyan, citecolor=black}

%% file: preamble_files/texcount_rules.tex

%% file: preamble_files/preamble_notation.tex
\usepackage{longtable}
\usepackage{xurl} %
\usepackage{lettrine}

\definecolor{labelA}{HTML}{1F5FA9}   %
\definecolor{labelB}{HTML}{C1531A}   %

\DeclareFieldFormat*{journaltitle}{\mkbibemph{#1}}

\tikzset{
    imap/.style={node distance=2cm and 3cm, >={Latex[length=5pt, width=5pt]}},
    lab/.style={font=\footnotesize},
    dep/.style={align=center, text width=4.4cm},
    mealli/.style={draw=black!50, text=black!50},
    mealliN/.style={text=black!50},
}

\newcommand{\arrowheadScale}{1.0}  %
\newcommand{\directedEdgeColor}{black}
\newcommand{\bidirectedEdgeColor}{black}
\newlength{\hnodesep}
\newlength{\vnodesep}
\def\ind{\perp\!\!\!\perp}
\newcommand{\notind}{\not\!\perp\!\!\!\perp}

\newcommand{\ct}{\widetilde{c}}

\newcommand{\pd}{p_d}

\newcommand{\mis}[1]{1^{\text{mis}}}
\newcommand{\obs}[1]{1^{\text{obs}}}

\newcommand{\Sobs}{S^{\text{obs}}}
\newcommand{\Smis}{S^{\text{mis}}}
\newcommand{\sobs}{s^{\text{obs}}}
\newcommand{\smis}{s^{\text{mis}}}

\newcommand{\supp}{\operatorname{support}}

\newcommand{\pMAR}[1]{p^{\text{MAR}}_{#1}} %

\newcommand{\Xbar}[1]{\overline{X}}
\newcommand{\XbarOne}[1]{\overline{X}^{(1)}}

\renewcommand{\xi}{x^i}

\newcommand{\m}{\mathcal{M}} %
\newcommand{\indM}{\ind_{p}}
\newcommand{\notindM}{\notind_{p}}

\newcommand{\An}{\operatorname{An}} %
\newcommand{\Pa}{\operatorname{Pa}} %
\newcommand{\pa}{\operatorname{pa}} %
\newcommand{\Z}{Z_{V,W}}

\newcommand{\MARS}{\text{MAR}(S) }
\newcommand{\MNARS}{\text{MNAR}(S) }
\newcommand{\VMARS}{\text{VMAR}(S) }

%% file: figures/c_inactivity_dags.tex
\begin{figure}[H]
\centering
\tikzset{
  ell/.style={draw, fill=white, inner sep=2pt, line width=1.5pt, shape=ellipse},
  every path/.style={>=Stealth, line width=1.2pt, >={Stealth[scale=\arrowheadScale]}},
  inact/.style={->, dotted, line width=1.2pt, color=blue, >=stealth}
}
\begin{subfigure}[b]{0.23\textwidth}
\centering
\begin{tikzpicture}[scale=0.8, transform shape]
\node[name=A, ell]{$A$};
\node[name=B, ell, right=\hnodesep of A]{$B$};
\node[name=D, ell, right=\hnodesep of B]{$D$};
\draw[->, line width=1.2pt, >=stealth]
(D) edge (B);
\draw[inact]
(A) edge (B);
\end{tikzpicture}
\caption{Direct interaction}
\end{subfigure}
\hfill
\begin{subfigure}[b]{0.36\textwidth}
\centering
\begin{tikzpicture}[scale=0.8, transform shape]
\node[name=A, ell]{$A$};
\node[name=B, ell, right=\hnodesep of A]{$B$};
\node[name=K, ell, right=\hnodesep of B]{$K$};
\node[name=D, ell, right=\hnodesep of K]{$D$};
\draw[->, line width=1.2pt, >=stealth]
(K) edge (B)
(K) edge (D);
\draw[inact]
(A) edge (B);
\end{tikzpicture}
\caption{Coarsened partial determinism}
\end{subfigure}
\hfill
\begin{subfigure}[b]{0.38\textwidth}
\centering
\begin{tikzpicture}[scale=0.8, transform shape]
\node[name=D, ell]{$D$};
\node[name=A, ell, right=\hnodesep of D]{$A$};
\node[name=B, ell, right=\hnodesep of A]{$B$};
\draw[->, line width=1.2pt, >=stealth]
(D) edge (A);
\draw[inact]
(A) edge (B);
\end{tikzpicture}
\caption{Restriction on the support of $A$}
\end{subfigure}
\caption{\label{fig:c_inactivity_dags}Three causal mechanisms that can generate c-inactivity. These are not exhaustive. In each panel, the blue dotted edge $A \to B$ is c-inactive in context $D = d$. (a) $D$ interacts with $A$ in $f_B$ such that $A$ does not affect $B$ when $D = d$. (b) $D$ is not a parent of $B$, but $D = d$ implies $K = k$, and $K$ interacts with $A$ in $f_B$ such that $A$ does not affect $B$ when $K = k$. (c) $D = d$ confines $A$ to values across which $f_B$ does not vary (e.g., a single value).}
\end{figure}

%% file: figures/tikz_monotone_MAR.tex
\begin{figure}[!htbp]
\centering
\tikzset{
  ell/.style={draw,fill=white, inner sep=2pt, line width=1.5pt, font=\Large},
  swig vsplit={gap=3pt, line color right=red},
  every path/.style={>=Stealth, line width=1.2pt, >={Stealth[scale=\arrowheadScale]}},
  csi/.style={->, line width=1.2pt, dotted, color=blue, >=stealth},
  determ/.style={->, line width=1.2pt, >=stealth},
  determlab/.style={midway, below=10pt, font=\small, inner sep=1pt, text=black}
}%
\begin{subfigure}[t]{0.45\textwidth}
\centering
\begin{tikzpicture}[scale=0.68, transform shape]

\node[name=L1, ell, shape=ellipse, fill={rgb:black,1;white,2}]{$L_1$};
\node[name=L2, ell, shape=ellipse, right=40mm of L1, fill={rgb:black,1;white,2}]{$L_2$};
\node[name=L3, ell, shape=ellipse, right=40mm of L2, fill={rgb:black,1;white,2}]{$L_3$};

\node[name=RL1, ell, shape=ellipse, below=20mm of L1]{$R_{L_1}$};
\node[name=RL2, ell, shape=ellipse, below=20mm of L2]{$R_{L_2}$};
\node[name=RL3, ell, shape=ellipse, below=20mm of L3]{$R_{L_3}$};

\draw[->,line width=1.2pt,>=stealth]
(L1) edge (L2)
(L2) edge (L3);

\draw[->,line width=1.2pt,>=stealth]
(L1) edge[out=20,in=160] (L3);

\draw[csi]
(L1) edge node[midway, color=blue, sloped, below, font=\normalsize] {$R_{L_1} = 0$} (RL2);

\draw[csi]
(L1) edge node[midway, sloped, color=blue, below, font=\normalsize] {$R_{L_2} = 0$} (RL3);

\draw[csi]
(L2) edge node[midway, sloped, color=blue, below, font=\normalsize] {$R_{L_2} = 0$} (RL3);

\draw[determ]
(RL1) -- node[determlab] {$R_{L_1} = 0 \Rightarrow R_{L_2} = 0$} (RL2);

\draw[csi]
(RL2) -- node[pos=0.3, above, color=blue, font=\normalsize] {$R_{L_1} = 0$}
         node[determlab] {$R_{L_2} = 0 \Rightarrow R_{L_3} = 0$} (RL3);
\end{tikzpicture}
\caption{$G$}
\end{subfigure}
\hfill%
\begin{subfigure}[t]{0.45\textwidth}
\centering
\begin{tikzpicture}[scale=0.68, transform shape]

\node[name=L1, ell, shape=ellipse, fill={rgb:black,1;white,2}]{$L_1$};
\node[name=L2, ell, shape=ellipse, right=40mm of L1, fill={rgb:black,1;white,2}]{$L_2$};
\node[name=L3, ell, shape=ellipse, right=40mm of L2, fill={rgb:black,1;white,2}]{$L_3$};

\node[name=RL1, ell, shape=ellipse, below=20mm of L1]{$R_{L_1}$};
\node[name=RL2, ell, shape=ellipse, below=20mm of L2]{$R_{L_2}$};
\node[name=RL3, ell, shape=ellipse, below=20mm of L3]{$R_{L_3}$};

\draw[->,line width=1.2pt,>=stealth]
(L1) edge (L2)
(L2) edge (L3);

\draw[->,line width=1.2pt,>=stealth]
(L1) edge[out=20,in=160] (L3);

\draw[determ]
(RL1) -- node[determlab] {$R_{L_1} = 0 \Rightarrow R_{L_2} = 0$} (RL2);
\end{tikzpicture}
\caption{$G[\{R_{L_1} = 0\}]$}
\end{subfigure}

\vspace{0.6cm}

\begin{subfigure}[t]{0.45\textwidth}
\centering
\begin{tikzpicture}[scale=0.68, transform shape]

\node[name=L1, ell, shape=ellipse, fill={rgb:black,1;white,2}]{$L_1$};
\node[name=L2, ell, shape=ellipse, right=40mm of L1, fill={rgb:black,1;white,2}]{$L_2$};
\node[name=L3, ell, shape=ellipse, right=40mm of L2, fill={rgb:black,1;white,2}]{$L_3$};

\node[name=RL1, ell, shape=ellipse, below=20mm of L1]{$R_{L_1}$};
\node[name=RL2, ell, shape=ellipse, below=20mm of L2]{$R_{L_2}$};
\node[name=RL3, ell, shape=ellipse, below=20mm of L3]{$R_{L_3}$};

\draw[->,line width=1.2pt,>=stealth]
(L1) edge (L2)
(L2) edge (L3);

\draw[->,line width=1.2pt,>=stealth]
(L1) edge[out=20,in=160] (L3);

\draw[->,line width=1.2pt,>=stealth]
(L1) edge (RL2);

\draw[determ]
(RL1) -- node[determlab] {$R_{L_1} = 0 \Rightarrow R_{L_2} = 0$} (RL2);
\draw[determ]
(RL2) -- node[determlab] {$R_{L_2} = 0 \Rightarrow R_{L_3} = 0$} (RL3);
\end{tikzpicture}
\caption{$G[\{R_{L_2} = 0\}]$}
\end{subfigure}
\hfill%
\begin{subfigure}[t]{0.45\textwidth}
\centering
\begin{tikzpicture}[scale=0.68, transform shape]

\node[name=L1, ell, shape=ellipse, fill={rgb:black,1;white,2}]{$L_1$};
\node[name=L2, ell, shape=ellipse, right=40mm of L1, fill={rgb:black,1;white,2}]{$L_2$};
\node[name=L3, ell, shape=ellipse, right=40mm of L2, fill={rgb:black,1;white,2}]{$L_3$};

\node[name=RL1, ell, shape=ellipse, below=20mm of L1]{$R_{L_1}$};
\node[name=RL2, ell, shape=ellipse, below=20mm of L2]{$R_{L_2}$};
\node[name=RL3, ell, shape=ellipse, below=20mm of L3]{$R_{L_3}$};

\draw[->,line width=1.2pt,>=stealth]
(L1) edge (L2)
(L2) edge (L3);

\draw[->,line width=1.2pt,>=stealth]
(L1) edge[out=20,in=160] (L3);

\draw[->,line width=1.2pt,>=stealth]
(L1) edge (RL2)
(L1) edge (RL3)
(L2) edge (RL3);

\draw[determ]
(RL1) -- node[determlab] {$R_{L_1} = 0 \Rightarrow R_{L_2} = 0$} (RL2);
\draw[determ]
(RL2) -- node[determlab] {$R_{L_2} = 0 \Rightarrow R_{L_3} = 0$} (RL3);
\end{tikzpicture}
\caption{$G[\{R_{L_3} = 0\}] = G$}
\end{subfigure}
\caption{\label{fig:monotone_MAR}DAGs depicting monotone MAR with no complete variables, with and without pruning. Annotations below edges between members of $R$: determinism due to monotone missingness. Blue dotted edges are annotated with their labels, i.e., contexts in which they are c-inactive; each is also c-inactive in any context that forces the annotated value. %
}
\end{figure}

%% file: figures/tikz_vmar_vmnar.tex
\begin{figure}[!htbp]
\centering
\begin{subfigure}[t]{0.31\textwidth}
\centering
\begin{tikzpicture}[scale=1, transform shape]
  \tikzset{
    ell/.style={draw,fill=white, inner sep=2pt, line width=1.5pt},
    swig vsplit={gap=3pt, line color right=red},
    every path/.style={>=Stealth, line width=1.2pt, >={Stealth[scale=\arrowheadScale]}}
  }
  \node[name=L1, ell, shape=ellipse]{$L_1$};
  \node[name=L2,  ell, shape=ellipse, right=\hnodesep of L1, fill={rgb:black,1;white,2}]{$L_2$};
  \node[name=L3,  ell, shape=ellipse, right=\hnodesep of L2, fill={rgb:black,1;white,2}]{$L_3$};
  \node[name=RL2, ell, shape=ellipse, below=\vnodesep of L2]{$R_{L_2}$};
  \node[name=RL3, ell, shape=ellipse, below=\vnodesep of L3]{$R_{L_3}$};
  \draw[->,line width=1.2pt,>=stealth]
    (L1) edge (L2)
    (L2) edge (L3)
    (L1) edge (RL2)
    (L1) edge (RL3);
  \draw[->,line width=1.2pt,>=stealth]
    (L1) edge[out=30,in=150] (L3);
  \draw[->, dotted, line width=1.2pt, color=red, >=stealth]
    (L2) edge (RL2)
    (L2) edge (RL3)
    (L3) edge (RL2)
    (L3) edge (RL3);
\end{tikzpicture}
\caption{}
\end{subfigure}\hfill
\begin{subfigure}[t]{0.31\textwidth}
\centering
\begin{tikzpicture}[scale=1, transform shape]
  \tikzset{
    ell/.style={draw, fill=white, inner sep=2pt, line width=1.5pt},
    every path/.style={>=Stealth, line width=1.2pt, >={Stealth[scale=1.0]}}
  }
  \node[name=L1,  ell, shape=ellipse, fill={rgb:black,1;white,2}]{$L_1$};
  \node[name=U1,  ell, shape=ellipse, right=\hnodesep of L1]{$U_1$};
  \node[name=L2,  ell, shape=ellipse, right=\hnodesep of U1]{$L_2$};
  \node[name=U2,  ell, shape=ellipse, below=\vnodesep of U1]{$U_2$};
  \node[name=RL1, ell, shape=ellipse, below=\vnodesep of L1]{$R_{L_1}$};
  \draw[->, dotted, line width=1.2pt, color=red, >=stealth]
    (U1) edge (L1)
    (U1) edge (L2)
    (U2) edge (L2)
    (U2) edge (RL1);
\end{tikzpicture}
\caption{}
\end{subfigure}\hfill
\begin{subfigure}[t]{0.31\textwidth}
\centering
\begin{tikzpicture}[scale=1, transform shape]
  \tikzset{
    ell/.style={draw, fill=white, inner sep=2pt, line width=1.5pt},
    every path/.style={>=Stealth, line width=1.2pt, >={Stealth[scale=1.0]}}
  }
  \node[name=L1,  ell, shape=ellipse, fill={rgb:black,1;white,2}]{$L_1$};
  \node[name=L2,  ell, shape=ellipse, right=\hnodesep of L1]{$L_2$};
  \node[name=RL1, ell, shape=ellipse, below=\vnodesep of L1]{$R_{L_1}$};
  \draw[<->, dotted, line width=1.2pt, color=red, >=stealth]
    (L1) edge (L2)
    (L2) edge (RL1);
\end{tikzpicture}
\caption{}
\end{subfigure}
\caption{Variables shaded in gray are incomplete. With $S := L$ and $\{U_1, U_2\}$ latent: (a) each red dotted edge is a VMAR-violating path; and (b) the red dotted path $L_1 \gets U_1 \to L_2 \gets U_2 \to R_{L_1}$ is a VMAR-violating path. (c) is the latent projection (Section \ref{sec:connections_other_lit}) of (b) with $U_1, U_2$ marginalized out.}\label{fig:VMAR_VMNAR_examples}
\end{figure}

%% file: figures/mnar_implication_map.tex
\begin{figure}[!htbp]
\centering
\begin{subfigure}[t]{\textwidth}
\centering
\begin{tikzpicture}[imap,
  free/.style={},           %
  assume/.style={dashed}]   %

\node (mnar)   {MNAR($S$)};
\node (vl)     [left=of mnar, dep] {Variable-level dependence};
\node (mvp)    [below=of vl] {MAR-violating path};
\node (mc)     at (mnar |- mvp) {VMAR-violating path};
\node (pmar)   [right=2.2cm of mnar] {$\pMAR{S}(s) \ne p(s)$};

\draw[->, assume] ([yshift= 3pt]mnar.west) -- node[above, lab] {(1) weak int.} ([yshift= 3pt]vl.east);
\draw[->, free]   ([yshift=-3pt]vl.east)   -- ([yshift=-3pt]mnar.west);

\draw[->, free]   ([xshift= 3pt]vl.south)  -- ([xshift= 3pt]mvp.north);
\draw[->, assume] ([xshift=-3pt]mvp.north) -- node[left, lab] {(2) c-faith.} ([xshift=-3pt]vl.south);

\draw[->, free]   ([yshift= 3pt]mvp.east) -- ([yshift= 3pt]mc.west);
\draw[->, assume] ([yshift=-3pt]mc.west)  -- node[below, lab, align=center] {(3) CSCI \\ restr.} ([yshift=-3pt]mvp.east);

\draw[->, assume] ([yshift= 3pt]mnar.east) -- node[above, lab] {(4) no canc.} ([yshift= 3pt]pmar.west);
\draw[->, free]   ([yshift=-3pt]pmar.west) -- ([yshift=-3pt]mnar.east);

\draw[->, free]   (mnar.south) -- (mc.north);

\end{tikzpicture}
\end{subfigure}

\vspace{2em}

\begin{subfigure}[t]{\textwidth}
\centering
\begin{tikzpicture}[imap,
  free/.style={},           %
  assume/.style={dashed}]   %

\node (mar)    {MAR($S$)};
\node (vl)     [left=of mar, dep] {No variable-level dependence};
\node (mvp)    [below=of vl] {No MAR-violating path};
\node (mc)     at (mar |- mvp) {No VMAR-violating path};
\node (pmar)   [right=2.2cm of mar] {$\pMAR{S}(s) = p(s)$};

\draw[<-, assume] ([yshift= 3pt]mar.west) -- node[above, lab] {(1) weak int.} ([yshift= 3pt]vl.east);
\draw[<-, free]   ([yshift=-3pt]vl.east)  -- ([yshift=-3pt]mar.west);

\draw[<-, free]   ([xshift= 3pt]vl.south)  -- ([xshift= 3pt]mvp.north);
\draw[<-, assume] ([xshift=-3pt]mvp.north) -- node[left, lab] {(2) c-faith.} ([xshift=-3pt]vl.south);

\draw[<-, free]   ([yshift= 3pt]mvp.east) -- ([yshift= 3pt]mc.west);
\draw[<-, assume] ([yshift=-3pt]mc.west)  -- node[below, lab, align=center] {(3) CSCI \\ restr.} ([yshift=-3pt]mvp.east);

\draw[<-, assume] ([yshift= 3pt]mar.east)  -- node[above, lab] {(4) no canc.} ([yshift= 3pt]pmar.west);
\draw[<-, free]   ([yshift=-3pt]pmar.west) -- ([yshift=-3pt]mar.east);

\draw[->, free]   (mc.north) -- (mar.south);

\end{tikzpicture}
\end{subfigure}
\caption{Implications among MNAR($S$) and related conditions (top) and their negations (bottom). Numbers on edges correspond to parts of Theorem \ref{thm:mnar_equivalent}. Dashed edges require the indicated assumptions: weak int. = weak intersection (Assumption \ref{ass:weak_intersection}); c-faith. = c-faithfulness (Definition \ref{def:c_faithfulness}); CSCI restr. = no CSCI in missingness contexts (Assumption \ref{ass:no_csci}); no canc. = no perfect cancellations in the MAR law (Supplement, Assumption \ref{supp-ass:no_mar_law_cancellations}). Solid edges hold in every compatible SCM obeying complete-case positivity. If $Z$ is a set of complete auxiliary variables and $S := V \cup Z$, then $\pMAR{S}(s) \ne p(s)$ can be replaced by $\pMAR{V \cup Z}(v) \ne p(v)$ in the top panel and $\pMAR{S}(s) = p(s)$ be replaced by $\pMAR{V \cup Z}(v) = p(v)$ in the bottom panel (Corollary \ref{coro:aux_equivalent}); the dashed edge (4) then also requires no perfect cancellations over the auxiliaries (Supplement, Assumption \ref{supp-ass:no_margin_cancellations}).}\label{fig:mnar_implication_map} 
\end{figure}

%% file: figures/tikz_admgs_aux_sets.tex
\begin{figure}[htbp!]
\centering
\begin{tikzpicture}[baseline=(current bounding box.center),
                    node distance=10mm and 10mm]
    \sansmath
    \tikzset{
        line width=1.2pt,
        ell/.style={draw, fill=white, inner sep=2pt, line width=1.5pt},
        every path/.style={>=Stealth, line width=1.2pt, >={Stealth[scale=\arrowheadScale]}},
        bidirected/.style={Latex-Latex},
        swig vsplit={gap=3pt, line color right=\bidirectedEdgeColor}
    }
    \node[name=V1, ell, shape=ellipse] {$V_1$};
    \node[name=W1, ell, shape=ellipse, right=of V1] {$W_1$};
    \node[name=U1, ell, shape=ellipse, right=of W1] {$U_1$};
    \node[name=V2, ell, shape=ellipse, right=of U1, fill={rgb:black,1;white,2}] {$V_2$};
    \node[name=RV2, ell, shape=ellipse, below=14mm of V2] {$R_{V_2}$};
    \node[name=W3, ell, shape=ellipse] at ($(V2)!0.5!(RV2) + (6mm,0)$) {$W_2$};
    \node[name=U2, ell, shape=ellipse] at ($(W1)!0.5!(RV2)$) {$U_2$};
    \draw[->, dotted, line width=1.2pt, >=stealth, color=red]
        (U1) edge (W1)
        (U2) edge (W1)
        (U1) edge (V2)
        (U2) edge (RV2);
    \draw[->, line width=1.2pt, >=stealth, color=\directedEdgeColor]
        (W3) edge (V2)
        (W3) edge (RV2);
    \draw[->,line width=1.2pt,>=stealth,color=\directedEdgeColor]
    (V1) edge[out=20,in=160] (V2)
    (V1) edge[out=320,in=180] (RV2);
\end{tikzpicture}
\caption{\label{fig:suff_aux_set} A DAG $G$ illustrating Theorem \ref{thm:suff_set_general}, with target variables $V := \{ V_1, V_2 \}$ and candidate auxiliary variables $W := \{ W_1, W_2 \}$. Gray shading: incomplete variables. Red dotted path: VMAR-violating path when including $W_1$. $U_1$ and $U_2$ are latent.}
\end{figure}

%% file: figures/applied_example_dags.tex
\begin{figure}[!htbp]
\centering
\resizebox{.55\linewidth}{!}{%
\begin{tikzpicture}
\tikzset{
    line width=1.2pt,
    ell/.style={draw, fill=white, inner sep=2pt, line width=1.5pt, shape=ellipse},
    every path/.style={>=Stealth, line width=1.2pt, >={Stealth[scale=1.4]}},
    swig vsplit={gap=3pt, line color right=red}
}
\node[name=C,  ell] {Sex};
\node[name=A,  ell, right=10mm of C] {$\text{ID}_1$};
\node[name=U1, ell, above right=10mm and 10mm of A] {$U_1$};
\node[name=U2, ell, below right=10mm and 10mm of A] {$U_2$};
\path let \p1 = (U1), \p2 = (U2) in
    coordinate (MidU) at ($(\p1)!.5!(\p2)$);
\node[name=D,  ell, right=10mm of MidU] {Income};
\node[name=Y1, ell, fill={rgb:black,1;white,2}, right=10mm of D] {$\text{ID}_2$};
\path let \p1 = (Y1), \p2 = (U2) in
    coordinate (RYcoord) at (\x1, \y2);
\node[name=RY, ell, at=(RYcoord)] {$R_{\text{ID}_2}$};
\draw[->, dotted, color=red]
    (U1) edge (D)
    (U2) edge (D)
    (U1) edge (Y1)
    (U2) edge (RY); 
\draw[->, color=\directedEdgeColor]
    (A)  edge (U1)
    (A)  edge (U2);
\draw[->, color=\directedEdgeColor]    (A)  to[out=330, in=210] (Y1);
\draw[->, color=\directedEdgeColor] (C) to[out=50,  in=180] (U1);
\draw[->, color=\directedEdgeColor] (C) to[out=310, in=180] (U2);
\end{tikzpicture}
}
\caption{\label{fig:applied_example_dag}
DAG $G$ representing the applied example. Gray shading: Incomplete variables. Red dotted path: VMAR-violating path when including income. $\text{ID}_1$: Sexual identity at baseline (exposure; $V_1$). $\text{ID}_2$: Sexual identity at follow-up (outcome; $V_2$). $U_1$: Stigmatization. $U_2$: Parental leave. Sex and income are complete candidate auxiliary variables; $U_1$ and $U_2$ are latent.}
\end{figure}

%% file: figures/results_index_table.tex
\begin{table}[H]
\centering
\singlespacing
\footnotesize
\setlength{\tabcolsep}{4pt}
\renewcommand{\arraystretch}{1.1}
\begin{tabular}{@{}>{\raggedright\arraybackslash}p{2.0cm}>{\raggedright\arraybackslash}p{7.6cm}>{\raggedright\arraybackslash}p{5.0cm}@{}}
\toprule
\textbf{Result} & \textbf{Description} & \textbf{Relies on} \\
\midrule

\multicolumn{3}{@{}l}{\textbf{Assumptions}} \\[1pt]

Ass.\ \ref{ass:cc_positivity}
& Complete-case positivity
& --- \\

Ass.\ \ref{ass:no_margin_cancellations}
& No perfect cancellations over the auxiliaries
& --- \\

Ass.\ \ref{ass:no_mar_law_cancellations}
& No perfect cancellations in the MAR law
& Def.\ \ref{main-def:eMAR} \\

Ass.\ \ref{ass:weak_intersection}
& Weak intersection
& --- \\

Ass.\ \ref{main-ass:no_csci}
& No CSCI in missingness contexts
& Def.\ \ref{main-def:pruned_DAG} \\

\midrule
\multicolumn{3}{@{}l}{\textbf{Definitions}} \\[1pt]

Def.\ \ref{main-def:eMAR}
& MAR and MNAR
& --- \\

Def.\ \ref{main-def:c_inactive}
& c-inactivity
& --- \\

Def.\ \ref{main-def:pruned_DAG}
& Pruned DAG $G[d]$
& Def.\ \ref{main-def:c_inactive} \\

Def.\ \ref{main-def:c_faithfulness}
& c-faithfulness
& Def.\ \ref{main-def:pruned_DAG} \\

Def.\ \ref{main-def:VMAR}
& VMAR, VMNAR, and VMAR-violating paths
& --- \\

Def.\ \ref{main-def:mar_violating_path}
& MAR-violating paths
& Def.\ \ref{main-def:pruned_DAG} \\

Def.\ \ref{main-def:variable_dependence}
& Variable-level dependence
& --- \\

Def.\ \ref{main-def:suff_aux_set}
& Sufficient auxiliary set
& Def.\ \ref{main-def:VMAR} \\

Def.\ \ref{def:minimal_vmar_path}
& Minimal VMAR-violating paths
& Def.\ \ref{main-def:VMAR} \\

Def.\ \ref{def:projection}
& Latent projection
& --- \\

\midrule
\multicolumn{3}{@{}l}{\textbf{Supporting results}} \\[1pt]

Lem.\ \ref{lm:pruning_soundness}
& Markov property with c-inactivity
& Def.\ \ref{main-def:c_inactive}, \ref{main-def:pruned_DAG} \\

Lem.\ \ref{lm:product_support_implies_weak_intersection}
& Selecting a single member (product support implies weak intersection)
& --- \\

Lem.\ \ref{lm:general_path_truncation_helper}
& Path truncation helper
& --- \\

Lem.\ \ref{lm:constructing_a_dsep_set}
& Conditional d-separating set among a subset of nodes
& --- \\

Prop.\ \ref{prop:mar_fnl_correct}
& Correctness of the MAR-assuming functional
& Ass.\ \ref{ass:cc_positivity}, \ref{ass:no_mar_law_cancellations} (last statement only); Def.\ \ref{main-def:eMAR} \\

Prop.\ \ref{prop:minimal_vmar_paths}
& Minimal VMAR-violating paths in the latent projection
& Def.\ \ref{main-def:VMAR}, \ref{def:projection}; Lem.\ \ref{lm:general_path_truncation_helper} \\

\midrule
\multicolumn{3}{@{}l}{\textbf{Main results}} \\[1pt]

Thm.\ \ref{thm:mnar_variable_level}
& Variable-level dependence $\Rightarrow$ MNAR($S$); converse under weak intersection
& Ass.\ \ref{ass:weak_intersection} (converse); Def.\ \ref{main-def:eMAR}, \ref{main-def:variable_dependence} \\

Thm.\ \ref{thm:mnar_mechanistic}
& Variable-level dependence $\Rightarrow$ MAR-violating path; converse under c-faithfulness
& Def.\ \ref{main-def:c_faithfulness} (converse), \ref{main-def:mar_violating_path}, \ref{main-def:variable_dependence}; Lem.\ \ref{lm:pruning_soundness} \\

Thm.\ \ref{thm:mnar_paths}
& MAR-violating path $\Rightarrow$ VMAR-violating path; converse under no CSCI in missingness contexts
& Ass.\ \ref{main-ass:no_csci} (converse); Def.\ \ref{main-def:VMAR}, \ref{main-def:mar_violating_path} \\

Thm.\ \ref{main-thm:mnar_equivalent}(4)
& MAR-assuming functional incorrect $\Rightarrow$ MNAR($S$); converse under no perfect cancellations in the MAR law
& Ass.\ \ref{ass:cc_positivity}, \ref{ass:no_mar_law_cancellations} (converse); Prop.\ \ref{prop:mar_fnl_correct} \\

Lem.\ \ref{main-coro:incomplete_vars_cant_fix_mnar}
& Incomplete variables cannot restore VMAR
& Def.\ \ref{main-def:VMAR} \\

Lem.\ \ref{lm:margin_correct}
& Correctness for the target law
& Ass.\ \ref{ass:no_margin_cancellations} \\

Coro.\ \ref{main-coro:aux_equivalent}
& MAR-assuming identification with complete auxiliaries
& Ass.\ \ref{main-ass:no_csci}; Def.\ \ref{main-def:c_faithfulness}; Lem.\ \ref{lm:margin_correct}; Thm.\ \ref{thm:mnar_variable_level} (forward only), \ref{thm:mnar_mechanistic}, \ref{thm:mnar_paths}, \ref{main-thm:mnar_equivalent}(4) \\

Thm.\ \ref{thm:suff_set_general}
& A sufficient auxiliary set for partially unknown DAGs
& Def.\ \ref{main-def:suff_aux_set}; Lem.\ \ref{main-coro:incomplete_vars_cant_fix_mnar}, \ref{lm:constructing_a_dsep_set} \\

\bottomrule
\end{tabular}
\caption{\label{tb:results_index} Index of all theoretical results and their dependencies, i.e., other results or assumptions used to establish the result in the first column. ``Converse'' marks an assumption needed only for the converse direction.
}
\end{table}

%% file: figures/proof_dependency_flowchart.tex
 
\ifx\flowchip\undefined
\DeclareRobustCommand{\flowchip}[1]{%
  \raisebox{-0.2ex}{\mbox{\setlength{\fboxsep}{0pt}\setlength{\fboxrule}{0.6pt}%
    \fcolorbox{black}{#1}{\rule{0pt}{6pt}\rule{6pt}{0pt}}}}}
\fi
 
\begin{figure}[!ht]
\centering
 
\colorlet{assText}{blue!70!black}   %
\newcommand{\deplab}[1]{{\fontsize{15.5}{18}\selectfont\textbf{#1}}}
\newcommand{\depsub}[1]{\\[2.5pt]{\fontsize{14}{16.5}\selectfont #1}}
\pgfmathsetmacro{\flowTipScale}{2.79*\arrowheadScale}
 
\resizebox{\textwidth}{!}{%
\begin{tikzpicture}[
  auto=false, node distance=0pt, every path/.style={},
  box/.style={draw, rounded corners=2pt, text width=4.8cm, align=center,
              inner sep=5pt, minimum height=2.4cm,
              execute at begin node={\parindent=0pt \parskip=0pt
                \hyphenpenalty=10000 \exhyphenpenalty=0}},
  supp/.style={box, line width=0.9pt, fill=white},          %
  main/.style={box, line width=3pt, fill=white},            %
  ass/.style={box, line width=0.9pt, fill=white, draw=assText, text=assText,
              text width=4.2cm, minimum height=1.3cm},
  assM/.style={ass, line width=3pt},                        %
  arr/.style={draw, -{Stealth[scale=\flowTipScale]}, line width=0.7pt, shorten >=1pt, rounded corners=5pt},
  aarr/.style={arr, draw=assText},
  darr/.style={arr, dashed},
  daarr/.style={aarr, dashed},
]
 
\def\aA{0}  \def\aB{5.0} \def\aC{10.0} \def\aD{17.0} \def\aE{22.5} \def\aF{28.5}
\def\yz{0} \def\yo{-3.9} \def\yt{-7.8} \def\yth{-12.0} \def\yfo{-16.2}
\def\xTfour{3.0} \def\xLSone{13.5} \def\xCoro{15.75}
 
\node[assM] (A0i) at (\aA,\yz)
  {\deplab{Assumption \ref{ass:cc_positivity}}\depsub{Complete-case positivity}};
 
\node[ass] (A0iii) at (\aB,\yz)
  {\deplab{Assumption \ref{ass:no_mar_law_cancellations}}\depsub{No perfect cancellations in the MAR law}};
 
\node[assM] (A1) at (\aC,\yz)
  {\deplab{Assumption \ref{ass:weak_intersection}}\depsub{Weak intersection}};
 
\node[assM] (CF) at (\aD,\yz)
  {\deplab{Definition \ref{main-def:c_faithfulness}}\depsub{c-faithfulness}};
 
\node[assM] (A2) at (\aE,\yz)
  {\deplab{Assumption \ref{main-ass:no_csci}}\depsub{No CSCI in missingness contexts}};
 
\node[ass] (A0ii) at (\aF,\yz)
  {\deplab{Assumption \ref{ass:no_margin_cancellations}}\depsub{No perfect cancellations over the auxiliaries}};
 
\node[supp] (PS1) at (\aA,\yo)
  {\deplab{Proposition \ref{prop:mar_fnl_correct}}%
   \depsub{Correctness of the MAR-assuming functional}};
 
\node[supp] (LS1) at (\xLSone,\yo)
  {\deplab{Lemma \ref{lm:pruning_soundness}}%
   \depsub{Markov property with c-inactivity}};
 
\node[main] (T4) at (\xTfour,\yt)
  {\deplab{Theorem \ref{main-thm:mnar_equivalent}(4)}%
   \depsub{MNAR($S$) $\Leftrightarrow$ MAR-assuming functional incorrect}};
 
\node[main] (T1) at (\aC,\yt)
  {\deplab{Theorem \ref{thm:mnar_variable_level}}%
   \depsub{Variable-level dependence $\Leftrightarrow$ MNAR($S$)}};
 
\node[main] (T2) at (\aD,\yt)
  {\deplab{Theorem \ref{thm:mnar_mechanistic}}%
   \depsub{MAR-violating path $\Leftrightarrow$ variable-level dependence}};
 
\node[main] (T3) at (\aE,\yt)
  {\deplab{Theorem \ref{thm:mnar_paths}}%
   \depsub{MAR-violating path $\Leftrightarrow$ VMAR-violating path}};
 
\node[main] (L1) at (\aF,\yt)
  {\deplab{Lemma \ref{lm:margin_correct}}%
   \depsub{Correctness for the target law}};
 
\node[main] (C1) at (\xCoro,\yth)
  {\deplab{Corollary \ref{main-coro:aux_equivalent}}%
   \depsub{MAR-assuming identification with complete auxiliaries}};
 
\node[supp] (LS3) at (\xTfour,\yfo)
  {\deplab{Lemma \ref{lm:general_path_truncation_helper}}%
   \depsub{Path truncation helper}};
 
\node[supp] (PS2) at (\aC,\yfo)
  {\deplab{Proposition \ref{prop:minimal_vmar_paths}}%
   \depsub{Minimal VMAR-violating paths in the latent projection}};
 
\node[supp] (LS4) at (\aE,\yfo)
  {\deplab{Lemma \ref{lm:constructing_a_dsep_set}}%
   \depsub{Conditional d-separating set among a subset of nodes}};
 
\node[main] (LV) at (16.25,\yfo)
  {\deplab{Lemma \ref{main-coro:incomplete_vars_cant_fix_mnar}}%
   \depsub{Incomplete variables cannot restore VMAR}};
 
\node[main] (TH2) at (\aF,\yfo)
  {\deplab{Theorem \ref{thm:suff_set_general}}%
   \depsub{A sufficient auxiliary set for partially unknown DAGs}};
 
\draw[aarr] (A0i.south)   -- (PS1.north);
\draw[daarr] (A0iii.south) -- (A0iii.south |- T4.north);
\draw[daarr] (A1.south)    -- (T1.north);
\draw[daarr] (CF.south)    -- (T2.north);
\draw[daarr] (A2.south)    -- (T3.north);
\draw[daarr] (A0ii.south)  -- (L1.north);
 
\draw[arr] (PS1.south) -- (T4.north west);
\draw[darr] (LS1.south) -- (T2.north west);
 
\draw[arr]  (T4.south)       |- (C1.west);
\draw[darr] (T1.south)       -- (C1.north west);
\draw[darr] (T2.south)       -- (T2.south |- C1.north);
\draw[darr] (T3.south)       -- (C1.north east);
\draw[arr]  (L1.south)       |- (C1.east);
 
\draw[darr] (LS3.east) -- (PS2.west);
\draw[arr] (LS4.east) -- (TH2.west);
\draw[arr] (LV.south) -- ++(0,-0.9) -| (TH2.south);
 
\end{tikzpicture}%
}

\caption[Dependency structure of the results]{Dependency structure of the theoretical results. Each edge indicates that the child relies on the parent; only direct dependences are drawn. Dashed edges: The statement in the child node is an equivalence, and only one direction depends on the parent at the tail of the dashed edge. Thick borders mark results and assumptions stated in the main text; thin borders mark those appearing only in the Supplement. Blue boxes are assumptions.}\label{fig:proof_dependency_flowchart} %
\end{figure}

%% file: figures/mnar_implication_map_with_mealli.tex
\begin{figure}[!htbp]
\centering
\begin{tikzpicture}[imap,
  free/.style={},           %
  assume/.style={dashed}]   %

\node (mnar)   {MNAR($S$)};
\node (vl)     [left=of mnar, dep] {Variable-level dependence};
\node (mvp)    [below=of vl] {MAR-violating path};
\node (mc)     [below=of mvp] {VMAR-violating path};
\node (pmar)   [right=2.2cm of mnar] {$\pMAR{S}(s) \ne p(s)$};
\node (joint)  at (mnar |- mc) [mealliN] {$S^- \notindM R(S) \mid S^+$};
\node (mealli) at ($(mnar |- mvp) + (5cm, 0)$) [mealliN, align=center, text width=4cm] {$\exists \, S^* \in S^-, R^* \in R(S)$: \\ $S^* \notindM R^* \mid S \setminus \{S^*\}$};

\draw[->, assume] ([yshift= 3pt]mnar.west) -- node[above, lab] {(1) weak int.} ([yshift= 3pt]vl.east);
\draw[->, free]   ([yshift=-3pt]vl.east)   -- ([yshift=-3pt]mnar.west);

\draw[->, free]   ([xshift= 3pt]vl.south)  -- ([xshift= 3pt]mvp.north);
\draw[->, assume] ([xshift=-3pt]mvp.north) -- node[left, lab] {(2) c-faith.} ([xshift=-3pt]vl.south);

\draw[->, free]   ([xshift= 3pt]mvp.south) -- ([xshift= 3pt]mc.north);
\draw[->, assume] ([xshift=-3pt]mc.north)  -- node[left, lab, align=right] {(3) CSCI \\ restr.} ([xshift=-3pt]mvp.south);

\draw[->, assume] ([yshift= 3pt]mnar.east) -- node[above, lab] {(4) no canc.} ([yshift= 3pt]pmar.west);
\draw[->, free]   ([yshift=-3pt]pmar.west) -- ([yshift=-3pt]mnar.east);

\draw[->, free]   (mnar.south west) -- (mc.north east);

\draw[->, mealli, free]   ([yshift= 3pt]joint.west) -- ([yshift= 3pt]mc.east);
\draw[->, mealli, assume] ([yshift=-3pt]mc.east)    -- node[below, lab, mealliN] {std.\ faith.} ([yshift=-3pt]joint.west);

\draw[->, mealli, free] (mnar) -- (joint);

\draw[->, mealli, free]   ([xshift=-2pt, yshift= 2pt]mealli.south west) -- ([xshift=-2pt, yshift= 2pt]joint.north east);
\draw[->, mealli, assume] ([xshift= 2pt, yshift=-2pt]joint.north east) to node[below, sloped, lab, mealliN] {weak int., indep.\ $R(S)$} ([xshift= 2pt, yshift=-2pt]mealli.south);

\draw[->, mealli, assume] (mealli.north west) -- node[above right, lab, mealliN] {indep.\ $R(S)$, pos.} (mnar.south east);

\draw[->, mealli, assume, rounded corners] (mc.south) -- ++(0,-0.8)
  -| node[pos=0.25, below, lab, mealliN] {std.\ faith.} ([xshift=0.6cm]mealli.east) -- (mealli.east);

\end{tikzpicture}
\caption{\label{fig:mnar_implication_map_with_mealli} Implications among MNAR($S$) and related conditions, with a comparison to Mealli \& Rubin (2015).\autocite{mealli2015clarifying} Numbers on edges correspond to parts of Theorem \ref{main-thm:mnar_equivalent} in the main text. Dashed edges require the indicated assumptions: weak int. = weak intersection (Assumption \ref{ass:weak_intersection}); c-faith. = c-faithfulness (Definition \ref{main-def:c_faithfulness}); CSCI restr. = no CSCI in missingness contexts (Assumption \ref{main-ass:no_csci} in the main text); no canc. = no perfect cancellations in the MAR law (Assumption \ref{ass:no_mar_law_cancellations}). Solid edges hold in every compatible SCM obeying complete-case positivity. The diagonal black edge is the standard result that VMAR($S$) implies MAR($S$) in every compatible SCM (equivalently, MNAR($S$) implies that a VMAR-violating path exists); it is not part of Theorem \ref{main-thm:mnar_equivalent}. Gray: comparison with Theorem 1 of Mealli \& Rubin (2015)\autocite{mealli2015clarifying} (discussed in Section \ref{main-sec:connections_other_lit} of the main text), whose condition is the rightmost node. std.\ faith. = faithfulness of the distribution to $G$; indep.\ $R(S)$ = $R_i \indM (R(S) \setminus \{R_i\}) \mid S$ for all $i$; pos. = $p(R_{S^*} = 0 \mid s) > 0$ wherever the dependence occurs.}
\end{figure}

%% file: figures/table_simulation_dags.tex
\begin{table}[!htbp]
\centering
\resizebox{0.97\columnwidth}{!}{%
\begin{tabular}{@{}c c l l l@{}}
\toprule
 & \textbf{Latent projection} & \makecell[c]{\textbf{Target} \\ \textbf{variables } $V$} & \makecell[c]{\textbf{Auxiliary} \\ \textbf{variables} $Z$} & \makecell[c]{\textbf{VMAR-violating path between} \\ \textbf{$V^-$ and $R(V \cup Z)$}} \\
\midrule
(a) & \multirow{3}{*}{\raisebox{-0.5\height}{\input{figures/table1_dags/table1_dag_c}}}
    & \multirow{2}{*}{$\{A\}$} & $\emptyset$ & None \\[0.5em]
    & & & $\{B\}$ & $R_A \leftrightarrow B \leftrightarrow A$ \\[0.5em]
\cmidrule(l){3-5}
    & & $\{A, B\}$ & $\emptyset$ & $R_A \leftrightarrow B \leftrightarrow A$ \\
\addlinespace[1em]
\midrule
\addlinespace[1em]
(b) & \multirow{2}{*}{\raisebox{-0.5\height}{\input{figures/table1_dags/table1_dag_d}}}
    & $\{A\}$ & $\emptyset$ & None \\[0.5em]
\cmidrule(l){3-5}
    & & $\{A, B\}$ & $\emptyset$ & $R_A \leftrightarrow B$ \\
\addlinespace[1em]
\midrule
\addlinespace[1em]
(c) & \multirow{5}{*}{\raisebox{-0.5\height}{\input{figures/table1_dags/table1_dag_e}}}
    & \multirow{3}{*}{$\{B\}$} & $\emptyset$ & $R_B \gets A \to B$ \\[0.5em]
    & & & $\{A\}$ & None \\[0.5em]
    & & & $\{A, C\}$ & $R_B \leftrightarrow C \leftrightarrow B$ \\[0.5em]
\cmidrule(l){3-5}
    & & \multirow{2}{*}{$\{A, B\}$} & $\emptyset$ & None \\[0.5em]
    & & & $\{C\}$ & $R_B \leftrightarrow C \leftrightarrow B$ \\
\addlinespace[0.5em]
\bottomrule
\end{tabular}
}
\caption{\label{tb:simulation_dags}Latent projections of DAGs used in the simulation study, illustrating sufficient and insufficient choices of complete auxiliary variables. For compactness, associations caused by latent common causes ($U$) of variables are depicted as bidirected edges.}
\end{table}

%% file: figures/table1_dags/table1_dag_c.tex
\begin{tikzpicture}[baseline=(current bounding box.center)]
        \sansmath %
        \tikzset{
        line width=1.2pt,
        ell/.style={draw,fill=white, inner sep=2pt, line width=1.5pt},
        every path/.style={>=Stealth, line width=1.1pt, >={Stealth[scale=\arrowheadScale]}},
        swig vsplit={gap=3pt, line color right=red}
    };
        \node[name=A1, ell, shape=ellipse, fill={rgb:black,1;white,2}]{$A$};
        \node[name=B, ell, shape=ellipse, right=10mm of A1]{$B$};
        \node[name=RA, ell, shape=ellipse, below=5mm of A1]{$R_A$};
        \draw[->,line width=1.2pt,>=stealth, color=\bidirectedEdgeColor]
        (A1) edge (B)
        (B) edge (A1)
        (RA) edge (B)
        (B) edge (RA);
\end{tikzpicture}

%% file: figures/table1_dags/table1_dag_d.tex
\begin{tikzpicture}[baseline=(current bounding box.center)]
        \sansmath %
            \tikzset{
        line width=1.2pt,
        ell/.style={draw,fill=white, inner sep=2pt, line width=1.5pt},
        every path/.style={>=Stealth, line width=1.2pt, >={Stealth[scale=\arrowheadScale]}},
        swig vsplit={gap=3pt, line color right=red}
    };
        \node[name=A1, ell, shape=ellipse, fill={rgb:black,1;white,2}]{$A$};
        \node[name=B1, ell, shape=ellipse, right=10mm of A1, fill={rgb:black,1;white,2}]{$B$};
        \node[name=RA, ell, shape=ellipse, below=5mm of A1]{$R_A$};
        \node[name=RB, ell, shape=ellipse, below=5mm of B1]{$R_B$};
        \draw[->,line width=1.2pt,>=stealth, color=\bidirectedEdgeColor]
        (A1) edge (B1)
        (B1) edge (A1)
        (RA) edge (B1)
        (B1) edge (RA);
        \end{tikzpicture}

%% file: figures/table1_dags/table1_dag_e.tex
 \begin{tikzpicture}[baseline=(current bounding box.center)]
        \sansmath %
        \tikzset{
        line width=1.2pt,
        ell/.style={draw,fill=white, inner sep=2pt, line width=1.5pt},
        every path/.style={>=Stealth, line width=1.2pt, >={Stealth[scale=\arrowheadScale]}},
        swig vsplit={gap=3pt, line color right=red}
    };
        \node[name=A, ell, shape=ellipse]{$A$};
        \node[name=D, ell, shape=ellipse, right=10mm of A]{$C$};
        \node[name=B1, ell, shape=ellipse, right=10mm of D, fill={rgb:black,1;white,2}]{$B$};
        \node[name=RB, ell, shape=ellipse, below=5mm of B1]{$R_B$};
        \draw[->,line width=1.2pt,>=stealth, color=\bidirectedEdgeColor]
        (D) edge (B1)
        (B1) edge (D)
        (D) edge (RB)
        (RB) edge (D);
        \draw[->,line width=1.2pt,>=stealth, color=\directedEdgeColor]
        (A)[out=330, in=180] edge (RB)
        (A)[out=30, in=150] edge (B1);
    \end{tikzpicture}

%% file: figures/forest_plot_applied_example.tex
\begin{figure}[!htbp]
    \centering

    \begin{subfigure}{0.8\linewidth}
        \centering
        \caption{Among males}
        \label{fig:forest_male}
        \vspace{1.5em}
        \includegraphics[width=\linewidth]{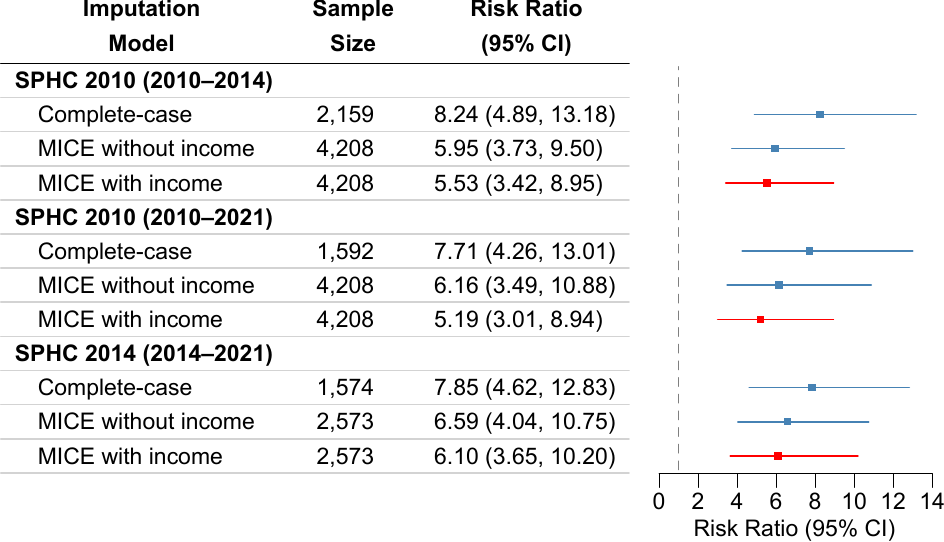}
    \end{subfigure}

    \vspace{1em}

    \begin{subfigure}{0.8\linewidth}
        \centering
        \label{fig:forest_female}
         \caption{Among females}
        \vspace{1.5em}
        \includegraphics[width=\linewidth]{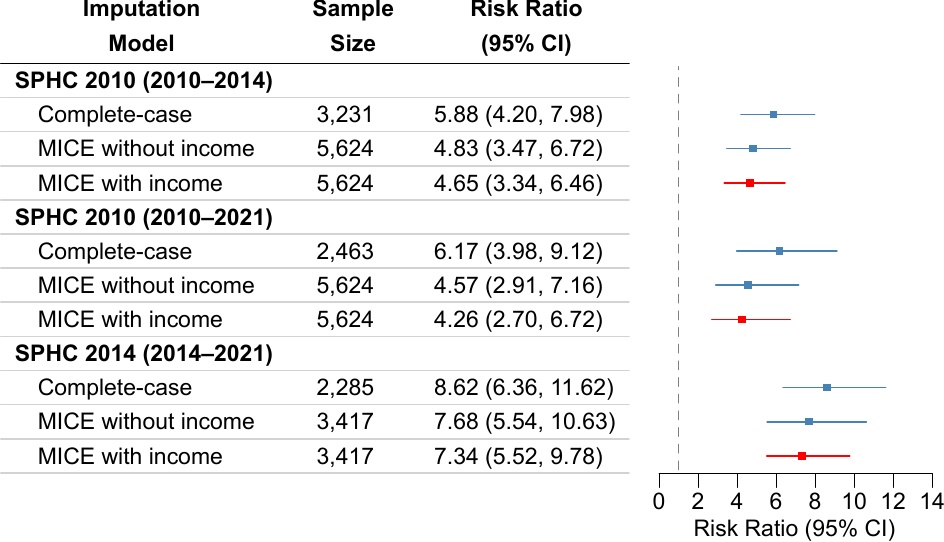}
    \end{subfigure}

    \caption{Estimated associations between non-heterosexual identity at baseline and change in sexual identity at follow-up, presented by cohort and imputation model and stratified by sex. The years in parentheses indicate the follow-up period for each cohort. MICE without income included age, country of birth, antidepressant use, and anxiolytic use as auxiliary variables, whereas MICE with income additionally included income. Abbreviations: CI, confidence interval; MICE, multiple imputation by chained equations; SPHC, Stockholm Public Health Cohort.}
    \label{fig:forest_plots}

\end{figure}